\documentclass[twocolumn,english,aps,prb,superscriptaddress,bibnotes,amsmath,amssymb,floatfix]{revtex4-1}
\usepackage[colorlinks=true,citecolor=blue,linkcolor=magenta]{hyperref}
\usepackage[utf8]{inputenc}
\usepackage[english]{babel}
\usepackage[T1]{fontenc}
\usepackage{textcomp}
\usepackage[markup=nocolor, authormarkupposition=left]{changes} 
\usepackage{soul}
\usepackage{url}
\usepackage{makecell}
\usepackage{booktabs}
\usepackage{upgreek}
\setaddedmarkup{{\color{red}#1}}
\usepackage{graphicx}
\usepackage{epstopdf}
\usepackage{threeparttable}
\graphicspath{{./figures/}}
\makeatletter
 
\newcommand{\Rmnum}[1]{\expandafter\@slowromancap\romannumeral #1@} 
\makeatother
\usepackage{multirow}

\pdfcatalog{/OpenAction [1 0 R /FitH null]}

\begin{document}
\title{A photonic integrated comb engine for ultracold quantum gases}

\author{Wei Sun}
\thanks{These authors contributed equally to this work.}
\affiliation{International Quantum Academy, Shenzhen 518048, China}

\author{Xiaoying Yan}
\thanks{These authors contributed equally to this work.}
\affiliation{International Quantum Academy, Shenzhen 518048, China}
\affiliation{Shenzhen Institute for Quantum Science and Engineering, Southern University of Science and Technology, Shenzhen 518055, China}

\author{Jinbao Long}
\thanks{These authors contributed equally to this work.}
\affiliation{International Quantum Academy, Shenzhen 518048, China}

\author{Sanli Huang}
\thanks{These authors contributed equally to this work.}
\affiliation{International Quantum Academy, Shenzhen 518048, China}
\affiliation{Hefei National Laboratory, University of Science and Technology of China, Hefei 230088, China}

\author{Zhixin Duan}
\thanks{These authors contributed equally to this work.}
\affiliation{International Quantum Academy, Shenzhen 518048, China}
\affiliation{Shenzhen Institute for Quantum Science and Engineering, Southern University of Science and Technology, Shenzhen 518055, China}

\author{Hao Tan}
\affiliation{International Quantum Academy, Shenzhen 518048, China}
\affiliation{Hefei National Laboratory, University of Science and Technology of China, Hefei 230088, China}

\author{Zhenyuan Shang}
\affiliation{International Quantum Academy, Shenzhen 518048, China}
\affiliation{Shenzhen Institute for Quantum Science and Engineering, Southern University of Science and Technology, Shenzhen 518055, China}

\author{Zeying Zhong}
\affiliation{International Quantum Academy, Shenzhen 518048, China}
\affiliation{Shenzhen Institute for Quantum Science and Engineering, Southern University of Science and Technology, Shenzhen 518055, China}

\author{Jiahao Sun}
\affiliation{International Quantum Academy, Shenzhen 518048, China}
\affiliation{Shenzhen Institute for Quantum Science and Engineering, Southern University of Science and Technology, Shenzhen 518055, China}

\author{Yue Hu}
\affiliation{International Quantum Academy, Shenzhen 518048, China}
\affiliation{Shenzhen Institute for Quantum Science and Engineering, Southern University of Science and Technology, Shenzhen 518055, China}

\author{Shichang Li}
\affiliation{International Quantum Academy, Shenzhen 518048, China}
\affiliation{Shenzhen Institute for Quantum Science and Engineering, Southern University of Science and Technology, Shenzhen 518055, China}

\author{Baoqi Shi}
\affiliation{International Quantum Academy, Shenzhen 518048, China}

\author{Yi-Han Luo}
\affiliation{International Quantum Academy, Shenzhen 518048, China}

\author{Shuyi Li}
\affiliation{International Quantum Academy, Shenzhen 518048, China}

\author{Chen Shen}
\affiliation{International Quantum Academy, Shenzhen 518048, China}
\affiliation{Qaleido Photonics, Shenzhen 518048, China}

\author{Hanqing Liu}
\affiliation{State Key Laboratory of Optoelectronic Materials and Devices, Institute of Semiconductors, Chinese Academy of Sciences, Beijing 100083, China}
\affiliation{Center of Materials Science and Optoelectronics Engineering, University of Chinese Academy of Sciences, Beijing 100083, China}

\author{Xiangbin Su}
\affiliation{State Key Laboratory of Optoelectronic Materials and Devices, Institute of Semiconductors, Chinese Academy of Sciences, Beijing 100083, China}
\affiliation{Center of Materials Science and Optoelectronics Engineering, University of Chinese Academy of Sciences, Beijing 100083, China}

\author{Haiqiao Ni}
\affiliation{State Key Laboratory of Optoelectronic Materials and Devices, Institute of Semiconductors, Chinese Academy of Sciences, Beijing 100083, China}
\affiliation{Center of Materials Science and Optoelectronics Engineering, University of Chinese Academy of Sciences, Beijing 100083, China}

\author{Zhichuan Niu}
\affiliation{International Quantum Academy, Shenzhen 518048, China}
\affiliation{State Key Laboratory of Optoelectronic Materials and Devices, Institute of Semiconductors, Chinese Academy of Sciences, Beijing 100083, China}
\affiliation{Center of Materials Science and Optoelectronics Engineering, University of Chinese Academy of Sciences, Beijing 100083, China}

\author{Shengjun Yang}
\affiliation{International Quantum Academy, Shenzhen 518048, China}
\affiliation{Shenzhen Institute for Quantum Science and Engineering, Southern University of Science and Technology, Shenzhen 518055, China}

\author{Junqiu Liu}
\email[]{liujq@iqasz.cn}
\affiliation{International Quantum Academy, Shenzhen 518048, China}
\affiliation{Hefei National Laboratory, University of Science and Technology of China, Hefei 230088, China}

\maketitle

\noindent\textbf{Cold atoms underpin quantum sensing, simulation and computation, but their coherent control demands highly stable optical fields whose generation, referencing and power scaling remain formidable integration challenges. 
While photonic integrated circuits have yielded compact visible lasers and high-$Q$ microresonators have enabled chip-scale optical frequency combs, these crucial technologies have largely remained functionally fragmented. 
Consequently, the coherent manipulation of ultracold quantum gases using a fully integrated laser-comb source has yet to be realized. 
Here we demonstrate a scalable, hybrid-integrated microcomb engine at 780~nm that seamlessly bridges frequency synthesis, atomic referencing and power amplification to achieve quantum state control of a Bose--Einstein condensate. 
By self-injection locking of electrically driven III--V lasers to high-$Q$ Si$_3$N$_4$ microresonators, we generate coherent platicon microcombs featuring 20- and 100-GHz mode spacings. 
Absolute referencing of the microcomb to an $^{85}$Rb transition actively suppresses long-term frequency drift from over 200 MHz to the 100-kHz level across 2,000 s. 
A selected comb tooth is subsequently injection-amplified to 102 mW, entirely preserving the microcomb's pristine coherence and stability. 
We utilize this synthesized field to construct an optical lattice, drive coherent two-photon Raman transitions, and prepare stationary spin--orbit-coupled and Raman-lattice states within an $^{87}$Rb condensate. 
By providing a synchronized optical grid, this atom-referenced microcomb can allow multiple optical-control channels to scale without a proportional multiplication of independent frequency references. 
Our work establishes a fully integrated frequency-synthesis architecture essential for realizing deployable, large-scale atomic quantum systems.
}

Cold atoms~\cite{Chu:98,Phillips:98,Cohen-Tannoudji:98} are central to quantum information science, offering precise energy levels, exceptional coherence times, robust environmental isolation, and high-fidelity optical control. 
The laser manipulation of these atoms drives a broad spectrum of quantum technologies.  
Cold-atom clocks and interferometers enable unparalleled precision in measuring time, inertial forces, gravity, and fundamental physical constants~\cite{Ludlow:15,Tino:14}. 
Ultracold quantum gases trapped in optical lattices or tweezer arrays can simulate programmable many-body Hamiltonians and topological phases~\cite{Bloch:12,Schafer:20}. 
Neutral atoms provide a scalable platform for quantum computation~\cite{Saffman:10}, while coherent atom--photon interfaces facilitate quantum memories, state transfer, and distributed quantum networking~\cite{Sangouard:11}. 

Translating these laboratory successes into field-deployable, networked, and spaceborne quantum systems requires overcoming substantial engineering hurdles. 
Although transportable optical lattice clocks and long-baseline atom interferometers push the boundaries of fundamental physics in demanding environments~\cite{Koller:17,Aveline:20,Abe:21}, scaling these applications escalates optical control into a severe system-engineering challenge. 
A practical cold-atom apparatus must reliably generate, stabilize, modulate, amplify, and route multiple mutually coherent optical fields~\cite{Hisai:19,Abe:21}. 
Conventional setups distribute these functions across macroscopic arrays of discrete lasers, reference cavities, and free-space optics.  This approach incurs prohibitive penalties in size, weight, power consumption, and cost (SWaP-C), and introduces acute vulnerabilities to alignment drift and mechanical vibration.

\begin{figure*}[t!]
\centering
\includegraphics[width=0.9\linewidth]{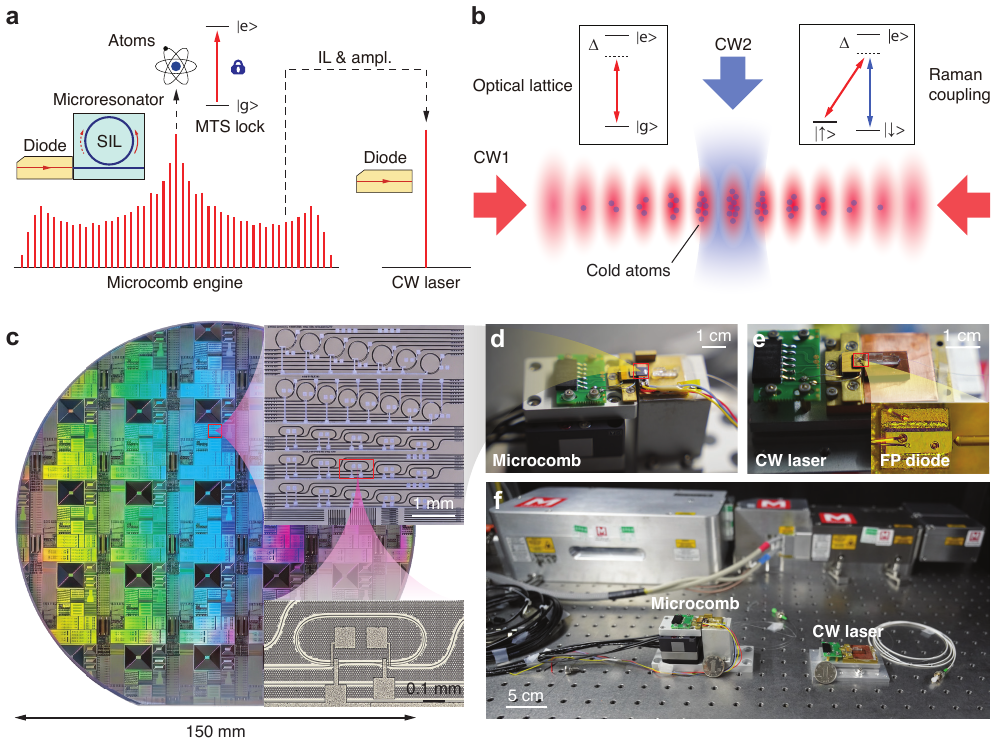}
\caption{
\footnotesize
\textbf{Integrated microcomb engine for cold-atom manipulation}.
\textbf{a,}
Generation of a platicon microcomb within a nonlinear microresonator edge-coupled to a laser diode. 
The continuous-wave (CW) pump is stabilized to an atomic transition $\lvert g\rangle\rightarrow\lvert e\rangle$ via a modulation-transfer spectroscopy (MTS) lock. 
A far-detuned comb line is filtered and amplified via injection locking (IL) into a secondary laser diode, creating a high-power, narrow-linewidth CW laser for downstream cold-atom manipulation.  
\textbf{b,}
Configuration for cold-atom manipulation. 
Counter-propagating beams (CW1) from the high-power diode form an optical lattice (left inset). 
A frequency-shifted secondary beam (CW2) from the same diode intersects CW1 to drive two-photon Raman coupling between the magnetic spin-up ($\lvert\uparrow\rangle$) and spin-down ($\lvert\downarrow\rangle$) states (right inset), where $\Delta$ represents the laser detuning from the atomic transition. 
\textbf{c,}
Wafer-scale fabrication of Si$_3$N$_4$ microresonators on a 150-mm substrate. 
Insets display an individual chip (top) and a single microresonator (bottom). 
\textbf{d,}
Photograph of the packaged microcomb module. 
\textbf{e,}
Photograph of the packaged CW-laser module. 
Inset shows the enclosed FP laser diode.
FP, Fabry--P\'erot.
\textbf{f,}
Size, weight, and complexity comparison of our packaged modules against a standard commercial Ti:sapphire laser commonly used in cold-atom experiments.
}
\label{Fig:1}
\end{figure*}

\begin{figure*}[t!]
\centering
\includegraphics[width=0.9\linewidth]{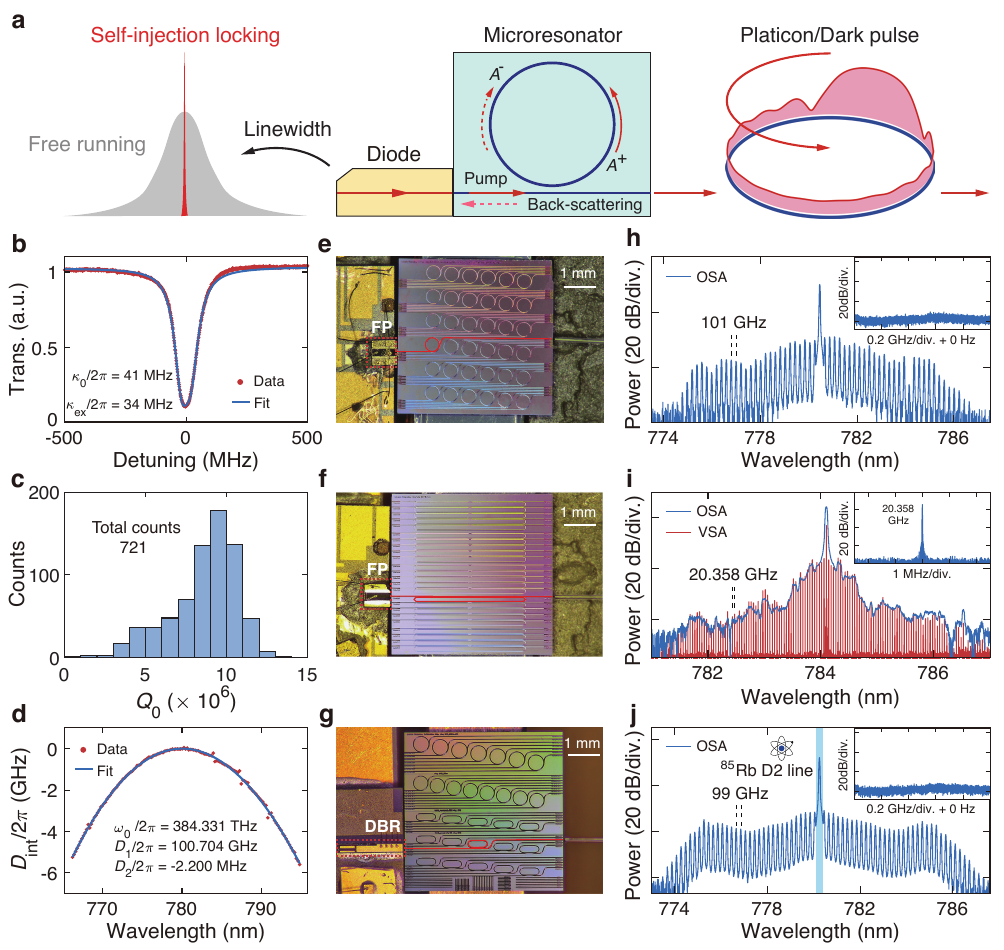}
\caption{
\footnotesize
\textbf{Microresonator characterization and platicon microcomb generation}.
\textbf{a,}
Self-injection locking (SIL) and platicon microcomb generation in the hybrid diode--microresonator chip.
$A^+$ is the forward propagating light in the microresonator, coupled from the diode pump.
Backscattering light, $A^-$, is induced and propagates backward into the diode, locking the diode laser frequency to the microresonator resonance and narrowing its linewidth.
With sufficient pump power, a platicon/dark pulse is generated in the microresonator via Kerr nonlinearity.
\textbf{b,}
Typical resonance profile of the high-$Q$ Si$_3$N$_4$ microresonators in the 780 nm band.  
Lorentzian fitting yields an intrinsic loss of $\kappa_0/2\pi = 41$ MHz and an external coupling rate of $\kappa_\mathrm{ex}/2\pi = 34$ MHz, corresponding to an intrinsic quality factor of $Q_0\approx9.4 \times 10^6$.
\textbf{c,}
Histogram of $Q_0$ values of 721 analyzed resonances, demonstrating a most probable value of $Q_0\approx9.5 \times 10^6$. 
\textbf{d,}
Typical integrated dispersion, $D_\mathrm{int}$, profile of the Si$_3$N$_4$ microresonators.
Fitting at $\omega_0/2\pi = 384.331$ THz (780.035 nm) extracts a free spectral range (FSR) of $D_1/2\pi=100.704$ GHz and normal group-velocity dispersion (GVD) of $D_2/2\pi=-2.200$ MHz, supporting platicon generation. 
\textbf{e--g,}
Hybrid-integration configurations for platicon microcomb generation via laser SIL.
An FP laser drives 101- and 20-GHz-FSR microresonators (panels \textbf{e, f}), while a DBR laser drives a 99-GHz-FSR microresonator (panel \textbf{g}). 
DBR, distributed-Bragg-reflector.
\textbf{h--j,}
Optical spectra of the platicon corresponding to panels \textbf{e--g}.
Spectra are measured using a commercial OSA (blue data) and our VSA (red data in panel \textbf{i})~\cite{Shi:25}.
Insets display electrical spectra measured by photodetecting the platicon pulse trains. 
The narrow beatnote centered at 20.358 GHz (panel \textbf{i}) and the absence of low-frequency amplitude noise (panels \textbf{h, j}) verify the intrinsic coherence of the platicons.
The blue-shaded region in panel \textbf{j} indicates the pump line locked to the $^{85}$Rb D2 transition for subsequent cold-atom experiments.
OSA, optical spectrum analyzer.
VSA, vector spectrum analyzer.
}
\label{Fig:2}
\end{figure*}

Photonic integrated circuits (PICs)~\cite{Thomson:16,Dutt:24} offer a compelling pathway to co-design these complex optical functions on a manufacturable, mechanically robust platform~\cite{Kitching:18,Newman:19}.  
While PIC technologies are firmly established at telecommunication wavelengths, their extension into the visible and near-infrared spectrum remains functionally fragmented~\cite{Tran:22,LuX:24,ZhangZ:23}. 
Current integrated laser--resonator sources exist largely as standalone device demonstrations~\cite{Castro:25,Isichenko:24,Prokoshin:24,Franken:21, Corato-Zanarella:23, Siddharth:25, Franken:25, Yu:19, LiuP:25, Loh:25,Chauhan:26,  Isichenko:26}, 
and atom-facing photonic circuits typically rely on externally generated light~\cite{Kohnen:11,Niffenegger:20,Mehta:20,Isichenko:23,ZhouX:24,Menon:24,Kodigala:24,Snijders:25,Christen:25}.  
Advanced coherent cold-atom control---such as the formation of optical lattices, atom interference, qubit rotation, and Hamiltonian engineering---has yet to be driven by fully chip-integrated laser sources. 
This level of control demands an exacting combination of spectral purity, wavelength stability, coherent frequency synthesis, power scalability, and low-noise operation. 
Consequently, an end-to-end platform bridging device fabrication and integration with frequency synthesis, power amplification, and coherent atomic operations remains a highly sought-after milestone.

Here, we demonstrate a photonic integrated frequency-comb engine operating in the 780 nm band that disciplines high-power laser diodes for coherent manipulation of ultracold quantum gases. 
Central to this engine is a Kerr-microresonator-based optical frequency comb (microcomb)~\cite{Kippenberg:18, Diddams:20} comprising a laser diode and a high-$Q$ silicon nitride (Si$_3$N$_4$) microresonator, both fabricated using mature III-V and CMOS foundry processes. 
By edge-coupling the diode to the microresonator and leveraging laser self-injection locking (SIL)~\cite{Kondratiev:23, Liang:15, Sun:25}, we generate a microcomb centered at 780 nm (Fig.~\ref{Fig:1}a).  
This microcomb is referenced to a rubidium (Rb) hyperfine transition via modulation-transfer spectroscopy (MTS). 
By isolating a selected comb line and injection-locking a secondary diode, we amplify the continuous-wave (CW) signal to 102 mW while perfectly preserving the comb's intrinsic coherence and stability. 
We subsequently deploy this amplified laser to construct optical lattices and drive two-photon Raman coupling within a Bose--Einstein condensate (BEC), successfully establishing a Raman-lattice configuration (Fig.~\ref{Fig:1}b). 
Packaged with essential electronic and optical routing components (Fig.~\ref{Fig:1}d, e), our integrated modules deliver distinct SWaP-C advantages over commercial Ti:sapphire lasers (Fig.~\ref{Fig:1}f) conventionally used in cold-atom platforms.

\noindent \textbf{Microresonator characterization.} 
The Si$_3$N$_4$ microresonator chips are fabricated using a CMOS-compatible foundry process on 150-mm multi-project wafers~\cite{Ye:23, Sun:25}. 
Comprehensive process details are presented in Methods.  
Figure~\ref{Fig:1}c showcases a full Si$_3$N$_4$ wafer alongside an individual photonic chip and a single microresonator.
Characterizing these devices with a visible-light vector spectrum analyzer (VSA)~\cite{Shi:25} enables the precise identification and fitting of resonances across the 766--795 nm wavelength range with sub-megahertz resolution and accuracy. 
Lorentzian fitting of a typical resonance profile (Fig.~\ref{Fig:2}b) yields an intrinsic loss of $\kappa_0/2\pi=41$ MHz, corresponding to an intrinsic quality factor of $Q_0=9.4\times10^6$. 
Statistical analysis of 721 evaluated resonances confirms a most probable $Q_0=9.5\times10^6$ (Fig.~\ref{Fig:2}c), translating to a linear optical propagation loss of just 6.3 dB/m within our Si$_3$N$_4$ waveguides. 
This exceptionally low optical loss in the near-visible band is critical for microcomb generation at low power thresholds, thereby ensuring compatibility with our III--V laser diodes.

Figure~\ref{Fig:2}d displays the measured integrated dispersion, $D_\mathrm{int}$, of the microresonator, defined as
\begin{equation}
D_\mathrm{int}(\mu)=\omega_\mu-\omega_0-D_1\mu=\sum_{n=2}^{\cdots}\frac{D_n}{n!}\mu^n
\end{equation}
where $\mu$ denotes the resonance mode number relative to the pump resonance $\mu=0$,  
and $\omega_\mu/2\pi$ is the frequency of the $\mu$-th resonance. 
The parameter $D_1/2\pi$ represents the free spectral range (FSR), 
$D_2$ characterizes the group-velocity dispersion (GVD), 
and terms $D_{n\geqslant3}$ describe higher-order dispersion.
The extracted negative value of $D_2$ indicates normal GVD, a necessary condition for platicon (dark-pulse) microcomb generation~\cite{Lobanov:15,Xue:15}. 
Characterization details are provided in Supplementary Information Note 1.

\noindent \textbf{Microcomb generation and stabilization. } 
To generate platicon microcombs, we employ three distinct hybrid-integrated diode--microresonator architectures:  
a custom-fabricated Fabry--P\'erot (FP) laser diode paired with a 101-GHz-FSR microresonator (Fig.~\ref{Fig:2}e), 
an FP diode with a 20-GHz-FSR microresonator (Fig.~\ref{Fig:2}f), 
and a commercial distributed-Bragg-reflector (DBR) diode with a 99-GHz-FSR microresonator (Fig.~\ref{Fig:2}g).
The fabrication process for the AlGaAs-based FP diodes is detailed in Methods. 
Edge-coupling efficiencies from the diodes into the Si$_3$N$_4$ bus waveguides via inverse tapers~\cite{ChenD:26} reach 61\% and 20\% for the FP and DBR diodes, respectively.
Characterization details of these diodes including their power, wavelength, beam quality and coupling efficiency are presented in Supplementary Information Notes 2 and 3.

Figure~\ref{Fig:2}a illustrates the principle of SIL and microcomb generation within these diode--microresonator configurations. 
By current-tuning the laser frequency into a microresonator resonance, light couples into the forward propagating field, $A^+$. 
Rayleigh backscattering within the microresonator subsequently injects light ($A^-$) back into the diode, triggering ultrafast SIL. 
This mechanism passively locks the laser frequency to the microresonator resonance~\cite{Kondratiev:23, Liang:15, Sun:25}, profoundly suppressing frequency noise. 
Once sufficient intracavity power is attained---enabled by the high microresonator $Q$---platicon formation is initiated. 
Details on the SIL process are provided in Supplementary Information Note~4.

Figures~\ref{Fig:2}h--j display the optical spectra of the resulting platicon microcombs.
While the 101- and 99-GHz-spaced microcombs (Figs.~\ref{Fig:2}h, j) are fully resolved by a standard optical spectrum analyzer (OSA), the OSA resolution is insufficient for the 20-GHz-spaced microcomb (Fig.~\ref{Fig:2}i). 
Instead, we employ our VSA~\cite{Shi:25}, featuring a 3-MHz frequency resolution, to unambiguously resolve the 20.358-GHz mode spacing.
The intrinsic coherence of these microcombs is validated by either the absence of low-frequency amplitude noise~\cite{Raja:19} or the presence of a pristine microwave beatnote upon direct photodetection and electrical spectrum analysis \cite{Sun:25}.
Numerical simulations are presented in Supplementary Information Note~5. 

\begin{figure*}[t!]
\centering
\includegraphics[width=0.9\linewidth]{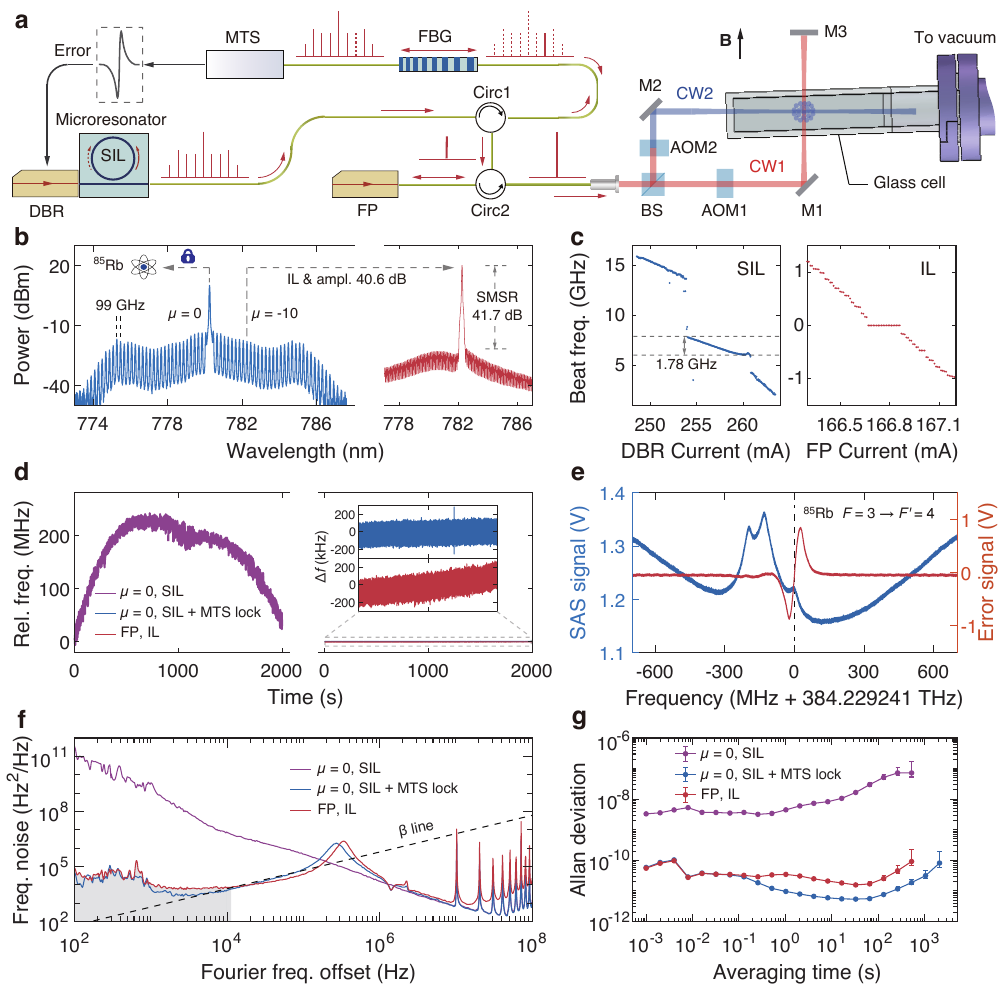}
\caption{
\footnotesize
\textbf{Frequency stabilization, power amplification, and noise suppression of the microcomb}.
\textbf{a,}
Schematic of the experimental setup for microcomb generation and stabilization (via SIL and MTS), amplification of a selected comb line (via IL), and free-space delivery to cold atoms. 
\textbf{b,}
Optical spectra of the SIL-generated, 99-GHz-spaced microcomb (left) and the injection-amplified CW laser (right).
Injection locking amplifies (IL \& ampl.) the $-10$th comb line to 102 mW (40.6 dB gain) with an SMSR of 41.7 dB. 
\textbf{c,}
Laser frequency tuning dynamics driven by diode currents during SIL (left) and IL (right). 
Vertical axes denote the respective beat frequencies of the DBR and FP lasers against a stable reference laser. 
The reference laser is generated by an external-cavity diode laser (Toptica, CTL 1550) locked to a commercial, self-referenced, fiber-based optical frequency comb and frequency-doubled to 780 nm. 
Horizontal axes denote the respective diode drive currents.
\textbf{d,}
Long-term frequency stability of the microcomb's SIL pump laser ($\mu=0$) and the FP laser.
The MTS lock (blue curve) suppresses the drift of the free-running SIL pump laser (purple curve) from $>200$ MHz over 2,000 s to the 100-kHz level. 
The injection-amplified FP laser directly inherits this stability (red curve).
\textbf{e,}
Saturated-absorption spectrum (SAS, blue curve) and MTS error (red curve) signals for the $^{85}$Rb $F=3 \to F'=4$ transition. 
\textbf{f, g,}
Frequency noise (panel \textbf{f}) and Allan deviation (panel \textbf{g}) measurements for the SIL and MTS-locked pump laser (blue curves) and the IL FP laser (red curves), demonstrating sub-30~kHz $\beta$-separation linewidths and fractional instability down to $1.6\times 10^{-11}$.
}
\label{Fig:3}
\end{figure*}

For downstream cold-atom manipulation, we tune the DBR laser (Fig.~\ref{Fig:2}g) into the wavelength band of the $^{85}$Rb D2 transition.
Figure~\ref{Fig:3}c left panel tracks the laser frequency evolution during the onset of SIL.
The achieved 1.78-GHz frequency range~\cite{Kondratiev:23} is sufficient for both interrogating and locking to atomic hyperfine transitions.
Figure~\ref{Fig:3}a outlines the experimental schematic for this frequency stabilization. 
The microcomb output is routed via a fiber circulator (Circ1) to a fiber Bragg grating (FBG). 
The transmitted light then enters an MTS setup, generating a saturated-absorption spectrum (SAS, blue curve in Fig.~\ref{Fig:3}e) alongside a sharp error signal corresponding to the $F=3 \to F'=4$ hyperfine transition of $^{85}$Rb (red curve in Fig.~\ref{Fig:3}e).  
Details concerning MTS are provided in Supplementary Information Note~6.
Crucially, because the microcomb is fully coherent, stabilizing the pump laser simultaneously disciplines all other comb lines and dramatically suppresses their frequency noise. 
This atomic referencing reduces the long-term pump-frequency drift from over 200 MHz down to the 100-kHz level across a 2,000-s measurement window (Fig.~\ref{Fig:3}d).

\begin{figure*}[t!]
\centering
\includegraphics[width=0.9\linewidth]{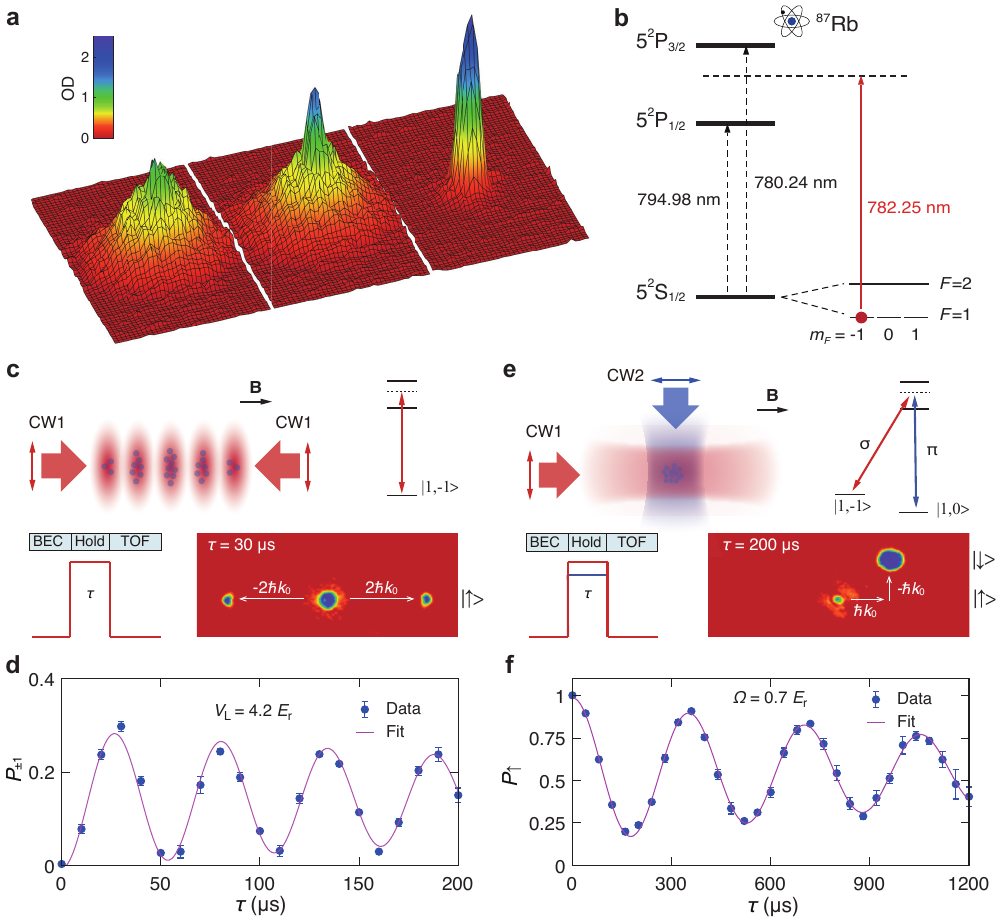}
\caption{
\footnotesize
\textbf{Optical lattice construction and Raman coupling calibrations}.
\textbf{a,}
Absorption images (left to right) detailing the emergence of a BEC from a thermal atomic cloud during evaporative cooling, evidenced by the formation of a bimodal density distribution. 
BEC, Bose--Einstein condensate.
\textbf{b,}
Energy-level diagram of $^{87}$Rb.
The $^{87}$Rb D1 and D2 transition wavelengths are 794.98 and 780.24 nm, respectively, with $F$ denoting the total atomic angular momentum and $m_F$ the corresponding magnetic quantum number. 
The BEC is initialized in the $\lvert F,m_F\rangle=\lvert1,-1\rangle$ state, and the amplified microcomb line provides the 782.25-nm driving field. 
\textbf{c,}
Kapitza--Dirac diffraction calibration of the optical lattice.
Retro-reflected CW1 beams of duration $\tau$ form an optical lattice and diffract the BEC into $\pm 2n\hbar k_{0}$ ($n \in \mathbb{N}^+$) momentum states. 
\textbf{d,} 
Fraction $P_{\pm 1}$ of diffracted atoms in the $\pm 2\hbar k_0$ momentum states versus pulse duration $\tau$. 
A damped sinusoidal fit yields a lattice depth of $V_{L} = 4.2~E_\mathrm{r}$, where $E_\mathrm{r}$ is the recoil energy. 
\textbf{e,}
Calibration of two-photon Raman coupling. 
Orthogonal beams CW1 and CW2 of duration $\tau$ drive Rabi oscillations through $\sigma$ and $\pi$ transitions between the $\lvert1,-1\rangle$ (spin-up, $\lvert\uparrow\rangle$) and $\lvert1,0\rangle$ (spin-down, $\lvert\downarrow\rangle$) states along a magnetic quantization axis $\mathbf{B}$, inducing a momentum transfer of ($\hbar k_0$, $-\hbar k_0$).
\textbf{f,}
Fraction $P_{\uparrow}$ of atoms in the spin-up state $\lvert\uparrow\rangle$ versus pulse duration $\tau$.
A damped sinusoidal fit of the Rabi oscillation yields a Raman coupling energy of $\Omega = 0.7~E_\mathrm{r}$.
}
\label{Fig:4}
\end{figure*}

\noindent \textbf{Injection amplification and free-space delivery.}
To minimize atomic heating and support operations such as spin-dependent or magic-wavelength optical lattices \cite{Heinz:20,Meng:23, Takamoto:05}, cold-atom manipulation typically requires light that is far-detuned from atomic resonances~\cite{WuZ:16}.  
Therefore, we isolate a far-detuned comb line ($\mu=-10$, detuned by approximately 990 GHz below the $\mu=0$ pump) using an FBG and inject it into a secondary FP diode via two fiber circulators (Circ1 and Circ2, Fig.~\ref{Fig:3}a).  
This injection locking (IL) securely anchors the FP laser emission to the selected comb tooth (Fig.~\ref{Fig:3}c right panel), boosting its optical power from $-20.5$ dBm (9~$\upmu$W) to 20.1 dBm (102 mW). 
This achieves a substantial optical gain of 40.6 dB while preserving a high side-mode suppression ratio (SMSR) of 41.7 dB (Fig.~\ref{Fig:3}b).
While a stand-alone far-detuned laser can provide the required optical power, obtaining absolute optical frequency with traceability to an atomic transition generally requires a complex, ultrastable, transfer cavity, as well as associated offset-lock chains or another frequency-comb link. 
Therefore, our microcomb acts here as a coherent frequency-transfer and synthesis network. 

We use a delayed self-heterodyne interferometer~\cite{Yuan:22} to characterize the frequency noise of both the pump DBR laser ($\mu = 0$) under SIL and MTS locking, and the FP laser under IL.
Figures~\ref{Fig:3}f, g compare the measured noise and drift performance, confirming the efficacy of MTS locking. 
The IL laser inherits the pristine spectral coherence and frequency stability of the disciplined microcomb, achieving an estimated intrinsic linewidth~\cite{DiDomenico:10} of 2.6 kHz, 
a $\beta$-separation linewidth~\cite{DiDomenico:10} of 23.6 kHz with a 100-Hz lower integration limit, 
and a $1/\pi$-integral linewidth~\cite{Isichenko:24} of 423.3 kHz, 
alongside a fractional frequency instability of $1.6 \times 10^{-11}$ at 33 s.
Measurement details on IL range,  laser frequency noise and Allan deviation are provided in Supplementary Information Notes~7 and 8.

This amplified, stable output is critical for a broad class of coherent atomic-control operations, including lattice- and Raman-based quantum simulation~\cite{Greiner:02,LinY:11,WuZ:16}, single-qubit control in neutral-atom processors~\cite{MaS:23}, and light-pulse atom interferometry~\cite{Kasevich:91}. 
For these applications, laser-frequency fluctuations remain negligible compared to the one-photon detuning, while the MTS lock actively suppresses technical noise in the crucial kilohertz-to-tens-of-kilohertz band---the exact scale that overlaps with lattice motional and Raman energies.
Atomic referencing is particularly valuable for wavelength-selective operations, such as tune-out and state-dependent lattices~\cite{Schmidt:16,Heinz:20,Meng:23} and magic-wavelength trapping in optical clocks~\cite{Katori:03,Takamoto:05,Ushijima:18,Kim:23}, where low-frequency noise must be paired with reproducible absolute frequencies. 
We estimate that transferring this stability to the electric-dipole (E1) magic frequency of $^{87}$Sr would bound lattice-frequency fluctuation and drift contributions to the clock transition below $10^{-18}$ (see Methods). 
Conversely, disabling the MTS lock increases frequency noise by orders of magnitude and restores a $>200$-MHz long-term drift, which would severely degrade phase-sensitive optical potentials via trap-position and differential-phase noise~\cite{Savard:97,LeGouet:07}. 

Finally, this stabilized, high-power beam is routed to free space and divided by a beam splitter (BS) into two paths (CW1 and CW2). 
Acousto-optic modulators (AOM1 and AOM2) frequency-shift these beams, creating a precise 7.332-MHz frequency difference, before they are directed into the cold-atom vacuum chamber by mirrors (M1 and M2). 
Retro-reflecting CW1 with another mirror (M3) establishes a periodic dipole potential, forming an optical lattice~\cite{Greiner:02}. 
Simultaneously, intersecting CW2 with CW1 drives two-photon Raman coupling between atomic states defined by a magnetic bias field $\mathbf{B}$, which serves as the quantization axis. 

\noindent \textbf{Quantum state manipulation and calibration.}
We leverage these stable optical fields to control a BEC of $^{87}$Rb, produced via sequential laser cooling and forced evaporative cooling~\cite{Anderson:95}. 
A BEC is a quantum-degenerate gas wherein a macroscopic fraction of atoms occupies the identical quantum state~\cite{Leggett:01}.
During forced evaporative cooling, systematically reducing the dipole-trap depth allows the most energetic atoms to escape, while elastic collisions rethermalize the remaining ensemble to progressively lower temperatures. 
Below a critical temperature threshold, a condensate forms. 
We experimentally confirm BEC onset by observing the emergence of a bimodal density distribution in time-of-flight (TOF) absorption imaging. 
Figure~\ref{Fig:4}a displays the optical depth (OD) of the ultracold atomic ensembles after a 25-ms TOF, clearly illustrating BEC formation.
The atoms are initially prepared in the $\lvert F,m_F\rangle=\lvert1,-1\rangle$ state via optical pumping.
The relevant $^{87}$Rb energy-level structure is presented in Fig.~\ref{Fig:4}b. 
782.25 nm denotes the operating wavelength of the IL laser relative to the $^{87}$Rb D1 (794.98 nm) and D2 (780.24 nm) transitions. 
Details concerning BEC production are described in Methods. 
Details on TOF imaging, OD and atom-number calculation are provided in Supplementary Information Note 9.

To calibrate the optical lattice depth $V_\mathrm{L}$, we employ Kapitza--Dirac (KD) diffraction~\cite{Gadway:09}.
The optical lattice formed by CW1 (Fig.~\ref{Fig:4}c) induces off-resonant dipole transitions from the $\lvert1,-1\rangle$ ground state to virtual excited states.
The resulting lattice potential is described by:
\begin{equation}
V_\mathrm{Latt}(x) = - V_\mathrm{L}\cos^2{(k_0x - \phi_\mathrm{L})},
\end{equation}
where $V_\mathrm{L} \propto E_1^2$ defines the lattice depth and is positive for the 782.25-nm lattice (red-detuned lattice), 
$E_1$ denotes the CW1 laser amplitude, 
$k_0=2\pi/\lambda_0$ is the wave number (with $\lambda_0 = 782.25$ nm), 
and $\phi_\mathrm{L}$ represents half the phase accumulated from the retro-reflecting loop.
The principle of KD diffraction is described in Methods. 

Pulsing this standing wave diffracts the condensate into ($\pm 2n\hbar k_0$, 0) momentum states, where $n \in \mathbb{N}^+$.
For a shallow lattice, diffraction orders of $n\geq2$ are negligible~\cite{Gadway:09}.
By monitoring the diffracted atomic fraction, $P_{\pm1}$, as a function of the pulse duration, $\tau$, and applying a damped sinusoidal fit (Fig.~\ref{Fig:4}d), we extract a lattice depth of $V_\mathrm{L}=4.2~E_\mathrm{r}$, where $E_\mathrm{r} = \hbar^2 k_0^2/(2m)$ is the recoil energy and $m$ is the atomic mass. 
A typical TOF absorption image acquired at $\tau = 30~\upmu\mathrm{s}$ is shown in Fig.~\ref{Fig:4}c.

Subsequently, by removing mirror M3, we calibrate the two-photon Raman coupling driven by the orthogonal CW1 and CW2 beams (Fig.~\ref{Fig:4}e).
With polarizations oriented perpendicular and parallel to $\mathbf{B}$, these beams drive $\sigma$ and $\pi$ transitions, respectively, coupling the $\lvert1,-1\rangle$ (spin-up $\lvert\uparrow\rangle$) and $\lvert1,0\rangle$ (spin-down $\lvert\downarrow\rangle$) states.
The spatial Raman coupling is given by:
\begin{equation}
\Omega_{\mathrm R} (x, y) = \Omega e^{i(k_0x + \phi_1)} e^{-i(k_0y + \phi_2)},
\end{equation}
where $\Omega \propto E_1 E_2$ is the Raman coupling energy, 
$E_2$ is the CW2 laser amplitude, 
$\phi_1$ and $\phi_2$ are the initial phases of the respective beams. 
By observing Rabi oscillations between the $\lvert\uparrow\rangle$ and $\lvert\downarrow\rangle$ states as a function of pulse duration $\tau$ (Fig.~\ref{Fig:4}e), we fit the spin-up fraction $P_{\uparrow}$ with a damped sinusoidal function (Fig.~\ref{Fig:4}f) and extract $\Omega=0.7~E_\mathrm{r}$.
A typical TOF absorption image acquired at $\tau = 200~\upmu\mathrm{s}$ is shown in Fig.~\ref{Fig:4}e.
The principle of Rabi oscillations is described in Methods. 
Details on the long-term evolution of the KD diffraction and Rabi oscillation, along with the atom loss, are provided in Supplementary Information Note~10. 

\begin{figure*}[t!]
\centering
\includegraphics[width=0.9\linewidth]{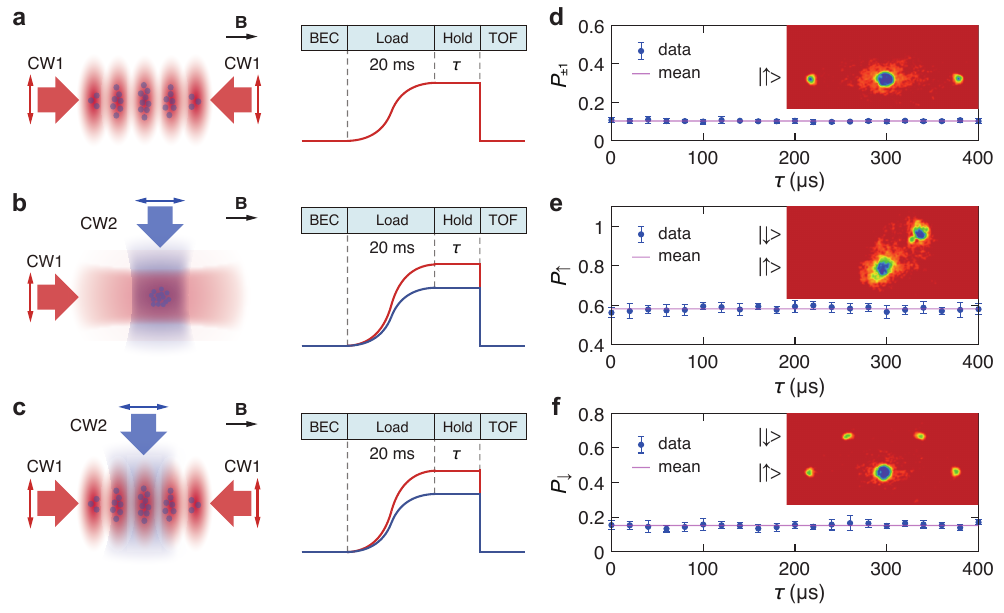}
\caption{
\footnotesize
\textbf{Adiabatic preparation of stationary quantum states}.
\textbf{a--c,}
Schematic protocols for loading the BEC into a one-dimensional optical lattice (panel \textbf{a}), a spin--orbit coupling configuration (panel \textbf{b}), and a Raman lattice (panel \textbf{c}). 
The laser intensity is adiabatically ramped along an S-shaped profile over 20 ms and held for time $\tau$ prior to TOF imaging. 
\textbf{d--f,}
Corresponding temporal evolution of the atomic populations. 
The invariant responses across varying hold times $\tau$ in the $\pm 2\hbar k_0$ momentum fraction (panel \textbf{d}), spin-up-state fraction (panel~\textbf{e}), and spin-down-state fraction (panel \textbf{f}) confirm stable loading of the atoms into the ground state.
Insets display the resulting momentum profiles acquired after TOF.
}
\label{Fig:5}
\end{figure*}

\noindent \textbf{Adiabatic preparation of stationary quantum states.}
Following calibration, we adiabatically ramp the CW1 power along an S-shaped profile over a 20-ms duration to load the BEC into the ground state of the optical lattice, explicitly avoiding significant atomic heating (Fig.~\ref{Fig:5}a).
Details of laser-induced atomic heating are provided in Supplementary Information Note 11.
While deep optical lattices are favoured for atomic clocks and quantum entanglement applications, intermediate-depth and shallow lattices are routinely utilized to simulate the Hubbard model~\cite{Hubbard:63}---a framework foundational for explaining complex many-body phenomena, including Mott-insulator, superfluid, and superconducting phases.
The one-dimensional standing wave in our experimental setup effectively forms a pancake optical lattice, serving as a fundamental building block for one- and higher-dimensional Hubbard models~\cite{Greiner:02,Spielman:07,Bloch:08,ShaoH:24}. 
Ground-state loading is verified by observing a constant diffracted momentum fraction, $P_{\pm1}$, which remains strictly independent of the hold time $\tau$ (Fig.~\ref{Fig:5}d). 

Applying this same adiabatic protocol, we successfully prepare stationary states within both spin--orbit coupling~\cite{LinY:11} and Raman-lattice~\cite{WangB:18} configurations (Fig.~\ref{Fig:5}b, c), evidenced by the stable, non-oscillating spin populations over time $\tau$ (Fig.~\ref{Fig:5}e, f). 
To further validate the adiabaticity of this process, we smoothly ramp down the laser intensities. 
A dominant zero-momentum condensate component is recovered with a residual thermal fraction, detailed in Supplementary Information Note~12.  
The reliable establishment of these configurations highlights the system's capacity to simulate complex quantum phases, including topological insulators and Weyl semimetals~\cite{WuZ:16,SongB:18,WangB:21}. 

The Raman-lattice configuration serves as a particularly stringent test of our microcomb engine's performance. 
Because the optical system simultaneously defines both the standing-wave lattice and the phase-coherent Raman coupling, the instantaneous Hamiltonian is highly sensitive to frequency-to-position conversion and differential Raman phase noise~\cite{Savard:97,LeGouet:07}.
Consequently, the low frequency noise of the IL laser provides a substantial stability margin, preserving the relative spatial phase and suppressing unwanted motional excitation during adiabatic loading.
Consistently, when the MTS locking is disabled, the frequency noise of the SIL microcomb engine becomes too high to support Raman-lattice ground-state preparation.
Under such free-running conditions, we observe strongly asymmetric momentum populations accompanied by severe heating and depletion of the condensate. 
Details are found in Supplementary Information Note 13.

\noindent \textbf{Summary and outlook.}
We have demonstrated a scalable, highly stable, microcomb--laser engine capable of driving optical lattices and spin--orbit coupling in ultracold atomic ensembles. 
Generated within integrated Si$_3$N$_4$ microresonators directly driven by a 780-nm CW laser diode, these microcombs provide versatile mode spacings of 20 and 100 GHz.
Both the Si$_3$N$_4$ and III--V diode chips were fabricated via high-volume foundry processes, bringing the estimated raw cost per engine below \$20. 
Through $^{85}$Rb hyperfine stabilization and subsequent injection locking, we extract amplified laser outputs with suppressed frequency noise and fractional instability that meet the stringent requirements of state-of-the-art optical lattice clocks, atom interferometers, and quantum simulators. 
Featuring a compact, hybrid-integrated footprint of just 29 mm$^2$, our fully packaged modules drastically reduce SWaP-C metrics, rendering them ideal for demanding field-deployable applications.
A comprehensive performance comparison between our microcomb engine and other integrated visible-light sources is supplied in Methods. 
In addition, a performance comparison between our microcomb engine and conventional lasers in cold-atom platforms, such as external-cavity diode lasers (ECDL), Ti:sapphire lasers and fiber lasers, is provided in Supplementary Information Note 14.

\begin{figure}[t!]
\centering
\includegraphics[width=0.9\columnwidth]{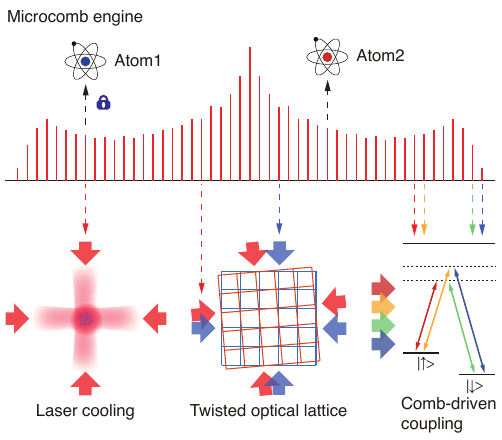}
\caption{
\footnotesize
\textbf{Microcomb engine for scalable cold-atom operations}.
An atom-referenced microcomb provides a common coherent frequency grid, used as frequency seeds for downstream optical channels. 
Selected comb teeth can reference or seed laser fields that drive laser cooling for the preparation of ultracold atoms.
Comb-seeded optical fields at tune-out wavelengths can generate multi-frequency optical potentials, such as twisted optical lattices~\cite{Meng:23}.
Mutually coherent comb teeth can drive transitions between atomic internal states, thereby enabling comb-driven coherent Raman couplings~\cite{Solaro:18,Hayes:10}. 
This scalable microcomb architecture allows the number of optical-control channels to increase without a proportional multiplication in independent frequency references.
}
\label{Fig:6}
\end{figure}

Beyond the single-tooth demonstration reported here, the intrinsic comb nature seamlessly supports frequency-multiplexed scaling.
As illustrated in Fig.~\ref{Fig:6}, an atom-referenced microcomb can simultaneously seed individual lasers for atom cooling and high-power lattice or Raman channels for coherent many-body manipulation. 
This architecture is uniquely advantageous for multi-frequency operations, such as the bichromatic tune-out lattices utilized in twisted-bilayer quantum simulations~\cite{Meng:23}, 
the 1.8-THz comb-driven Raman transition in Ca$^+$~\cite{Solaro:18}, 
the 12.6-GHz hyperfine-qubit control in $^{171}$Yb$^+$~\cite{Hayes:10}, 
and trichromatic coherent clock excitations~\cite{Carman:25}.
Furthermore, by substituting the pump diode, Si$_3$N$_4$ microcombs could be anchored to optical transitions in diverse atomic species---including Yb, Li, Na and metastable He---spanning a broad wavelength range from 399 nm~\cite{XuX:09,White:26} to 1.08 $\upmu$m~\cite{GongW:14}. 
In this scalable architecture, the number of optical control channels scales without a proportional multiplication of independent frequency references, effectively transforming the microcomb into a manufacturable frequency-distribution and synthesis engine for the entire cold-atom experimental sequence~\cite{Hisai:19,Abe:21}.

Benefiting from the ultralow optical loss and wide transparency window of Si$_3$N$_4$, PICs are rapidly bringing advanced atomic manipulation directly onto chips.
Most notably, much deeper optical lattices could be constructed by leveraging our microcomb to synchronize watt-level lasers, cementing this hybrid-integrated platform as a versatile engine for next-generation quantum simulation and sensing.
These sweeping advances signal a new technological paradigm where light generation, atomic referencing, frequency synthesis, modulation, amplification, and free-space delivery are co-designed holistically.
Ultimately, compact, low-cost, and coherent microcombs operating at visible wavelengths will facilitate the widespread deployment of integrated photonics for transportable atomic sensors, simulators, and clocks across diverse mobile platforms and into space.

\vspace{0.3cm}
\noindent \textbf{Methods}
\setcounter{figure}{0}
\renewcommand{\theHfigure}{ED.\arabic{figure}}

\medskip
\begin{footnotesize}

\begin{figure*}[!t]
\renewcommand{\figurename}{Extended Data Figure}
\centering
\includegraphics{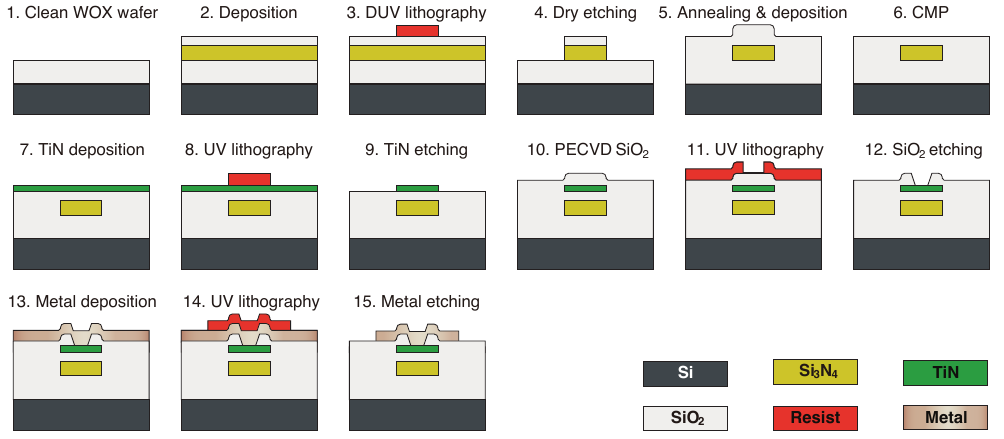}
\caption{
\footnotesize
\textbf{Fabrication process of the Si$_3$N$_4$ photonic integrated circuits.}
Schematic cross-sectional process flow detailing the fabrication of the Si$_3$N$_4$ photonic integrated circuits (PICs).
The process includes the deposition and subtractive patterning of the Si$_3$N$_4$ waveguides, high-temperature thermal annealing, SiO$_2$ top-cladding deposition, and subsequent planarization via chemical--mechanical polishing (CMP). 
This is followed by the fabrication of TiN thermo-optic heaters, plasma-enhanced chemical vapor deposition (PECVD) SiO$_2$ encapsulation, contact-via etching, and the final metal routing. 
Color-coded regions denote silicon (Si), silicon dioxide (SiO$_2$), silicon nitride (Si$_3$N$_4$), photoresist, titanium nitride (TiN), and metal layers.
}
\label{Fig:SiNfab}
\end{figure*}

\noindent \textbf{Fabrication process of Si$_3$N$_4$ PICs.}
The Si$_3$N$_4$ PICs and microresonators are fabricated on 150-mm wafers utilizing a subtractive process developed within our foundry platform~\cite{Ye:23, Sun:25}.
The fabrication process flow is presented in Extended Data Fig.~\ref{Fig:SiNfab}.
Fabrication begins with a cleaned silicon wafer with a 4-$\upmu$m-thick thermal wet SiO$_2$ layer.
A 300-nm-thick Si$_3$N$_4$ film is deposited via low-pressure chemical vapor deposition (LPCVD) and capped with a SiO$_2$ hardmask.
The waveguide pattern is defined using deep-ultraviolet scanner lithography (ASML PAS 850C, 110 nm resolution). 
The pattern is first etched into the SiO$_2$ hardmask via inductively coupled plasma (ICP) dry etching.
Following photoresist stripping with an O$_2$ plasma, a secondary dry-etch step transfers the pattern into the underlying Si$_3$N$_4$ core layer.
To mitigate optical absorption caused by residual hydrogen impurities from the LPCVD process, the patterned Si$_3$N$_4$ wafer undergoes thermal annealing at 1200 $^{\circ}$C in a nitrogen ambient~\cite{HuY:26}.
A 3-$\upmu$m-thick SiO$_2$ top cladding is then deposited and subjected to an additional annealing to further reduce hydrogen-induced optical absorption.
The upper SiO$_2$ surface is planarized by chemical--mechanical polishing.

Next, a 200-nm-thick titanium nitride (TiN) film is deposited onto the planarized SiO$_2$ surface. 
The TiN layer is patterned via photolithography and dry etching to construct resistive thermo-optic heaters.
These heaters are encapsulated within a plasma-enhanced chemical vapor deposition (PECVD) SiO$_2$ cladding.
Contact vias are defined by photolithography, and the PECVD SiO$_2$ is selectively etched to expose the TiN contact pads.
Finally, a top metal layer is deposited, lithographically patterned, and etched to establish the electrical interconnects to the integrated heaters.

\noindent \textbf{Fabrication process of FP laser diodes.}
The FP laser diodes are fabricated on an Al$_{0.08}$Ga$_{0.92}$As/Al$_{0.35}$Ga$_{0.65}$As multiple-quantum-well (MQW) wafer.
The fabrication process is illustrated in Extended Data Fig.~\ref{Fig:FP_fb}.
The MQW region with top p- and bottom n-type AlGaAs claddings is first grown by molecular beam epitaxy (MBE). 
The active region consists of three quantum-well layers and four barrier layers. 
Following MBE, maskless photolithography defines the cavity area using a 150-nm-thick SiO$_2$ layer as a hardmask.
Ridge waveguides are then formed by ICP etching.
Each fabricated ridge measures $800~\upmu$m in length, $2~\upmu$m in width, and $2~\upmu$m in height, supporting only the fundamental mode to ensure high electro-optical conversion efficiency and good beam quality.
A second epitaxial growth deposits an AlAs current-blocking layer. 
Subsequently, the SiO$_2$ hardmask is stripped using a diluted HF solution. 
A third epitaxy step grows a GaAs cap layer, enabling the direct deposition of metal electrodes. 
Electron-beam evaporation (EBV) is used to apply anti-reflection and high-reflection coatings on the cleaved facets. 
Pt/Ti/Pt/Au layers (5/10/10/200 nm) are deposited as the p-type contact metals, followed by rapid thermal annealing at $400~^{\circ}$C to reduce the contact resistance.
For the n-type electrode on the opposite side of the sample, maskless photolithography and EBV deposit Ni/AuGe/Ni/Au layers (5/100/35/300 nm), followed by rapid thermal annealing at $420~^{\circ}$C.
The MQW is located close to the p-side layer. 
To maintain a stable cladding temperature---thereby avoiding power limitation due to thermal rollover and ensuring good wavelength stability---the GaAs cap layer is fully covered with metal and bonded to a heat-dissipating substrate.

\begin{figure*}[b!]
\renewcommand{\figurename}{Extended Data Figure}
\centering
\includegraphics{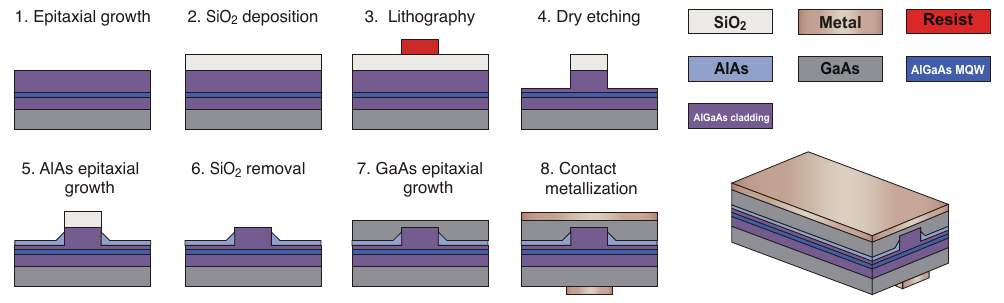}
\caption{
\footnotesize
\textbf{Fabrication of the FP laser diodes}.
Schematic overview illustrating the fabrication process, including ridge-waveguide patterning, secondary epitaxial growth of the AlAs current-blocking layer, final epitaxial growth of the GaAs cap layer, and subsequent contact metallization. 
The accompanying three-dimensional schematic illustrates the precise structural architecture of the fully processed laser-diode chip.
Color-coded regions denote aluminium arsenide (AlAs), gallium arsenide (GaAs), aluminium gallium arsenide (AlGaAs) multiple-quantum-well (MQW) and cladding, silicon dioxide (SiO$_2$), photoresist, and metal layers.
}
\label{Fig:FP_fb}
\end{figure*}

\noindent \textbf{Estimate of lattice-frequency-induced clock uncertainty.} 
To quantitatively evaluate the impact of lattice-laser frequency fluctuations on an optical lattice clock's performance, we analyze the frequency-dependent E1 lattice shift of the $^{87}$Sr clock transition near its magic wavelength.
Based on established lattice-light-shift models~\cite{Ushijima:18}, the frequency-dependent E1 contribution for atoms occupying the axial motional ground state is expressed as:
\begin{equation}
\Delta \nu_{\mathrm{LS}}^{(\mathrm{E1})} \simeq a\,\delta_\mathrm{L} \left( \frac{\sqrt{v}}{2}-v \right),
\end{equation}
where $a=1.859\times10^{-11}$ is an experimentally determined E1 sensitivity coefficient~\cite{Kim:23}, 
$\delta_\mathrm{L} = \nu_\mathrm{latt} - \nu_\mathrm{magic}$ represents the lattice laser's detuning from the E1 magic frequency, 
and $v=V_\mathrm{L}/E_\mathrm{r}^\mathrm{Sr}$ defines the lattice depth normalized to the $^{87}$Sr recoil energy.
At a typical operational lattice depth~\cite{Ushijima:18} of $v=72$, the corresponding sensitivity to lattice-frequency variations is $\left | a \left( \sqrt{v}/2-v \right) \right | = 1.26$ mHz/MHz.
Consequently, treating a conservative 100-kHz long-term lattice-frequency drift as a steady-state detuning yields a clock shift of 0.126 mHz, translating to a fractional uncertainty of $2.9\times10^{-19}$ relative to the $\simeq429$ THz clock frequency of $^{87}$Sr.
Similarly, treating our measured 23.6-kHz $\beta$-separation linewidth as a characteristic short-term detuning fluctuation results in a fractional shift of $6.9\times10^{-20}$.
If the measured fractional frequency instability of $1.6\times10^{-11}$ at 33 s were directly transferred to an 813-nm lattice laser, it corresponds to a roughly 5.9-kHz lattice-frequency fluctuation, bounding the frequency-induced clock shift to approximately $1.7\times10^{-20}$ at $v=72$.
By comparison, operating with the unlocked $\beta$-separation linewidth of 2.6~MHz would substantially amplify this uncertainty scale to \(\sim 7.6\times10^{-18}\). 
More critically, the unsuppressed $>200$ MHz long-term carrier drift observed in the absence of MTS locking would induce a dominant E1 fractional shift of $\sim 5.9\times10^{-16}$. 
These projections demonstrate that atomic referencing is essential for anchoring the lattice laser to the magic frequency condition over extended operational timescales. 

\noindent \textbf{BEC production and Raman-lattice experiments.}
The experimental apparatus for BEC is shown in Extended Data Fig.~\ref{Fig:BEC}.
Initially, $^{87}$Rb atoms are precooled within a two-dimensional magneto-optical trap (2D-MOT).
This continuous atomic beam is subsequently propelled by a pushing laser through a differential pumping tube and into an ultrahigh-vacuum science chamber, where the atoms are captured by a three-dimensional magneto-optical trap (3D-MOT).
Following a 4-s loading step, the 3D-MOT accumulates approximately $5\times10^8$ atoms. 
The 3D cooling cycle is sustained using an auxiliary repumping laser, which continuously drives the atoms from $\lvert5^{2}S_{1/2},F=1\rangle$ manifold back into the cooling cycle.
At this stage, the 3D-MOT atomic cloud exhibits a peak density of roughly $2.5\times10^{11}~\mathrm{cm}^{-3}$ and a temperature of approximately $184~\upmu\mathrm{K}$.
The 3D-MOT is then spatially compressed for 0.1 s by ramping the magnetic-field gradient, increasing the peak atomic density to $5.2\times10^{11}~\mathrm{cm}^{-3}$.
A subsequent 40-ms dark-MOT stage, followed by a 10-ms optical molasses stage, applies robust sub-Doppler cooling, yielding an ensemble of approximately $2\times10^8$ atoms at $10~\upmu$K.
Finally, the atoms are optically pumped into the $\lvert F=1,m_F=-1\rangle$ Zeeman sublevel---defined here as the spin-up state $\lvert\uparrow\rangle$---and efficiently loaded into a crossed 1064-nm optical dipole trap (ODT) to initiate evaporative cooling.

The crossed ODT configuration utilizes horizontally propagating beams that intersect precisely at the atomic cloud's center. 
To counteract gravity and provide robust vertical confinement, the beam waist is highly elliptical. 
The vertical waist measures approximately $55~\upmu$m, roughly one-third the size of the $182~\upmu$m horizontal waist. 
Forced evaporative cooling is executed by exponentially lowering the optical trap depth across six successive ramping stages. 
This comprehensive evaporation sequence spans 6.8 s and reliably produces a nearly pure BEC containing approximately $2\times10^5$ atoms at a final temperature of 30~nK.
This condensate serves as the starting point for all Raman-lattice experiments presented in this work.

During the Raman-lattice experiments, the crossed ODT is strictly maintained at its final evaporative trap depth to securely confine the condensate. 
The Raman-lattice optical potential is generated by intersecting a 782.25-nm one-dimensional (1D) optical lattice with an orthogonal Raman beam, both aligned entirely within the horizontal plane. 
The 1D lattice is constructed by retro-reflecting and refocusing the incident lattice beam using a dedicated mirror--lens pair. 
The lattice beam is linearly polarized along the $y$-axis. 
A bias magnetic field of about 10.4 G defines a quantization axis along the $x$-axis. 
The focused lattice beam possesses a waist of approximately $154~\upmu$m. 
The orthogonal Raman beam is linearly polarized along the $x$-axis and is shaped to a broader waist of approximately $300~\upmu$m to ensure highly uniform illumination across the condensate.

\begin{figure*}[t!]
\renewcommand{\figurename}{Extended Data Figure}
\centering
\includegraphics{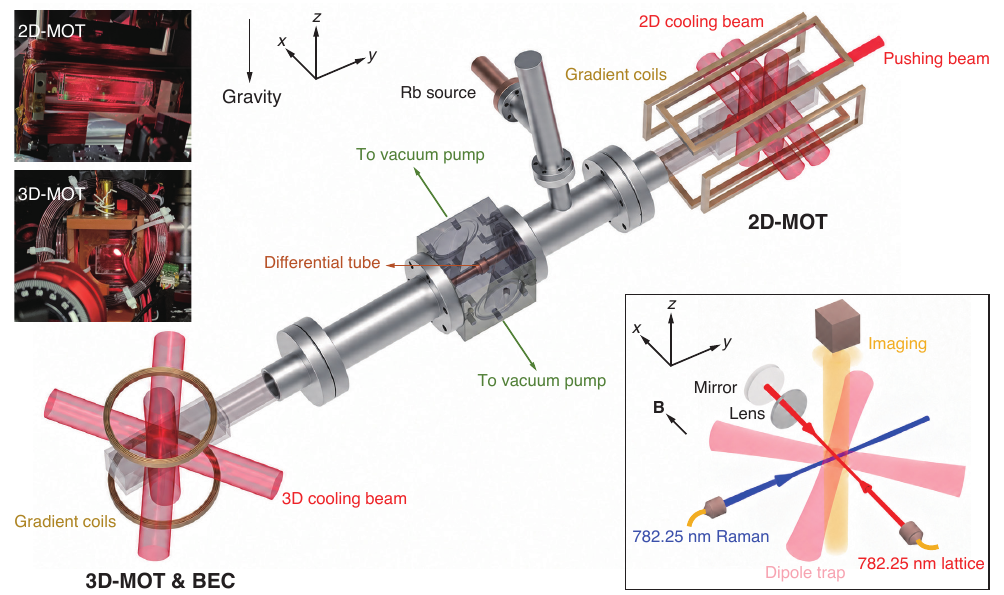}
\caption{
\footnotesize
\textbf{Experimental setup for BEC production and Raman-lattice configurations.}
$^{87}$Rb atoms are precooled within a 2D-MOT chamber and subsequently transported by a pushing beam through a differential pumping tube into the main science chamber for capture in a 3D-MOT. 
Subsequent sub-Doppler cooling and forced evaporative cooling within a crossed optical dipole trap reliably yield a nearly pure BEC. 
Left inset shows photographs of the 2D- and 3D-MOT regions. 
Right boxed inset graphically illustrates the precise optical configuration utilized for aligning the crossed dipole trap alongside the 782.25-nm Raman-lattice laser beams.
2D-MOT, two-dimensional magneto-optical trap.
3D-MOT, three-dimensional magneto-optical trap.
}
\label{Fig:BEC}
\end{figure*}

\noindent \textbf{KD diffraction and lattice-depth calibration.}
In ultracold atoms, a standing-wave optical potential coherently couples discrete momentum states~\cite{Morsch:06,Gadway:09} separated by $2\hbar k_0$.
Neglecting interatomic interactions and the external trapping confinement, the single-atom Hamiltonian is expressed as:
\begin{equation}
\hat{H} = \frac{\hat{p}_{x}^{2}}{2m} - V_{\mathrm L}(t)\cos^{2}(k_0x-\phi_\mathrm{L}),
\label{eq:KD_Hamiltonian}
\end{equation}
where $\hat{p}_{x}$ defines the momentum operator, 
and $V_{\mathrm L}(t)$ denotes the time-dependent lattice depth.
Defining the discrete momentum states as $p_n=2n\hbar k_0$ with $n \in \mathbb{Z}$, the time-evolving wavefunction is expanded as:
\begin{equation}
\left|\psi(t)\right\rangle = \sum_{n=-\infty}^{\infty} c_n(t)\left|2n\hbar k_0\right\rangle,
\qquad c_n(0)=\delta_{n0}.
\label{eq:KD_expansion}
\end{equation}
Applying the phase transformation \(c_n=e^{-2in\phi_\mathrm{L}}\widetilde{c}_n\) and subsequently dropping the tilde for clarity, the temporal evolution reduces to a set of coupled momentum-state equations:
\begin{equation}
i\hbar\dot{c}_n = \left(4n^2E_{\mathrm r}-\frac{V_{\mathrm L}(t)}{2}\right)c_n - \frac{V_{\mathrm L}(t)}{4} \left(c_{n-1}+c_{n+1}\right).
\label{eq:KD_coupled}
\end{equation}
For the measurements conducted here, atomic populations diffracted into higher-order momentum states $|n|\geqslant 2$ are negligible. 
Consequently, the system dynamics are accurately captured by truncating the basis to a three-state subspace corresponding to $n=0,\pm1$.
For a square pulse applied at a constant lattice depth $V_{\mathrm L}$, this non-interacting three-mode model yields an analytical solution for the fractional populations~\cite{Gadway:09}:
\begin{equation}
P_{\pm1}(\tau) = \frac{\chi_{\mathrm{KD}}^2} {\Delta_{\mathrm{KD}}^2+\chi_{\mathrm{KD}}^2} \sin^2\left(\frac{\omega_{\mathrm{KD}}\tau}{2}\right),
\label{eq:KD_population}
\end{equation}
where $P_{\pm1}=|c_{+1}|^2 + |c_{-1}|^2$ represents the total diffracted fraction, 
$\Delta_{\mathrm{KD}}=4E_{\mathrm r}/\hbar$ is the characteristic detuning, 
$\chi_{\mathrm{KD}}=V_{\mathrm L}/(\sqrt{2}\hbar)$ is the effective coupling rate, 
and $\omega_{\mathrm{KD}}= \sqrt{\Delta_{\mathrm{KD}}^2+\chi_{\mathrm{KD}}^2}$ represents the generalized Kapitza--Dirac oscillation frequency.
The lattice depth is directly extracted from this measured oscillation frequency via the relation:
\begin{equation}
V_{\mathrm L} = \sqrt{2}\hbar \sqrt{\omega_{\mathrm{KD}}^2 - \left(\frac{4E_{\mathrm r}}{\hbar}\right)^2}.
\label{eq:KD_depth}
\end{equation}
Experimentally, we determine $\omega_{\mathrm{KD}}$ by fitting the time-resolved diffraction data $P_{\pm1}(\tau)$ with a damped oscillatory function.
If driving deeper lattices where higher-order diffraction becomes significant, $V_{\mathrm L}$ is instead calibrated through exact numerical integration of the full, untruncated momentum-state equations  \cite{Gadway:09}.

\noindent \textbf{Two-photon Raman coupling and Rabi oscillations.}
The spin states $|\uparrow\rangle=|F=1,m_F=-1\rangle$ and $|\downarrow\rangle=|F=1,m_F=0\rangle$ are coherently coupled via a stimulated two-photon Raman transition driven collectively by orthogonal beams CW1 and CW2.
Because the single-photon laser detunings vastly exceed both the excited-state natural linewidths and the single-photon Rabi frequencies, the intermediate excited states can be adiabatically eliminated, yielding an effective two-level system~\cite{Brion:07,Dalibard:11}. 
The resulting spatially dependent Raman coupling strength is formulated as:
\begin{equation}
\Omega_{\mathrm R}(\mathbf r) = \Omega \exp\left\{i\left[(\mathbf k_1-\mathbf k_2)\cdot\mathbf r  +\phi_1-\phi_2 \right] \right\},
\end{equation}
where $\mathbf q=\mathbf k_1-\mathbf k_2 =k_0(\hat{\mathbf{x}}-\hat{\mathbf{y}})$  denotes the effective two-photon Raman wave vector, signifying a net momentum transfer of $\hbar\mathbf q$ per Raman process.
For a fixed one-photon detuning, polarization geometry, and set of atomic dipole matrix elements, the base coupling strength scales directly with the laser field amplitudes, $\Omega\propto E_1E_2$.
Operating in the momentum basis $\{|\uparrow,\mathbf p\rangle,|\downarrow,\mathbf p+\hbar\mathbf q\rangle\}$, where $\mathbf p \sim 0$ represents the initial center-of-mass momentum of the cold condensate, the effective Hamiltonian is given by~\cite{Brion:07}:
\begin{equation}
\hat H_{\mathrm R} = \frac{1}{2} 
\begin{pmatrix}
-\hbar\delta_{\mathbf p} & \Omega e^{-i\phi}\\
\Omega e^{i\phi} & \hbar\delta_{\mathbf p}
\end{pmatrix},
\qquad
\phi=\phi_1-\phi_2 ,
\end{equation}
Here, $\delta_{\mathbf p}$ represents the generalized detuning, encompassing the intrinsic two-photon detuning, atomic recoil shifts, Doppler shifts, and any differential AC Stark shifts induced by the lasers. 
Provided that interaction-induced mean-field shifts, spontaneous photon scattering, and background atom loss remain negligible over the interrogation duration, atoms initialized in the $|\uparrow\rangle$ state evolve dynamically under a square Raman pulse of duration $\tau$ according to the Rabi formulae~\cite{Foot:05}
\begin{equation}
P_{\downarrow}(\tau) = \frac{(\Omega/\hbar)^2}{\delta_{\mathbf{p}}^2+(\Omega/\hbar)^2}\sin^2\left(\frac{\omega_{\mathrm{R}}\tau}{2}\right),
\qquad
P_{\uparrow}(\tau)=1-P_{\downarrow}(\tau),
\end{equation}
where $\omega_{\mathrm{R}}=\sqrt{\delta_{\mathbf{p}}^2+(\Omega/\hbar)^2}$ is the generalized angular Rabi frequency. 
Under the condition of exact two-photon resonance $\delta_{\mathbf{p}}=0$, these expressions simplify to idealized, full-contrast oscillations:
\begin{equation}
P_{\uparrow}(\tau) = \cos^2\left(\frac{\Omega\tau}{2\hbar}\right),
\qquad
P_{\downarrow}(\tau) = \sin^2\left(\frac{\Omega\tau}{2\hbar}\right).
\end{equation}
Experimentally, the temporal evolution of the spin-up population $P_{\uparrow}(\tau)$ is fitted with a damped sinusoidal function.
The extracted oscillation frequency satisfies $\omega_{\mathrm{fit}}=\omega_{\mathrm R}$.
Consequently, the bare Raman coupling strength is isolated via the relation:
\begin{equation}
\Omega = \hbar\sqrt{\omega_{\mathrm{fit}}^2-\delta_{\mathbf p}^2},
\end{equation}
which, when perfectly on resonance, $\delta_{\mathbf p}=0$, yields the direct equivalence $\Omega=\hbar\omega_{\mathrm{fit}}$.

\noindent \textbf{Comparison with visible integrated light sources.} 
Table~\ref{tab:LasersCompare} comprehensively benchmarks our microcomb engine against a range of representative visible integrated light sources. 
While prior landmark demonstrations have independently addressed isolated milestones---such as integrated laser emission, chip-based reference cavities, standalone microcomb formation, or atom-facing photonic routing---our architectural platform natively unifies these functionalities. 
Specifically, we bridge foundry-fabricated active and passive photonics, hybrid laser--microresonator integration, absolute atomic frequency disciplining, specific high-power comb-line extraction, and subsequent coherent many-body atomic control into a single, cohesive experimental chain.

\begin{table*}[t]
\caption{
\footnotesize
\textbf{Comparison of visible integrated light sources.}
The architecture describes the integration of the primary source and excludes
downstream fiber-connected amplifiers.
``Hybrid'' and ``Heterogeneous'' denote hybrid integration and heterogeneous integration, respectively.
``Passive'' denotes a single passive chip pumped by an external bulk laser.
``Discrete'' denotes two passive chips connected by fibers and pumped by an external bulk laser.
$\Delta\nu_{\mathrm{i}}$, $\Delta\nu_{\mathrm{int}}$, and
$\Delta\nu_{\beta}$ denote the intrinsic, integral, and
$\beta$-separation linewidths, respectively.
``--'' denotes not reported.
``MHF'' denotes mode-hop-free.
Compared with CW lasers, the microcomb sources have an additional advantage of multiple-wavelength scalability.
}
\label{tab:LasersCompare}
\centering

\small
\setlength{\tabcolsep}{3.0pt}
\renewcommand{\arraystretch}{1.16}

\resizebox{\textwidth}{!}{%
\begin{threeparttable}

\begin{tabular}{lccccccc}
\toprule
Reference
& Source
& Integration
& Wavelength
& Validation
& Linewidth
& Stabilization
& MHF tuning
\\
\midrule

\textbf{This work}
& \makecell{Microcomb,\\
             20 \& 100 GHz}
& Hybrid
& 780 nm
& BEC control
& \makecell{2.6 kHz $\Delta\nu_{\mathrm{i}}$;\\
             23.6 kHz $\Delta\nu_{\beta}$;\\
             423.3 kHz $\Delta\nu_{\mathrm{int}}$}
& \makecell{$^{85}$Rb MTS;\\
             $1.6\times10^{-11}$ (33 s)}
& 1.78 GHz
\\

Yu et al.~\cite{Yu:19}
& \makecell{Microcomb, \\
             1 THz}
& Passive
& 1064 nm\tnote{a}
& --
& --
& Free running
& --
\\

Liu et al.~\cite{LiuP:25}
& \makecell{Microcomb, \\
             35 GHz}
& Passive
& 780 nm
& --
& --
& Free running
& --
\\

Loh et al.~\cite{Loh:25}
& CW laser
& Passive
& 674 nm
& $^{88}$Sr$^+$ clock
& --
& \makecell{Spiral cavity;\\
             $7.5\times10^{-14}$ (30 ms)}
& --
\\

Chauhan et al.~\cite{Chauhan:26}
& CW laser
& Discrete
& 674 nm
& $^{88}$Sr$^+$ qubit/clock
& \makecell{14 Hz $\Delta\nu_{\mathrm{i}}$;\\
             322 Hz $\Delta\nu_{\mathrm{int}}$}
& \makecell{Spiral cavity;\\
             $8.8\times10^{-13}$ (20 ms)}
& --
\\

Isichenko et al.~\cite{Isichenko:26}
& CW laser
& Discrete
& 780 nm
& $^{87}$Rb laser cooling
& --
& $^{87}$Rb SAS
& --
\\

Castro et al.~\cite{Castro:25}
& CW laser
& Heterogeneous
& 780 nm
& --
& $<6$ kHz $\Delta\nu_{\mathrm{i}}$
& $^{87}$Rb SAS
& $>100$ GHz
\\

Isichenko et al.~\cite{Isichenko:24}
& CW laser
& Hybrid
& 780 nm
& --
& \makecell{0.74 Hz $\Delta\nu_{\mathrm{i}}$;\\
             864 Hz $\Delta\nu_{\mathrm{int}}$}
& Free running
& 2.5 GHz
\\

Prokoshin et al.~\cite{Prokoshin:24}
& CW laser
& Hybrid
& 780 nm
& --
& $105$ Hz
& Free running
& --
\\

Franken et al.~\cite{Franken:21}
& CW laser
& Hybrid
& 684 nm
& --
& $2.3$ kHz $\Delta\nu_{\mathrm{i}}$
& Free running
& --
\\

Corato-Zanarella et al.~\cite{Corato-Zanarella:23}
& CW laser
& Hybrid
& 404--785 nm
& --
& Few-kHz $\Delta\nu_{\mathrm{i}}$
& Free running
& $\leq 33.9$ GHz
\\

Siddharth et al.~\cite{Siddharth:25}
& CW laser
& Hybrid
& 461 nm
& --
& $<30$ kHz $\Delta\nu_{\mathrm{i}}$
& Free running
& 125 MHz\tnote{b}
\\

Franken et al.~\cite{Franken:25}
& CW laser
& Hybrid
& 408 nm
& --
& $\sim300$ kHz $\Delta\nu_{\mathrm{i}}$
& Free running
& --
\\

\bottomrule
\end{tabular}

\begin{tablenotes}[flushleft]
\small
\item[a] Dispersive wave in 780-nm band.
\item[b] Or 900-MHz MHF tuning range for $<600$ kHz $\Delta\nu_{\mathrm{i}}$.
\end{tablenotes}

\end{threeparttable}%
}

\end{table*}


\noindent \textbf{Acknowledgments}: 
We thank Dengke Chen and Zhen Chen for assistance in chip design and characterization, and Bing Yang for the helpful suggestions.
This work is supported by National Natural Science Foundation of China (Grant No. U25D9005, 12404436, 12404417, 62405202, 12494601), 
Quantum Science and Technology--National Science and Technology Major Project (Grant No. 2023ZD0301500 and 2021ZD0300800), 
Shenzhen Science and Technology Program (Grant No. RCJC20231211090042078 and KQTD20251201091722036), 
and National Key R\&D Program of China (Grant No. 2024YFA1409300). 

\noindent \textbf{Author contributions}: 
J. Long, X. Y., W. S. and Shichang L.  performed the microcomb experiment. 
X. Y., Z. D., W. S. and S. Y. performed the BEC experiment.
Shuyi L. and Y.-H. L. designed the Si$_3$N$_4$ chips. 
S. H., Z. S., Z. Z., J. S., Y. H., B. S. and C. S. fabricated and characterized the Si$_3$N$_4$ chips. 
H. T., H. L., X. S., H. N. and Z. N. fabricated and characterized the laser diodes.
W. S., X. Y., J. Long, H. T. and J. Liu analyzed the data and wrote the manuscript, with the input from others.
J. Liu supervised the project.

\noindent \textbf{Disclosures} 
C. S. and J. Liu are co-founders of Qaleido Photonics, a start-up that is developing heterogeneous silicon nitride integrated photonics technologies. 
Others declare no conflicts of interest.

\noindent \textbf{Data Availability Statement}: 
The code and data used to produce the plots in this work will be released in the \texttt{Zenodo} repository upon publication of this work. 

\end{footnotesize}
\bibliographystyle{apsrev4-1}
\bibliography{bibliography}
\end{document}


\title{Supplementary Information to: A photonic integrated comb engine for ultracold quantum gases}

\author{Wei Sun}
\thanks{These authors contributed equally to this work.}
\affiliation{International Quantum Academy, Shenzhen 518048, China}

\author{Xiaoying Yan}
\thanks{These authors contributed equally to this work.}
\affiliation{International Quantum Academy, Shenzhen 518048, China}
\affiliation{Shenzhen Institute for Quantum Science and Engineering, Southern University of Science and Technology, Shenzhen 518055, China}

\author{Jinbao Long}
\thanks{These authors contributed equally to this work.}
\affiliation{International Quantum Academy, Shenzhen 518048, China}

\author{Sanli Huang}
\thanks{These authors contributed equally to this work.}
\affiliation{International Quantum Academy, Shenzhen 518048, China}
\affiliation{Hefei National Laboratory, University of Science and Technology of China, Hefei 230088, China}

\author{Zhixin Duan}
\thanks{These authors contributed equally to this work.}
\affiliation{International Quantum Academy, Shenzhen 518048, China}
\affiliation{Shenzhen Institute for Quantum Science and Engineering, Southern University of Science and Technology, Shenzhen 518055, China}

\author{Hao Tan}
\affiliation{International Quantum Academy, Shenzhen 518048, China}
\affiliation{Hefei National Laboratory, University of Science and Technology of China, Hefei 230088, China}

\author{Zhenyuan Shang}
\affiliation{International Quantum Academy, Shenzhen 518048, China}
\affiliation{Shenzhen Institute for Quantum Science and Engineering, Southern University of Science and Technology, Shenzhen 518055, China}

\author{Zeying Zhong}
\affiliation{International Quantum Academy, Shenzhen 518048, China}
\affiliation{Shenzhen Institute for Quantum Science and Engineering, Southern University of Science and Technology, Shenzhen 518055, China}

\author{Jiahao Sun}
\affiliation{International Quantum Academy, Shenzhen 518048, China}
\affiliation{Shenzhen Institute for Quantum Science and Engineering, Southern University of Science and Technology, Shenzhen 518055, China}

\author{Yue Hu}
\affiliation{International Quantum Academy, Shenzhen 518048, China}
\affiliation{Shenzhen Institute for Quantum Science and Engineering, Southern University of Science and Technology, Shenzhen 518055, China}

\author{Shichang Li}
\affiliation{International Quantum Academy, Shenzhen 518048, China}
\affiliation{Shenzhen Institute for Quantum Science and Engineering, Southern University of Science and Technology, Shenzhen 518055, China}

\author{Baoqi Shi}
\affiliation{International Quantum Academy, Shenzhen 518048, China}

\author{Yi-Han Luo}
\affiliation{International Quantum Academy, Shenzhen 518048, China}

\author{Shuyi Li}
\affiliation{International Quantum Academy, Shenzhen 518048, China}

\author{Chen Shen}
\affiliation{International Quantum Academy, Shenzhen 518048, China}
\affiliation{Qaleido Photonics, Shenzhen 518048, China}

\author{Hanqing Liu}
\affiliation{State Key Laboratory of Optoelectronic Materials and Devices, Institute of Semiconductors, Chinese Academy of Sciences, Beijing 100083, China}
\affiliation{Center of Materials Science and Optoelectronics Engineering, University of Chinese Academy of Sciences, Beijing 100083, China}

\author{Xiangbin Su}
\affiliation{State Key Laboratory of Optoelectronic Materials and Devices, Institute of Semiconductors, Chinese Academy of Sciences, Beijing 100083, China}
\affiliation{Center of Materials Science and Optoelectronics Engineering, University of Chinese Academy of Sciences, Beijing 100083, China}

\author{Haiqiao Ni}
\affiliation{State Key Laboratory of Optoelectronic Materials and Devices, Institute of Semiconductors, Chinese Academy of Sciences, Beijing 100083, China}
\affiliation{Center of Materials Science and Optoelectronics Engineering, University of Chinese Academy of Sciences, Beijing 100083, China}

\author{Zhichuan Niu}
\affiliation{International Quantum Academy, Shenzhen 518048, China}
\affiliation{State Key Laboratory of Optoelectronic Materials and Devices, Institute of Semiconductors, Chinese Academy of Sciences, Beijing 100083, China}
\affiliation{Center of Materials Science and Optoelectronics Engineering, University of Chinese Academy of Sciences, Beijing 100083, China}

\author{Shengjun Yang}
\affiliation{International Quantum Academy, Shenzhen 518048, China}
\affiliation{Shenzhen Institute for Quantum Science and Engineering, Southern University of Science and Technology, Shenzhen 518055, China}

\author{Junqiu Liu}
\email[]{liujq@iqasz.cn}
\affiliation{International Quantum Academy, Shenzhen 518048, China}
\affiliation{Hefei National Laboratory, University of Science and Technology of China, Hefei 230088, China}

\maketitle

\section{Characterization of microresonators in the 780-nm band}

\begin{figure*}[b!]
\renewcommand{\figurename}{Supplementary Figure}
\centering
\includegraphics{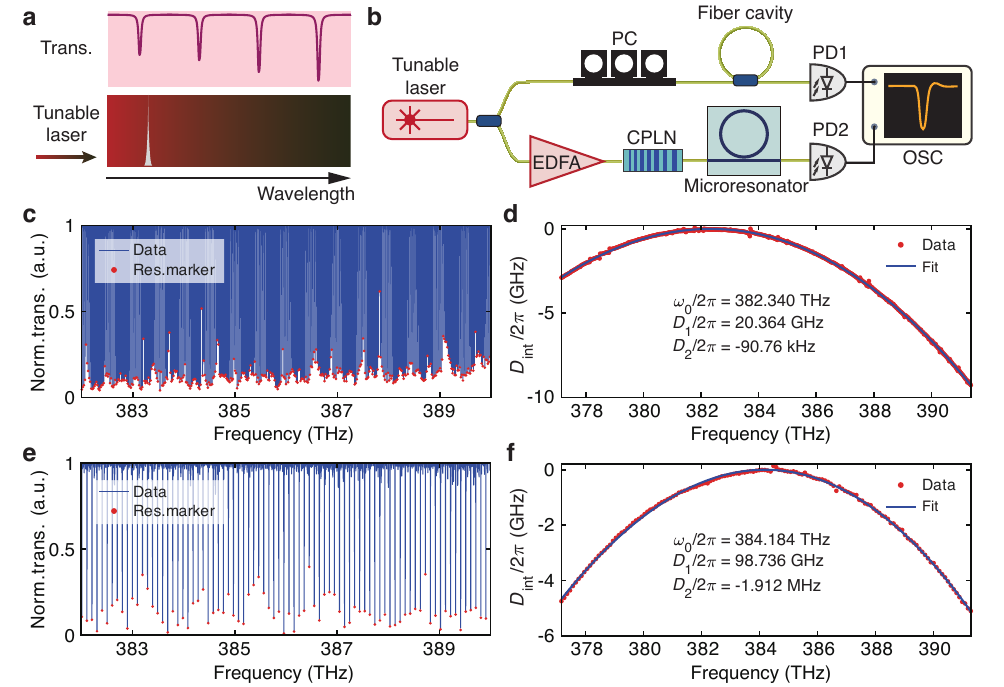}
\caption{
\textbf{Characterization of Si$_3$N$_4$ microresonators in the 780-nm band}.
\textbf{a,} 
Principle of broadband microresonator characterization using a frequency-calibrated swept laser.
\textbf{b,}
Experimental setup of the VSA in the 780-nm band.
A tunable laser is frequency-referenced by a calibrated fiber cavity, amplified by an EDFA and subsequently frequency-doubled in a CPLN waveguide to probe the microresonator.
PC, polarization controller.
PD, photodetector.
OSC, oscilloscope.
\textbf{c,}
Broadband transmission spectrum of the microresonator used in Fig. 2f in the main text.
Red markers indicate the identified cavity resonances.
\textbf{d,}
Integrated dispersion $D_{\mathrm{int}}/2\pi$ extracted from \textbf{c}.
The dispersion is expanded around a microresonator resonance $\omega_0/2\pi=382.340$ THz, yielding $D_1/2\pi=20.364$ GHz and $D_2/2\pi=-90.76$ kHz. 
\textbf{e,}
Broadband transmission spectrum of the microresonator used in Fig. 2g in the main text.
\textbf{f,}
Integrated dispersion $D_{\mathrm{int}}/2\pi$ extracted from \textbf{e}.
Referenced to a microresonator resonance $\omega_0/2\pi=384.184$ THz. 
The fit gives $D_1/2\pi=98.736$ GHz and $D_2/2\pi=-1.912$ MHz.
}
\label{Fig:780setup}
\end{figure*}

The Si$_3$N$_4$ microresonators employed for 780-nm microcomb generation are characterized using a near-visible-band vector spectrum analyzer (VSA) covering the 766--795~nm wavelength range (377.1--391.4 THz)~\cite{Shi:25}. 
The measurement principle and experimental configuration are shown in Supplementary Fig.~\ref{Fig:780setup}a, b.
A mode-hop-free tunable external-cavity diode laser (ECDL, Santec TSL-570) operating in the near-infrared is continuously swept across the measurement range and divided into two optical paths.
In the reference path, the laser transmission through a pre-calibrated fiber cavity with nearly uniform resonance spacing~\cite{Luo:24} is simultaneously recorded, providing a dense set of frequency markers during the sweep.
In the measurement path, the laser is amplified by an erbium-doped fiber amplifier (EDFA) and subsequently frequency doubled in a chirped periodically poled lithium niobate (CPLN) waveguide.
The generated second-harmonic field is coupled to the microresonator under test.
Because the fundamental and second-harmonic fields originate from the same swept laser, the calibrated frequency of the fiber-cavity signal can be mapped directly onto the near-visible probe frequency, enabling broadband reconstruction of the microresonator transmission spectrum.

The resonance frequencies are first identified from the calibrated transmission trace, as shown in Supplementary Fig.~\ref{Fig:780setup}c, e.
Supplementary Fig.~\ref{Fig:780setup}c and e show the transmission spectra of the microresonators in Fig. 2f and g in the main text, respectively.
While measurement is performed over the 377.1--391.4 THz band, only the 382--390 THz range is displayed for clarity.
In the transmission trace, each resonance is then locally fitted using the coupled-mode response~\cite{Li:13}
\begin{equation}
	T(\Delta\omega)	= \left|1-\frac{\kappa_{\rm ex} \left[i\Delta\omega+(\kappa_0+\kappa_{\rm ex})/2\right]}{\left[i\Delta\omega+(\kappa_0+\kappa_{\rm ex})/2\right]^2+\kappa_{\rm c}^2/4}\right|^2 ,
	\label{eq:SI_1}
\end{equation}
where $\Delta\omega = \omega_0 - \omega$ is the detuning of the microresonator resonance $\omega_0$ from the tunable laser frequency $\omega$, $\kappa_0$ is the intrinsic loss rate, $\kappa_{\rm ex}$ is the external coupling rate, and $\kappa_{\rm c}$ accounts for coupling between the clockwise and counter-clockwise cavity modes when resonance splitting is present.
For resonances without appreciable mode splitting, Eq.~\ref{eq:SI_1} reduces to the conventional Lorentzian cavity response.
A representative resonance fit alongside the extracted values for $\kappa_0/2\pi$ and $\kappa_{\rm{ex}}/2\pi$ are shown in Fig. 2b in the main text.
In addition, the intrinsic quality factor of each resonance is obtained from $Q_0=\omega_0/\kappa_0$.
The histogram of $Q_0$ is displayed in Fig. 2c in the main text, corresponding to a most probable value of $Q_0=9.5\times10^6$.

Supplementary Fig.~\ref{Fig:780setup}d and f show the dispersion over the full 377.1--391.4 THz span obtained from the transmission traces in Supplementary Fig.~\ref{Fig:780setup}c and e, respectively.
For the microresonator in Fig. 2f in the main text, referenced to a microresonator resonance at $\omega_0/2\pi=382.340$ THz, the fit in Supplementary Fig.~\ref{Fig:780setup}d yields $D_1/2\pi=20.364$ GHz and $D_2/2\pi=-90.76$ kHz.
For the microresonator in Fig. 2g in the main text, referenced to a microresonator resonance at $\omega_0/2\pi=384.184$ THz, Supplementary Fig.~\ref{Fig:780setup}f gives $D_1/2\pi=98.736$ GHz and $D_2/2\pi=-1.912$ MHz.
The negative $D_2$ values indicate normal group-velocity dispersion (GVD) for both devices.

\vspace{0.5cm}
\section{Power and wavelength of FP and DBR lasers}
The output power and emission wavelength of the FP diode are shown in Supplementary Fig.~\ref{Fig:S2}a. 
Above a threshold current of 40 mA, the FP diode's output power increases linearly, reaching 185 mW at 200 mA.
The emission wavelength spans 5.1 nm over the measured current range at 25 $^{\circ}\mathrm{C}$ and exhibits multiple discrete wavelength jumps owing to longitudinal-mode hopping. 
At a fixed current of 130 mA, the emission wavelength redshifts by 2.0 nm as the temperature increases from 20 to 30 $^{\circ}\mathrm{C}$, with additional discrete wavelength jumps, as shown in Supplementary Fig.~\ref{Fig:S2}c.

The corresponding characteristics of the DBR laser are shown in Supplementary Fig.~\ref{Fig:S2}b, d. 
As the current increases from 0 to 250 mA, the DBR laser's output power increases to 183 mW with a threshold current of 70 mA.
The emission wavelength range is 0.2 nm during current tuning at 25 $^{\circ}\mathrm{C}$.
At a fixed current of 130 mA, the emission wavelength redshifts smoothly by 0.5 nm as the temperature increases from 20 to 30 $^{\circ}$C.

\begin{figure*}[b!]
\renewcommand{\figurename}{Supplementary Figure}
\centering
\includegraphics{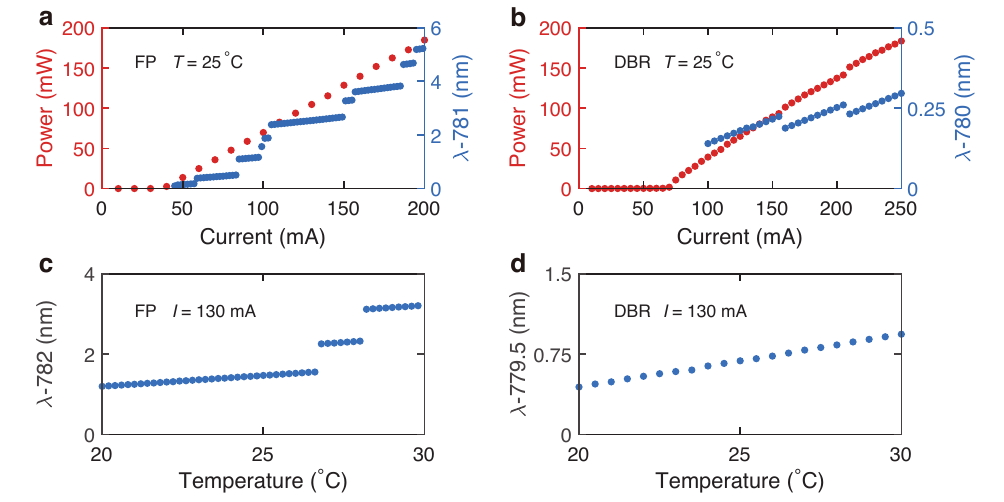}
\caption{
\textbf{Power and wavelength of FP and DBR lasers}.
\textbf{a, b,}
Measured output power (red circles) and emission wavelength (blue circles) of the FP laser (panel \textbf{a}) and the DBR laser (panel \textbf{b}) as functions of current at 25 $^{\circ}\mathrm{C}$.
\textbf{c, d,} 
Measured emission wavelength of the FP laser (panel \textbf{c}) and the DBR laser (panel \textbf{d}) as functions of temperature at a fixed current of 130 mA.
}
\label{Fig:S2}
\end{figure*}

\vspace{0.5cm}
\section{Laser beam quality and coupling efficiency characterization}
Laser diodes generally emit beams with approximately Gaussian transverse intensity profiles.
Owing to asymmetric optical confinement along the two transverse axes of a semiconductor laser cavity, the emitted beam has different diameters and divergence angles along two orthogonal axes, resulting in an elliptical profile.
The propagation behavior of a Gaussian beam differs between the near and far fields.
As illustrated in Supplementary Fig.~\ref{Fig:S4}a, the beam diameter $D$ increases during free-space propagation.
In the near field, the beam diameter varies nonlinearly with propagation distance.
In the far field, the beam diameter increases approximately linearly with propagation distance.
The divergence half-angle $\theta$ characterizes the rate of beam expansion in the far field.

\begin{figure*}[b!]
\renewcommand{\figurename}{Supplementary Figure}
\centering
\includegraphics{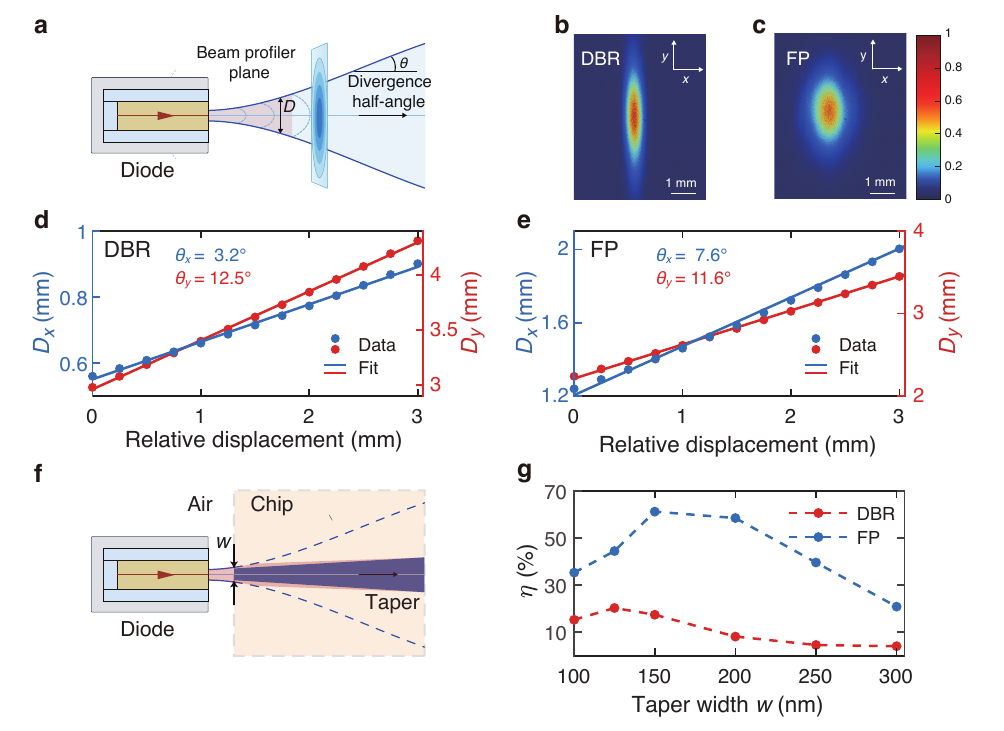}
\caption{
\textbf{Beam qualities and coupling efficiency of the FP and DBR lasers}.
\textbf{a,}
Schematic of the beam-divergence half-angle measurement. 
A beam profiler is placed on the laser propagation line, measuring the full width at half maximum (FWHM) beam diameter $D$ in the transverse plane.
\textbf{b, c,} 
Representative normalized transverse intensity profiles of the DBR (panel \textbf{b}) and FP (panel \textbf{c}) lasers. hgyhgyyyyyy
\textbf{d, e,}
Measured FWHM beam diameters $D_y$ (red) and $D_x$ (blue) as functions of relative displacement for the DBR (panel \textbf{d}) and FP (panel \textbf{e}) lasers. 
\textbf{f,}
Schematic of laser coupling between the laser diode and an on-chip waveguide.
The waveguide has an inverse taper with a tip width $w$ at the chip facet.
\textbf{g,}
Measured coupling efficiency $\eta$ as a function of taper width $w$ for the DBR (red) and FP (blue) lasers. 
}
\label{Fig:S4}
\end{figure*}

A beam profiler (CINOGY CinCam CMOS-1201-EL) is leveraged to measure the transverse beam diameters at different relative displacements.
Representative normalized intensity profiles of the DBR and FP lasers are shown in Supplementary Fig.~\ref{Fig:S4}b, c, respectively.
Both lasers exhibit elliptical, approximately Gaussian beam profiles.
The beam diameters $D_x$ and $D_y$ are determined from the full width at half maximum (FWHM) of the intensity profiles along the $x$ and $y$ directions.
For each transverse direction $i=x,y$, the FWHM divergence half-angle $\theta_i$ is extracted by fitting the far-field beam diameter using $D_i(z)=D_{i,0}+2z\tan\theta_i$, where $z$ is the relative displacement of the beam profiler along the propagation direction, $D_{i,0}$ is the fitted FWHM beam diameter at $z=0$, and the slope of the linear fit is $2\tan\theta_i$.
As shown in Supplementary Fig.~\ref{Fig:S4}d, the fitted divergence half-angles of the DBR laser are $\theta_x=3.2^\circ$ and $\theta_y=12.5^\circ$. 
Supplementary Fig.~\ref{Fig:S4}e shows that the corresponding divergence half-angles of the FP laser are $\theta_x=7.6^\circ$ and $\theta_y=11.6^\circ$.

Efficient coupling of the laser output into the on-chip waveguides is essential for integrating laser diodes with photonic chips. 
We employ inverse waveguide tapers at the chip facets to achieve mode matching between the laser-diode beam profile and the on-chip waveguide mode, as illustrated in Supplementary Fig.~\ref{Fig:S4}f.
Because the optical mode matching depends on both the beam profile of the laser diode and the taper geometry, the coupling efficiency $\eta$ varies with the taper width $w$.
Supplementary Fig.~\ref{Fig:S4}g shows the measured coupling efficiencies $\eta$ of the DBR and FP lasers as a function of the taper width $w$.
The maximum measured coupling efficiencies are approximately 20\% at $w=125$ nm for the DBR laser and 61\% at $w=150$ nm for the FP laser.


\vspace{0.5cm}
\section{Principle of SIL process}

We consider one diode-laser mode coupled to a pair of counterpropagating microresonator modes, neglecting Kerr and thermal resonance shifts~\cite{Kondratiev:17}.
The diode field $A_{\mathrm L}$ excites the forward microresonator field $A^+$, while Rayleigh backscattering couples part of the intracavity field into the backward field $A^-$, which returns to the laser and provides resonant optical feedback.
In a frame rotating at the generated angular frequency $\omega$, the linear microresonator equations are
\begin{equation}
\begin{aligned}
\frac{\mathrm d A^+}{\mathrm d t}&=-\left(\frac{\kappa}{2}+i\Delta\omega\right)A^+ + i\frac{\kappa_{\mathrm c}}{2}A^-+F,\\
\frac{\mathrm d A^-}{\mathrm d t}&=-\left(\frac{\kappa}{2}+i\Delta\omega\right)A^- + i\frac{\kappa_{\mathrm c}}{2}A^+,
\end{aligned}
\label{eq:SI_SIL_linear_fields}
\end{equation}
where $\Delta\omega=\omega_0-\omega$, $\omega_0$ is the unperturbed microresonator resonance frequency, $\kappa=\kappa_0+\kappa_{\mathrm{ex}}$ is the loaded linewidth, and $\kappa_{\mathrm c}/2$ is the coupling rate between the counterpropagating modes.
The driving field is
\begin{equation}
F=\sqrt{\kappa_{\mathrm{ex}}\kappa_{\mathrm L}}\,A_{\mathrm L}e^{i\phi_{\mathrm B}},
\end{equation}
where $\kappa_{\mathrm L}$ is the effective laser output-coupling rate and $\phi_{\mathrm B}$ is the one-way feedback-path phase, including fixed coupling phases.

For stationary fields, $\mathrm dA^\pm/\mathrm dt=0$, and Eq.~\eqref{eq:SI_SIL_linear_fields} becomes
\begin{equation}
\left(\frac{\kappa}{2}+i\Delta\omega\right)A^+ - i\frac{\kappa_{\mathrm c}}{2}A^-=F,\qquad
\left(\frac{\kappa}{2}+i\Delta\omega\right)A^- - i\frac{\kappa_{\mathrm c}}{2}A^+=0.
\label{eq:SI_SIL_stationary_fields}
\end{equation}
We define the normalized generation detuning and backscattering strength as
\begin{equation}
\zeta=\frac{2\Delta\omega}{\kappa}=\frac{2(\omega_0-\omega)}{\kappa},\qquad \beta=\frac{\kappa_{\mathrm c}}{\kappa}.
\end{equation}
The backward-field response is derived as
\begin{equation}
\frac{A^-}{F}=\frac{2}{\kappa}\frac{i\beta}{(1+i\zeta)^2+\beta^2}.
\label{eq:SI_SIL_linear_response}
\end{equation}
Equation~\eqref{eq:SI_SIL_linear_response} represents the complex frequency-dependent field that is backscattered by the microresonator and returned to the laser.

We next relate this optical feedback to the generated laser frequency.
The free-running laser--microresonator detuning is defined as
\begin{equation}
\xi=\frac{2(\omega_0-\omega_{\mathrm{fr}})}{\kappa}=\frac{2\delta\omega_{\mathrm L}+\gamma\alpha_{\mathrm g}}{\kappa},
\label{eq:SI_SIL_linear_detunings}
\end{equation}
where $\omega_{\mathrm{fr}}=\omega_{\mathrm L}-\alpha_{\mathrm g}\gamma/2$ is the free-running laser frequency including the cold laser-cavity resonance frequency $\omega_{\rm L}$ and the gain-induced frequency shift $\alpha_{\mathrm g}\gamma/2$, $\delta\omega_{\mathrm L}=\omega_0-\omega_{\mathrm L}$, $\gamma$ is the laser-mode loss rate, and $\alpha_{\mathrm g}$ is the linewidth-enhancement factor.
The difference between the free-running and generated detunings is
\begin{equation}
\xi-\zeta=\frac{2(\omega-\omega_{\mathrm{fr}})}{\kappa}.
\label{eq:SI_SIL_detuning_difference}
\end{equation}

The field returned from the microresonator produces a complex feedback rate $H=\sqrt{\kappa_{\mathrm{ex}}\kappa_{\mathrm L}}\,e^{i\phi_{\mathrm B}}A^- / A_{\mathrm L}$. 
To show how $H$ changes the laser frequency, we denote the saturated modal gain by $G$.
The stationary laser amplitude and phase equations can be written as
\begin{equation}
0=\frac{G-\gamma}{2}+\operatorname{Re}H,\qquad
\omega-\omega_{\mathrm L}=-\frac{\alpha_{\mathrm g}G}{2}-\operatorname{Im}H, 
\label{eq:SI_SIL_laser_phase}
\end{equation}
which yields
\begin{equation}
\omega-\omega_{\mathrm L}=-\frac{\alpha_{\mathrm g}\gamma}{2}+\alpha_{\mathrm g}\operatorname{Re}H-\operatorname{Im}H.
\end{equation}
Using $\omega_{\mathrm{fr}}=\omega_{\mathrm L}-\alpha_{\mathrm g}\gamma/2$, the feedback-induced frequency shift becomes
\begin{equation}
\omega-\omega_{\mathrm{fr}}=\alpha_{\mathrm g}\operatorname{Re}H-\operatorname{Im}H=-\operatorname{Im}\left[(1-i\alpha_{\mathrm g})H\right].
\label{eq:SI_SIL_feedback_frequency_shift}
\end{equation}
Thus, both the phase and amplitude of the returned field influence the laser frequency because the linewidth-enhancement factor couples gain fluctuations to phase fluctuations.

Using $F=\sqrt{\kappa_{\mathrm{ex}}\kappa_{\mathrm L}}\,A_{\mathrm L}e^{i\phi_{\mathrm B}}$ and Eq.~\eqref{eq:SI_SIL_linear_response}, the feedback rate becomes
\begin{equation}
H=\frac{2\kappa_{\mathrm{ex}}\kappa_{\mathrm L}}{\kappa}e^{i2\phi_{\mathrm B}}\frac{i\beta}{(1+i\zeta)^2+\beta^2}.
\label{eq:SI_SIL_feedback_response}
\end{equation}
The factor $e^{i2\phi_{\mathrm B}}$ accounts for the round-trip phase accumulated along the laser--microresonator feedback path.
Combining Eqs.~\eqref{eq:SI_SIL_detuning_difference}, \eqref{eq:SI_SIL_feedback_frequency_shift}, and \eqref{eq:SI_SIL_feedback_response} gives the implicit stationary tuning relation
\begin{equation}
\zeta=\xi+K_0\operatorname{Im}\left[(1-i\alpha_{\mathrm g})e^{i2\phi_{\mathrm B}}\frac{i\beta}{(1+i\zeta)^2+\beta^2}\right],
\qquad K_0=\frac{4\kappa_{\mathrm{ex}}\kappa_{\mathrm L}}{\kappa^2}.
\label{eq:SI_SIL_linear_tuning}
\end{equation}
Defining $K=2K_0\beta\sqrt{1+\alpha_{\mathrm g}^2}$ and $\Theta=2\phi_{\mathrm B}-\arctan\alpha_{\mathrm g}+\pi/2$ gives
\begin{equation}
\xi=\zeta+\frac{K}{2}\frac{2\zeta\cos\Theta-(1+\beta^2-\zeta^2)\sin\Theta}{(1+\beta^2-\zeta^2)^2+4\zeta^2}.
\label{eq:SI_SIL_linear_tuning_real}
\end{equation}
Equation~\eqref{eq:SI_SIL_linear_tuning_real} describes the stationary relation between the frequency of the free-running laser and the generated frequency under optical feedback.

On a stable branch, a small variation of the free-running detuning is related to the generated detuning by
\begin{equation}
\delta\xi=\mathcal{S}\,\delta\zeta,\qquad \mathcal{S}=\left.\frac{\mathrm d\xi}{\mathrm d\zeta}\right|_{\zeta_*},
\end{equation}
where $\zeta_*$ denotes the operating point.
Because $\delta\xi=-2\delta\omega_{\mathrm{fr}}/\kappa$ and $\delta\zeta=-2\delta\omega/\kappa$, the generated-frequency fluctuation is $\delta\omega=\delta\omega_{\mathrm{fr}}/\mathcal{S}$.
The stabilization coefficient $\mathcal S$ therefore quantifies the suppression of small free-running laser-frequency fluctuations.
Near the resonant operating point $\zeta_*=0$ and at optimum feedback phase $\Theta=0$, Eq.~\eqref{eq:SI_SIL_linear_tuning_real} reduces near resonance to
\begin{equation}
\xi\simeq\left[1+\frac{K}{(1+\beta^2)^2}\right]\zeta,
\end{equation}
and hence
\begin{equation}
\left.\mathcal{S}\right|_{\zeta_*=0,\,\Theta=0}=1+\frac{K}{(1+\beta^2)^2}.
\label{eq:SI_SIL_stabilization_coefficient}
\end{equation}
In the quasistatic white-frequency-noise limit, the frequency-noise power spectral density is reduced by $|\mathcal S|^{-2}$.
Since the Lorentzian linewidth is proportional to the white-frequency-noise level, the linewidth ratio is
\begin{equation}
\frac{\Delta\nu_{\mathrm{SIL}}}{\Delta\nu_{\mathrm{fr}}}\simeq\frac{1}{|\mathcal S|^2}.
\label{eq:SI_SIL_linewidth_reduction}
\end{equation}
Defining the loaded resonator quality factor $Q_{\mathrm m}$, loading fraction $\eta_{\mathrm{ex}}$, resonant amplitude reflection magnitude $\Gamma_{\mathrm m}$, and the diode-laser cavity quality factor $Q_\mathrm{d}$ (assuming $\gamma\simeq\kappa_\mathrm{L}$) by
\begin{equation}
Q_{\mathrm m}=\frac{\omega_0}{\kappa},\qquad \eta_{\mathrm{ex}}=\frac{\kappa_{\mathrm{ex}}}{\kappa},\qquad \Gamma_{\mathrm m}=\frac{2\eta_{\mathrm{ex}}\beta}{1+\beta^2},\qquad Q_{\mathrm d}\simeq\frac{\omega_0}{\kappa_{\mathrm L}},
\end{equation}
the linewidth ratio is
\begin{equation}
\frac{\Delta\nu_{\mathrm{SIL}}}{\Delta\nu_{\mathrm{fr}}}
\simeq\left[1+\frac{4\Gamma_{\mathrm m}\sqrt{1+\alpha_{\mathrm g}^2}}{1+\beta^2}\frac{Q_{\mathrm m}}{Q_{\mathrm d}}\right]^{-2}.
\label{eq:SI_SIL_linewidth_quality_factors}
\end{equation}
For weak backscattering ($\beta\ll1$) and strong stabilization, 
the linewidth ratio then reduces to
\begin{equation}
\frac{\Delta\nu_{\mathrm{SIL}}}{\Delta\nu_{\mathrm{fr}}}\simeq\frac{Q_{\mathrm d}^2}{Q_{\mathrm m}^2}\frac{1}{16\Gamma_{\mathrm m}^2(1+\alpha_{\mathrm g}^2)}.
\label{eq:SI_SIL_linewidth_quality_limit}
\end{equation}
At fixed $\Gamma_{\mathrm m}$, $Q_{\mathrm d}$, and $\alpha_{\mathrm g}$, the SIL linewidth therefore scales as $Q_{\mathrm m}^{-2}$.
Resonator fluctuations, feedback-path noise, finite propagation delay, and finite-bandwidth effects limit this ideal narrowing.


\vspace{0.5cm}
\section{Numerical simulation of SIL platicons}

To model self-injection-locked platicons in the 780-nm band, we start from a universal Kerr--thermal model developed in Ref.~\cite{LiS:25} and subsequently reduce it by eliminating dynamical variables to facilitate the numerical simulations.
The starting model describes the carrier density and optical field in the semiconductor laser, the forward- and backward-propagating fields in the microresonator, and the thermally induced resonance shift.

We introduce the following field rescalings
\begin{equation}
A_{\rm L}=\sqrt{\frac{\epsilon_0 n_{\rm L}^2V_{\rm act}}{\hbar\omega_{\rm L}}}\,E_{\rm L},\qquad A^\pm=\sqrt{\frac{\epsilon_0 n_0^2V_{\rm eff}}{2\hbar\omega_0}}\,E^\pm,
\label{eq:SI_source_field_normalizati4on}
\end{equation}
where $\epsilon_0$ is the vacuum permittivity, $\hbar$ is the reduced Planck constant, $E_{\rm L}$ is the electric-field amplitude in the semiconductor laser cavity, and $E^\pm$ are the forward- and backward-propagating electric-field amplitudes in the microresonator.
The quantities $n_{\rm L}$ and $V_{\rm act}$ are the refractive index and effective optical mode volume of the laser active region, respectively, while $n_0$ and $V_{\rm eff}$ are the corresponding refractive index and effective mode volume of the microresonator mode.
With these rescaled field amplitudes, the starting equations can be written as
\begin{align}
\frac{\partial N}{\partial t}&=\frac{I_\mathrm{inj}}{eV_{\rm act}}-\frac{1}{V_{\rm act}}\left[G_N(N-N_0)+G_0\right]|A_{\rm L}|^2-\kappa_NN,
\label{eq:SI_source_carrier}\\
\frac{\partial A_{\rm L}}{\partial t}&=\frac{1}{2}(1+i\alpha_{\rm g})G_N(N-N_0)A_{\rm L}+i e^{i(\Phi_{\rm fr}(t)+\phi_{\rm fb,L})}\kappa_{\rm L\leftarrow R}A^-_0,
\label{eq:SI_source_laser}\\
\frac{\partial A^+_\mu}{\partial t}&=-\left(\frac{\kappa}{2}+i\sum_{n=2}^{\infty}\frac{D_n}{n!}\mu^n-iT_{\rm h}\right)A^+_\mu+ig\left(\mathcal F[|A^+|^2A^+]_\mu+2A^+_\mu\sum_{\mu'}|A^-_{\mu'}|^2\right)\notag\\
&\quad+i\frac{\kappa_{\rm c}}{2}A^-_\mu+i e^{i(-\Phi_{\rm fr}(t)+\phi_{\rm fb,R})}\kappa_{\rm R\leftarrow L}A_{\rm L}\delta_{\mu0},
\label{eq:SI_source_forward}\\
\frac{\partial A^-_\mu}{\partial t}&=-\left(\frac{\kappa}{2}+i\sum_{n=2}^{\infty}\frac{D_n}{n!}\mu^n-iT_{\rm h}\right)A^-_\mu+ig\left(\mathcal F[|A^-|^2A^-]_\mu+2A^-_\mu\sum_{\mu'}|A^+_{\mu'}|^2\right)+i\frac{\kappa_{\rm c}}{2}A^+_\mu,
\label{eq:SI_source_backward}\\
\frac{\partial T_{\rm h}}{\partial t}&=\frac{1}{\tau_{\rm h}}\left[K_T\sum_\mu\left(|A^+_\mu|^2+|A^-_\mu|^2\right)-T_{\rm h}\right].
\label{eq:SI_source_thermal}
\end{align}
Here, $N$ is the carrier density in the active region of the semiconductor laser, $I_\mathrm{inj}$ is the injection current, and $e$ is the elementary charge.
The quantity $N_0$ denotes the carrier density at the free-running operating point.
The carrier-dependent gain is linearized around $N_0$ as $G(N)\simeq G_0+G_N(N-N_0)$, where $G_0=G(N_0)$ is the gain at the free-running operating point and $G_N=\left.\frac{\partial G}{\partial N}\right|_{N_0}$ is the differential gain coefficient.
The parameter $\kappa_N$ is the carrier-density decay rate.
The quantity $A_{\rm L}$ is the rescaled laser-cavity field, while $A^\pm_\mu$ are the forward- and backward-propagating fields of the $\mu$-th microresonator mode.
The integer $\mu$ denotes the relative mode number with respect to the pumped resonance, with $\mu=0$ corresponding to the directly driven mode and $A^-_0$ denoting its backward-propagating component.
The Kronecker delta $\delta_{\mu0}$ therefore restricts direct laser excitation to the pumped mode.
The operator $\mathcal F[\cdots]_\mu$ denotes the $\mu$-th Fourier component of the corresponding angular-domain intracavity field.
The coefficient $D_n$ is the $n$-th-order dispersion coefficient in the expansion of the microresonator resonance frequencies around the pumped mode.
The Kerr nonlinear coupling coefficient is
\begin{equation}
g=\frac{\hbar\omega_0^2cn_2}{n_0^2V_{\rm eff}},
\label{eq:SI_Kerr_coefficient}
\end{equation}
where $c$ is the speed of light in vacuum and $n_2$ is the Kerr nonlinear refractive index. 
The quantity $T_{\rm h}$ denotes the thermally induced shift of the microresonator resonance angular frequency.
The coefficient $K_T$ relates the summed intracavity modal quantity $\sum_\mu(|A^+_\mu|^2+|A^-_\mu|^2)$ to the corresponding thermal resonance shift, and $\tau_{\rm h}$ is the thermal relaxation time.
The laser-to-resonator and resonator-to-laser field-coupling coefficients are
\begin{equation}
\kappa_{\rm R\leftarrow L}=\sqrt{\frac{\kappa_{\rm ex}T_{\rm L}T_{\rm cR}}{\tau_{\rm L}}},\qquad \kappa_{\rm L\leftarrow R}=\sqrt{\frac{\kappa_{\rm ex}T_{\rm L}T_{\rm cL}}{\tau_{\rm L}}},
\label{eq:SI_source_couplings}
\end{equation}
where $T_{\rm L}$ is the power transmission of the laser output facet, $T_{\rm cR}$ is the power-coupling efficiency from the laser cavity to the microresonator, $T_{\rm cL}$ is the power-coupling efficiency from the microresonator back into the laser cavity, and $\tau_{\rm L}$ is the round-trip time of the laser cavity.
The fixed propagation phases associated with these two coupling directions are denoted by $\phi_{\rm fb,R}$ and $\phi_{\rm fb,L}$, respectively.
Because the laser and microresonator frequencies differ by only a small fraction of the optical carrier frequency, the phase difference accumulated over the short feedback path is negligible, and we therefore use a common one-way feedback phase $\phi_{\rm s}$ such that $\phi_{\rm fb,L}\simeq\phi_{\rm fb,R}\equiv\phi_{\rm s}$.
$\Phi_\mathrm{fr}(t) = \int_0^t \Delta_\mathrm{fr}(t^{'}) \mathrm dt'$ is the accumulated relative phase with $\Delta_\mathrm{fr}\equiv\omega_{\rm fr}-\omega_0$ being the free-running laser–microresonator detuning.
Accordingly, the normalized free-running laser--microresonator detuning defined in the preceding section is
\begin{equation}
\xi=\frac{2(\omega_0-\omega_{\rm fr})}{\kappa}=-\frac{2\Delta_{\rm fr}}{\kappa}.
\label{eq:SI_source_to_xi}
\end{equation}

We next normalize time and optical field amplitudes according to
\begin{equation}
\tilde t=\frac{\kappa t}{2},\qquad a_{\rm L}=\sqrt{\frac{2g}{\kappa}}A_{\rm L},\qquad a_\mu=\sqrt{\frac{2g}{\kappa}}A^+_\mu,\qquad b_\mu=\sqrt{\frac{2g}{\kappa}}A^-_\mu.
\label{eq:SI_simulation_normalization}
\end{equation}
We further define
\begin{equation}
d_n=\frac{2D_n}{\kappa},\qquad T=\frac{2T_{\rm h}}{\kappa},\qquad \beta=\frac{\kappa_{\rm c}}{\kappa},
\label{eq:SI_normalized_resonator_parameters}
\end{equation}
and
\begin{equation}
\widetilde\kappa_{\rm R\leftarrow L}=\frac{2\kappa_{\rm R\leftarrow L}}{\kappa},\qquad \widetilde\kappa_{\rm L\leftarrow R}=\frac{2\kappa_{\rm L\leftarrow R}}{\kappa}.
\label{eq:SI_normalized_couplings}
\end{equation}
For convenience, we introduce the normalized source-model detuning
\begin{equation}
\delta_{\rm fr}\equiv\frac{2\Delta_{\rm fr}}{\kappa}=\frac{2(\omega_{\rm fr}-\omega_0)}{\kappa}=-\xi.
\label{eq:SI_normalized_source_detuning}
\end{equation}
The same accumulated phase is $\Phi_{\rm fr}(\tilde t)=\int_0^{\tilde t}\delta_{\rm fr}(\tilde t')\,\mathrm d\tilde t'$, so $\dot\Phi_{\rm fr}=\delta_{\rm fr}$ even during a frequency scan.
After this normalization, the laser-field equation becomes
\begin{equation}
\frac{\partial a_{\rm L}}{\partial\tilde t}=\mathcal G(1+i\alpha_{\rm g})a_{\rm L}+i e^{i(\Phi_{\rm fr}(\tilde t)+\phi_{\rm s})}\widetilde\kappa_{\rm L\leftarrow R}b_0,
\label{eq:SI_normalized_laser}
\end{equation}
where
\begin{equation}
\mathcal G=\frac{G_N(N-N_0)}{\kappa}
\label{eq:SI_normalized_gain}
\end{equation}
is the real normalized carrier-dependent net gain.

For the present simulations, we focus on the SIL--Kerr dynamics and the dispersion-dependent platicon spectra rather than thermal transient dynamics.
We therefore neglect the thermal resonance shift by setting $T=0$ and retain only the second-order dispersion, such that $d_{n\geq3}=0$.
The normalized forward- and backward-propagating modal equations then reduce to
\begin{align}
\frac{\partial a_\mu}{\partial\tilde t}&=-\left(1+i\frac{d_2}{2}\mu^2\right)a_\mu+i\left(\mathcal F[|a|^2a]_\mu+2a_\mu\sum_{\mu'}|b_{\mu'}|^2\right)+i\beta b_\mu+i e^{i(-\Phi_{\rm fr}(\tilde t)+\phi_{\rm s})}\widetilde\kappa_{\rm R\leftarrow L}a_{\rm L}\delta_{\mu0},
\label{eq:SI_modal_forward_reduced}\\
\frac{\partial b_\mu}{\partial\tilde t}&=-\left(1+i\frac{d_2}{2}\mu^2\right)b_\mu+i\left(\mathcal F[|b|^2b]_\mu+2b_\mu\sum_{\mu'}|a_{\mu'}|^2\right)+i\beta a_\mu.
\label{eq:SI_modal_backward_reduced}
\end{align}
We introduce the angular-domain fields
\begin{equation}
a(\tilde t,\vartheta)=\sum_\mu a_\mu e^{i\mu\vartheta},\qquad b(\tilde t,\vartheta)=\sum_\mu b_\mu e^{i\mu\vartheta},
\label{eq:SI_angular_fields}
\end{equation}
with inverse transformations
\begin{equation}
a_\mu=\frac{1}{2\pi}\int_0^{2\pi}a(\tilde t,\vartheta)e^{-i\mu\vartheta}\,\mathrm d\vartheta,\qquad b_\mu=\frac{1}{2\pi}\int_0^{2\pi}b(\tilde t,\vartheta)e^{-i\mu\vartheta}\,\mathrm d\vartheta.
\label{eq:SI_inverse_angular_fields}
\end{equation}
Here, $\vartheta\in[0,2\pi)$ is the azimuthal angular coordinate in the microresonator.
Under this convention,
\begin{equation}
-\frac{i d_2}{2}\mu^2a_\mu\longleftrightarrow\frac{i d_2}{2}\frac{\partial^2a}{\partial\vartheta^2},
\end{equation}
while the Kerr four-wave-mixing term transforms locally as
\begin{equation}
\mathcal F[|a|^2a]_\mu\longleftrightarrow|a|^2a.
\end{equation}
Parseval's identity gives
\begin{equation}
\sum_\mu|a_\mu|^2=\frac{1}{2\pi}\int_0^{2\pi}|a|^2\,\mathrm d\vartheta,\qquad \sum_\mu|b_\mu|^2=\frac{1}{2\pi}\int_0^{2\pi}|b|^2\,\mathrm d\vartheta.
\label{eq:SI_parseval}
\end{equation}
Thus, the cross-phase-modulation term for either propagation direction depends on the mean intracavity power in the opposite direction.
Only the $\mu=0$ component of the backward field enters the feedback to the semiconductor laser,
\begin{equation}
b_0=\frac{1}{2\pi}\int_0^{2\pi}b(\tilde t,\vartheta)\,\mathrm d\vartheta.
\label{eq:SI_backward_zero_mode}
\end{equation}

We next eliminate the fast laser-amplitude dynamics.
Writing
\begin{equation}
a_{\rm L}=r e^{i\theta_{\rm L}},
\label{eq:SI_laser_polar_form}
\end{equation}
where $r=|a_{\rm L}|$ is the normalized laser-field amplitude and $\theta_{\rm L}$ is its feedback-induced phase, Eq.~\eqref{eq:SI_normalized_laser} gives
\begin{equation}
\frac{\dot r}{r}+i\dot\theta_{\rm L}=\mathcal G(1+i\alpha_{\rm g})+\mathcal H_{\rm fb},
\label{eq:SI_laser_amplitude_phase}
\end{equation}
where a dot denotes differentiation with respect to $\tilde t$ and
\begin{equation}
\mathcal H_{\rm fb}=i e^{i(\Phi_{\rm fr}(\tilde t)+\phi_{\rm s})}\widetilde\kappa_{\rm L\leftarrow R}\frac{b_0}{a_{\rm L}}.
\label{eq:SI_feedback_Q}
\end{equation}
Separating the real and imaginary parts yields
\begin{equation}
\frac{\dot r}{r}=\mathcal G+\operatorname{Re}\mathcal H_{\rm fb},\qquad \dot\theta_{\rm L}=\alpha_{\rm g}\mathcal G+\operatorname{Im}\mathcal H_{\rm fb}.
\label{eq:SI_laser_real_imaginary}
\end{equation}
For laser-amplitude relaxation much faster than the microresonator-field dynamics, we use the adiabatic approximation $\dot r/r\simeq0$, giving
\begin{equation}
\mathcal G=-\operatorname{Re}\mathcal H_{\rm fb}.
\end{equation}
The carrier-dependent gain is thereby eliminated from the laser-phase equation,
\begin{equation}
\dot\theta_{\rm L}=\operatorname{Im}\mathcal H_{\rm fb}-\alpha_{\rm g}\operatorname{Re}\mathcal H_{\rm fb}=\operatorname{Im}\left[(1-i\alpha_{\rm g})\mathcal H_{\rm fb}\right].
\label{eq:SI_laser_phase_reduced}
\end{equation}
We additionally neglect the residual feedback-induced variation of $r$, so that the pump magnitude is treated as constant during the microresonator-field simulation.
Consequently, the carrier equation no longer needs to be integrated explicitly, while the feedback-induced laser-frequency shift is retained through Eq.~\eqref{eq:SI_laser_phase_reduced}.

We now transform the two microresonator fields to the frame of the generated laser by defining
\begin{equation}
a=e^{i\chi}\psi^+,\qquad b=e^{i\chi}\psi^-,
\label{eq:SI_phase_transformation}
\end{equation}
with
\begin{equation}
\chi=\theta_{\rm L}-\Phi_\mathrm{fr}(\tilde t)+\phi_{\rm B},\qquad \phi_{\rm B}=\phi_{\rm s}+\frac{\pi}{2}.
\label{eq:SI_phase_definition}
\end{equation}
The additional phase shift $\pi/2$ absorbs the prefactor $i$ in the laser-to-resonator coupling term.
We define the real normalized pump amplitude as
\begin{equation}
f=\widetilde\kappa_{\rm R\leftarrow L}r.
\label{eq:SI_normalized_pump}
\end{equation}
Using
\begin{equation}
\frac{\partial a}{\partial\tilde t}=e^{i\chi}\left(\frac{\partial\psi^+}{\partial\tilde t}+i\dot\chi\,\psi^+\right),\qquad \frac{\partial b}{\partial\tilde t}=e^{i\chi}\left(\frac{\partial\psi^-}{\partial\tilde t}+i\dot\chi\,\psi^-\right),
\end{equation}
the angular-domain equations become
\begin{align}
\frac{\partial\psi^+}{\partial\tilde t}&=-(1+i\dot\chi)\psi^++\frac{i d_2}{2}\frac{\partial^2\psi^+}{\partial\vartheta^2}+i\left(|\psi^+|^2+2P^-\right)\psi^++i\beta\psi^-+f,
\label{eq:SI_transformed_forward}\\
\frac{\partial\psi^-}{\partial\tilde t}&=-(1+i\dot\chi)\psi^-+\frac{i d_2}{2}\frac{\partial^2\psi^-}{\partial\vartheta^2}+i\left(|\psi^-|^2+2P^+\right)\psi^-+i\beta\psi^+,
\label{eq:SI_transformed_backward}
\end{align}
where
\begin{equation}
P^\pm=\frac{1}{2\pi}\int_0^{2\pi}|\psi^\pm(\tilde t,\vartheta)|^2\,\mathrm d\vartheta.
\label{eq:SI_mean_powers}
\end{equation}
The quantities $P^+$ and $P^-$ are the normalized mean intracavity powers of the forward- and backward-propagating fields, respectively.
The phase transformation does not change the intracavity intensities, so $P^\pm$ are identical to the corresponding mean powers before the transformation.

The backward pump-mode component in the transformed frame is
\begin{equation}
\overline{\psi}^{-}\equiv\frac{1}{2\pi}\int_0^{2\pi}\psi^-(\tilde t,\vartheta)\,\mathrm d\vartheta,
\label{eq:SI_backward_pump_mode}
\end{equation}
where the overbar denotes the azimuthal average rather than complex conjugation.
Because $b_0=e^{i\chi}\overline{\psi}^{-}$, Eqs.~\eqref{eq:SI_feedback_Q}, \eqref{eq:SI_phase_definition}, and \eqref{eq:SI_normalized_pump} give
\begin{equation}
\mathcal H_{\rm fb}=\widetilde\kappa_{\rm L\leftarrow R}\widetilde\kappa_{\rm R\leftarrow L}e^{i2\phi_{\rm B}}\frac{\overline{\psi}^{-}}{f}.
\label{eq:SI_feedback_transformed}
\end{equation}
We therefore define
\begin{equation}
K_0\equiv\widetilde\kappa_{\rm L\leftarrow R}\widetilde\kappa_{\rm R\leftarrow L}=\frac{4\kappa_{\rm ex}T_{\rm L}\sqrt{T_{\rm cL}T_{\rm cR}}}{\kappa^2\tau_{\rm L}}.
\label{eq:SI_K0_source}
\end{equation}
Equivalently, by defining the effective laser coupling rate
\begin{equation}
\kappa_{\rm L}\equiv\frac{T_{\rm L}\sqrt{T_{\rm cL}T_{\rm cR}}}{\tau_{\rm L}},
\end{equation}
Eq.~\eqref{eq:SI_K0_source} becomes
\begin{equation}
K_0=\frac{4\kappa_{\rm ex}\kappa_{\rm L}}{\kappa^2},
\label{eq:SI_K0_consistent}
\end{equation}
which is identical to the definition used in the preceding section.

The phase rate of the transformed frame is
\begin{equation}
\dot\chi=\dot\theta_{\rm L}-\delta_{\rm fr}=\xi+\dot\theta_{\rm L},
\label{eq:SI_generation_detuning_intermediate}
\end{equation}
where $\delta_{\rm fr}=-\xi$ has been used.
Identifying the normalized generated laser--microresonator detuning as
\begin{equation}
\zeta\equiv\dot\chi=\frac{2(\omega_0-\omega)}{\kappa},
\end{equation}
where $\omega$ is the generated laser frequency under self-injection locking, and substituting Eqs.~\eqref{eq:SI_laser_phase_reduced} and \eqref{eq:SI_feedback_transformed}, we obtain
\begin{equation}
\zeta=\xi+K_0\operatorname{Im}\left[(1-i\alpha_{\rm g})e^{i2\phi_{\rm B}}\frac{\overline{\psi}^{-}}{f}\right].
\label{eq:SI_SIL_detuning_reduced}
\end{equation}

The reduced equations used in the numerical simulations are therefore
\begin{align}
\frac{\partial\psi^+}{\partial\tilde t}&=-(1+i\zeta)\psi^++\frac{i d_2}{2}\frac{\partial^2\psi^+}{\partial\vartheta^2}+i\left(|\psi^+|^2+2P^-\right)\psi^++i\beta\psi^-+f,
\label{eq:SI_SIL_forward_reduced}\\
\frac{\partial\psi^-}{\partial\tilde t}&=-(1+i\zeta)\psi^-+\frac{i d_2}{2}\frac{\partial^2\psi^-}{\partial\vartheta^2}+i\left(|\psi^-|^2+2P^+\right)\psi^-+i\beta\psi^+,
\label{eq:SI_SIL_backward_reduced}\\
\zeta&=\xi+K_0\operatorname{Im}\left[(1-i\alpha_{\rm g})e^{i2\phi_{\rm B}}\frac{\overline{\psi}^{-}}{f}\right].
\label{eq:SI_SIL_frequency_feedback}
\end{align}
These equations retain the bidirectional Kerr dynamics, Rayleigh backscattering, and nonlinear SIL frequency feedback, while the carrier density, laser-amplitude dynamics, and thermal variable no longer need to be propagated explicitly.
This reduction substantially decreases the numerical complexity compared with direct integration of the complete Kerr--thermal model.

\begin{table}[h!]
\renewcommand{\tablename}{Supplementary Table}
\centering
\caption{\textbf{Normalized parameters used in the
SIL-platicon simulations}.}
\label{SILsimulation_Parameter}

\footnotesize
\setlength{\tabcolsep}{5pt}
\renewcommand{\arraystretch}{1.3}

\begin{tabular*}{\textwidth}{
    @{\extracolsep{\fill}}|c|c|c|c|c|c|c|
}
\hline
\textbf{Parameter}
& $\phi_{\rm B}$
& $f$
& $\beta$
& $K_0$
& $\alpha_{\rm g}$
& $d_2$
\\
\hline
\textbf{Value}
& 2.74
& 2.93
& 0.99
& 195.19
& $-6.05$
& $-0.036$ or $-0.003$
\\
\hline
\end{tabular*}
\end{table}

\begin{figure*}[t!]
\renewcommand{\figurename}{Supplementary Figure}
\centering
\includegraphics{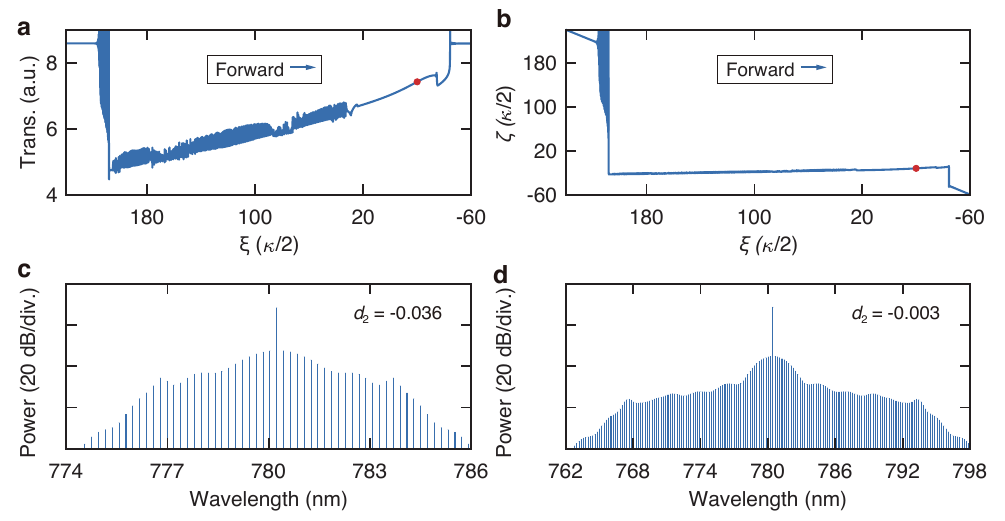}
\caption{
\textbf{Numerical simulation of SIL platicons in the 780-nm band}.
\textbf{a,} Simulated microresonator transmission during forward tuning (decreasing the free-running detuning $\xi$).
\textbf{b,} Simulated generation detuning $\zeta$ during forward tuning.
\textbf{c,} Simulated SIL platicon spectrum for $d_2=-0.036$, corresponding to the stable operating point marked by the red symbols in panels \textbf{a,b}.
\textbf{d,} Simulated SIL platicon spectrum for $d_2=-0.003$.
The spectral span increases substantially as $|d_2|$ approaches zero within the normal-GVD regime.
}
\label{Fig:SILsimulation}
\end{figure*}

We consider the normal-GVD regime, $d_2<0$, in which coherent platicon states can be supported, and use the normalized parameters listed in Supplementary Table~\ref{SILsimulation_Parameter}.
Supplementary Fig.~\ref{Fig:SILsimulation}a shows the simulated microresonator transmission during forward tuning (decreasing the free-running detuning $\xi$).
As the free-running laser frequency approaches the microresonator resonance, the transmission changes sharply as the system enters the SIL regime.
The corresponding generated detuning $\zeta$ is shown in Supplementary Fig.~\ref{Fig:SILsimulation}b.
Within the SIL range, $\zeta$ varies only weakly over a substantially broader scan of the free-running detuning $\xi$, reflecting the frequency stabilization produced by resonant optical feedback.
When the scan exceeds the locking range, the system leaves the SIL branch and the generated detuning changes suddenly.

To investigate the influence of microresonator dispersion on the spectral span of the 780-nm platicon microcomb, we simulate two cases with $d_2=-0.036$ and $d_2=-0.003$.
For $d_2=-0.036$, the platicon spectrum extracted at the stable operating point indicated by the red symbols in Supplementary Fig.~\ref{Fig:SILsimulation}a,b is shown in Supplementary Fig.~\ref{Fig:SILsimulation}c.
When the magnitude of the normal dispersion is reduced to $|d_2|=0.003$, corresponding to $d_2=-0.003$, the simulated spectrum broadens substantially, as shown in Supplementary Fig.~\ref{Fig:SILsimulation}d. 
As $|d_2|$ decreases, the dispersion-induced phase mismatch between microresonator modes is reduced, allowing four-wave mixing to populate modes farther from the pumped resonance and thereby extending the platicon spectrum.
These simulations indicate that dispersion engineering can substantially broaden 780-nm SIL platicon microcombs and provide access to a wider range of wavelengths relevant to atomic-photonic applications.

\newpage
\vspace{0.5cm}
\section{Atom-referenced stabilization of the microcomb}

To frequency-stabilize the microcomb pump line to an atomic reference, we use modulation transfer spectroscopy (MTS) on the $^{85}\mathrm{Rb}$ D$_2$ transition $\left|5^{2}\mathrm{S}_{1/2},F=3\right\rangle\rightarrow\left|5^{2}\mathrm{P}_{3/2},F'=4\right\rangle$.
The pump line is defined as $\mu=0$ and has a wavelength of approximately 780.24 nm. 
We denote the pump angular frequency by $\omega_{\rm p}$ and the atomic transition angular frequency by $\omega_{\rm a}$.
Their angular-frequency detuning is defined as
\begin{equation}
\Delta_{\rm a}=\omega_{\rm p}-\omega_{\rm a}.
\label{eq:MTS_atomic_detuning}
\end{equation}
The electro-optic modulator (EOM) phase-modulates the pump beam at $f_{\rm m}=5.667~\mathrm{MHz}$, corresponding to the angular modulation frequency $\omega_{\rm m}=2\pi f_{\rm m}$.
The complex representation of the phase-modulated pump field is
\begin{equation}
E_{\rm p}(t)=E_{\rm p,0}\exp\left[i\omega_{\rm p}t+i\beta_{\rm m}\sin(\omega_{\rm m}t)\right],
\qquad
\omega_{\rm m}=2\pi f_{\rm m},
\label{eq:MTS_phase_modulation}
\end{equation}
where $E_{\rm p,0}$ is the complex amplitude of the incident pump field and $\beta_{\rm m}$ is the phase-modulation index.

Using the Jacobi--Anger expansion, Eq.~\eqref{eq:MTS_phase_modulation} becomes
\begin{equation}
E_{\rm p}(t)=E_{\rm p,0}\sum_{n=-\infty}^{+\infty}J_n(\beta_{\rm m})e^{i(\omega_{\rm p}+n\omega_{\rm m})t},
\label{eq:MTS_sidebands}
\end{equation}
where $J_n(\beta_{\rm m})$ is the $n$-th-order Bessel function of the first kind.
The phase-modulated field therefore consists of the carrier at $\omega_{\rm p}$ and sidebands separated from the carrier by integer multiples of $\omega_{\rm m}$.

Near the Rb resonance, the phase-modulated pump and the counterpropagating unmodulated probe interact nonlinearly in the atomic vapor~\cite{Shirley:82}.
Degenerate four-wave mixing transfers the pump modulation to the probe, generating optical sidebands in the probe direction.
Beating between adjacent spectral components of the probe produces a photodetector-current component at the modulation frequency~\cite{McCarron:08}.
Retaining the dc term and the component at $\omega_{\rm m}$, the detected photocurrent can be written as
\begin{equation}
i(t)=i_{\rm dc}+2\operatorname{Re}\left[\widetilde{i}_{\rm m}(\Delta_{\rm a})e^{i\omega_{\rm m}t}\right]+\cdots,
\label{eq:MTS_photocurrent}
\end{equation}
where $i_{\rm dc}$ is the dc photocurrent and $\widetilde{i}_{\rm m}(\Delta_{\rm a})$ is the complex amplitude of the photocurrent component at $\omega_{\rm m}$.
The ellipsis denotes higher-harmonic components that are rejected by the subsequent phase-sensitive detection.

The mixer reference is written as $\cos(\omega_{\rm m}t+\phi_{\rm d})$, where $\phi_{\rm d}$ is the demodulation phase.
After mixing and low-pass filtering, the error signal can be expressed as
\begin{equation}
V_{\rm err}(\Delta_{\rm a})=G_{\rm d}\operatorname{Re}\left[\widetilde{i}_{\rm m}(\Delta_{\rm a})e^{-i\phi_{\rm d}}\right]+V_{\rm off},
\label{eq:MTS_demodulation}
\end{equation}
where $G_{\rm d}$ represents the overall conversion gain of the photodetection, mixer, and low-pass electronics, and $V_{\rm off}$ is the residual electronic offset.
The demodulation phase is adjusted to select the dispersive quadrature of the MTS response, and the error-signal offset is adjusted such that the zero crossing corresponds to $\Delta_{\rm a}=0$.
Expanding Eq.~\eqref{eq:MTS_demodulation} to first order around the zero crossing gives
\begin{equation}
V_{\rm err}(\Delta_{\rm a})\simeq S_{\rm MTS}\Delta_{\rm a},
\qquad
S_{\rm MTS}=\left.\frac{\partial V_{\rm err}}{\partial\Delta_{\rm a}}\right|_{\Delta_{\rm a}=0},
\label{eq:MTS_error_signal}
\end{equation}
where $S_{\rm MTS}$ is the local error-signal slope at the zero crossing.
The feedback loop maintains $V_{\rm err}\simeq0$ and thereby stabilizes the pump line near the center of the selected Rb transition.

\begin{figure*}[t!]
\renewcommand{\figurename}{Supplementary Figure}
\centering
\includegraphics{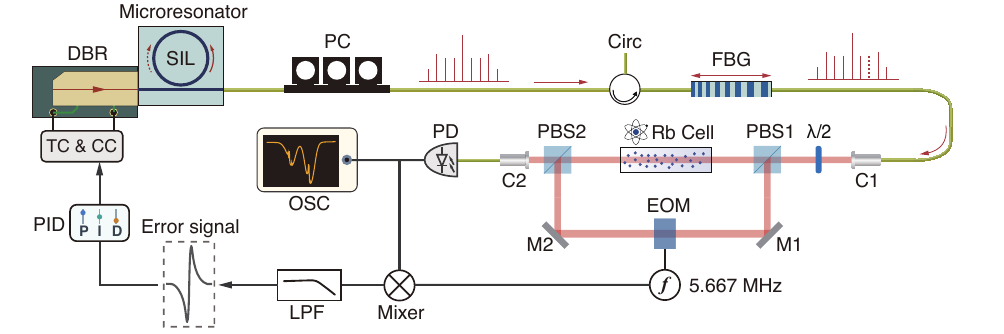}
\caption{
\textbf{Atom-referenced stabilization of the microcomb pump using MTS.}
The SIL microcomb is routed into the Rb cell after passing through a polarization controller (PC), a circulator (Circ), a fiber Bragg grating (FBG), and a collimator (C1).
A half-wave plate ($\lambda/2$) and a polarizing beam splitter (PBS1) split the FBG-filtered microcomb into two branches---a probe beam and a pump beam.
The pump beam is phase-modulated  by an EOM driven by a 5.667~MHz RF signal, and then routed into the Rb cell by PBS2. 
The counterpropagating probe beam is detected by a photodetector (PD).
The PD output is demodulated in a mixer by the same 5.667 MHz driver but with additional phase shift, generating an error signal. 
After passing a low-pass filter (LPF), the error signal is fed back to control the DBR drive current by a proportional--integral--derivative controller (PID). 
M, mirror. 
OSC, oscilloscope. 
TC, temperature controller. 
CC, current controller.
}
\label{Fig:MTS}
\end{figure*}

The experimental setup is shown in Supplementary Fig.~\ref{Fig:MTS}.
The DBR laser is self-injection-locked to the microresonator and generates a microcomb with a pump wavelength of approximately 780.24 nm.
The microcomb output enters the optical circulator and is then directed to a reflective fiber Bragg grating (FBG).
The FBG reflects the $-10$th comb line at 782.25 nm back to the circulator for injection amplification.
All remaining comb lines are transmitted through the FBG and enter the MTS setup.
The MTS response is dominated by the 0th pump line, which has the highest optical power and is the only comb line near resonance with the selected Rb transition.
A half-wave plate and a polarizing beam splitter (PBS1) divide the transmitted comb light into pump and probe beams.
The unmodulated probe beam passes directly through the Rb cell and reaches the photodetector through PBS2.
The pump beam is phase-modulated by the EOM and enters the Rb cell in the direction opposite to the probe beam.
The pump and probe beams overlap while counterpropagating through the Rb cell, thereby generating the modulation-transfer signal.
The photodetector output is sent to both an oscilloscope and a mixer.
The mixer uses the 5.667 MHz EOM-driving signal with additional phase shift $\phi_\mathrm{d}$.
After low-pass filtering (LPF), the error signal is processed by a proportional--integral--derivative (PID) controller and fed back directly to control the DBR drive current.
This feedback stabilizes the pump line near the $^{85}\mathrm{Rb}$ D$_2$ $F=3\rightarrow F'=4$ hyperfine transition.

\vspace{0.5cm}
\section{Injection locking of the FP and DBR lasers}
To characterize the injection-locking behavior of the FP and DBR lasers, we measure their heterodyne beat signals using the setup shown in Supplementary Fig.~\ref{Fig:S7}a.
A mode-hop-free tunable ECDL (Toptica CTL-1550) is amplified by an erbium-doped fiber amplifier (EDFA) and frequency doubled in a CPLN waveguide.
The generated 780-nm laser is divided into injection and reference paths.
The injection laser is directed into the laser diode through an optical circulator, whereas the reference light is frequency shifted by 80 MHz using an acousto-optic modulator (AOM).
The laser-diode output returns through the optical circulator and is then combined with the frequency-shifted reference laser.
The combined laser is directed onto a photodetector, and the resulting beat signal is recorded using an electrical spectrum analyzer (ESA, Rohde \& Schwarz FSW43).

During the measurement, the injected optical frequency is fixed, while the diode drive current is scanned to tune the free-running laser frequency across the injected frequency. 
The measured radio-frequency beat note represents the frequency difference between the laser diode and the frequency-shifted reference laser.
Outside the locking region, the beat frequency varies with the diode current. 
As the current tunes the free-running laser-diode frequency into a proper range, the laser diode locks to the injected laser at the injection frequency and the beat frequency remains fixed at 80 MHz.

Supplementary Fig.~\ref{Fig:S7}b shows the beat-frequency maps obtained by scanning the FP-laser drive current at different injection powers.
The enlarged regions in the insets show injection-locking current intervals of 0.4, 0.2 and 0.1 mA at injection powers of 50, 10 and 5 $\upmu\mathrm{W}$, respectively.
The locking range increases with the injected optical power.
For the DBR laser, the corresponding injection-locking current intervals are 0.8, 0.5 and 0.1 mA at injection powers of 10, 5 and 1 $\upmu\mathrm{W}$, respectively, as shown in Supplementary Fig.~\ref{Fig:S7}c.
At the common injection powers of 10 and 5 $\upmu\mathrm{W}$, the DBR laser exhibits a wider injection-locking current interval than the FP laser under the measured conditions.

\begin{figure*}[t!]
\renewcommand{\figurename}{Supplementary Figure}
\centering
\includegraphics{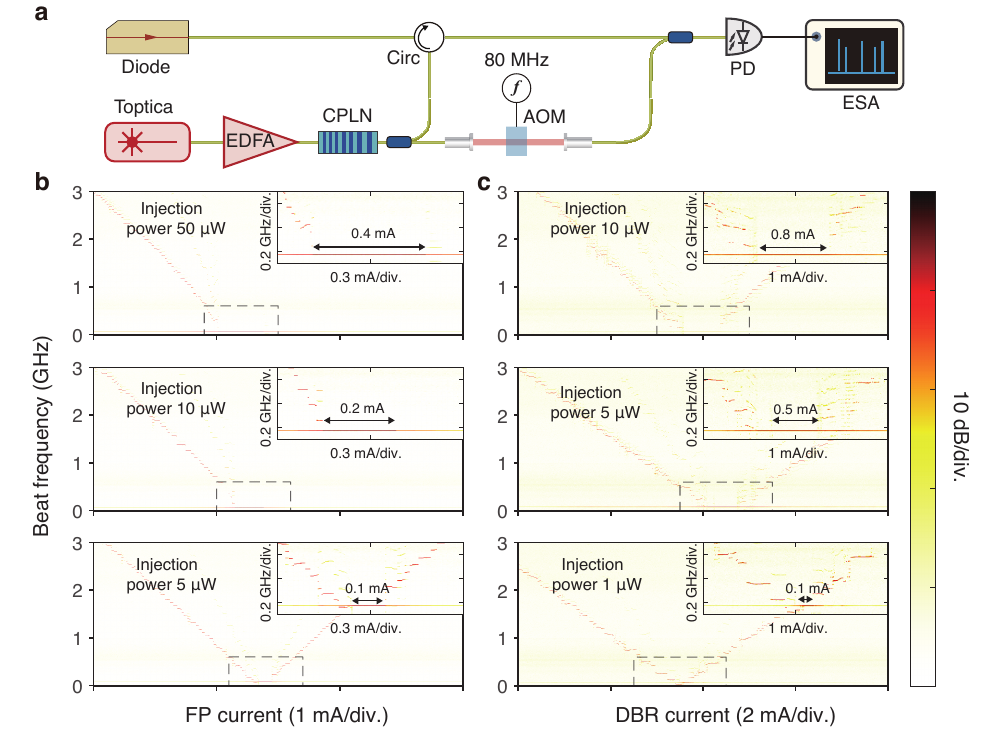}
\caption{
\textbf{Injection locking of the FP and DBR lasers}.
\textbf{a,}
Experimental setup for heterodyne characterization of injection locking. 
The frequency-doubled Toptica laser is divided into injection and reference beams, with the reference frequency shifted by 80 MHz.
The injection beam is routed into a diode laser by a circulator (Circ).
The diode laser beam is routed to beat the reference beam.
EDFA, erbium-doped fiber amplifier.
CPLN, chirped periodically poled lithium niobate.
AOM, acousto-optic modulator.
PD, photodetector.
ESA, electrical spectrum analyzer.
\textbf{b, c,}
Beat-frequency maps recorded while scanning the drive currents of the FP laser (panel \textbf{b}) and DBR laser (panel \textbf{c}) at different injection powers. 
The insets show enlarged views of the injection-locking regions marked by the dashed boxes. 
}
\label{Fig:S7}
\end{figure*}

\vspace{0.5cm}
\section{Frequency noise and Allan deviation measurements}

The frequency noise of the laser under test is measured using a delayed self-heterodyne interferometer (DSHI), whereas the Allan deviation is calculated from a heterodyne beat-frequency time series recorded using a frequency counter.
The experimental setup is shown in Supplementary Fig.~\ref{Fig:S8}.
For the frequency-noise measurement, the laser under test is polarization-adjusted using a fiber polarization controller (PC) and then passed through an optical isolator.
A half-wave plate and a polarizing beam splitter (PBS) divide the light into two optical paths.
One path is frequency-shifted by an AOM driven at 80 MHz, whereas the other passes through a 20-m fiber delay.
The two beams are recombined by a 50:50 fiber coupler and detected by a photodetector.
The resulting radio-frequency beat note is centered at 80 MHz, and its phase fluctuations contain the delayed difference of the laser phase noise.

\begin{figure*}[b!]
\renewcommand{\figurename}{Supplementary Figure}
\centering
\includegraphics{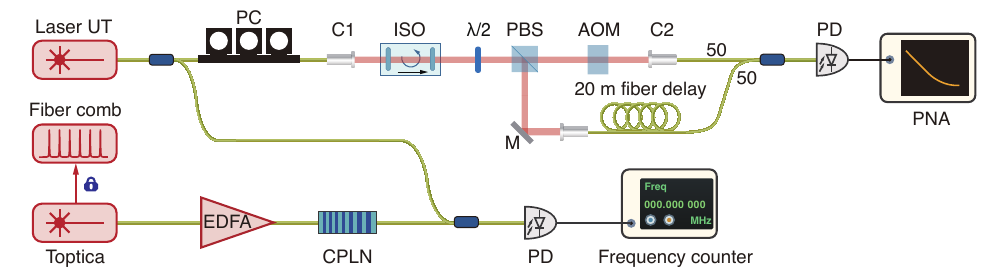}
\caption{
\textbf{Experimental setup for frequency-noise and Allan-deviation measurements.}
The laser under test (Laser UT) operating in 780 nm band is split into two branches---one branch beam is frequency-shifted by an AOM and the other one is delayed by a 20-m fiber, constituting a delayed self-heterodyne interferometer for frequency noise characterization.
For Allan-deviation measurement, a Toptica laser operating in 1560 nm band is frequency-locked on a fiber comb, and then is amplified by an EDFA and frequency doubled by a CPLN to beat the laser under test.
A frequency counter records the beat signal data.
PC, polarization controller.
C1 and C2, collimators.
ISO, optical isolator. 
$\lambda/2$, half-wave plate.
M, mirror.
PBS, polarizing beam splitter.
AOM, acousto-optic modulator. 
PD, photodetector.
PNA, phase-noise analyzer.
EDFA, erbium-doped fiber amplifier.
CPLN, chirped periodically poled lithium niobate.
}
\label{Fig:S8}
\end{figure*}

Let $\phi(t)$ denote the phase fluctuation of the laser under test and $\tau_{\rm d}$ the differential group delay between the two interferometer arms.
Ignoring the constant phase associated with the AOM frequency shift, the fluctuating part of the DSHI beat phase is
\begin{equation}
\Delta\phi(t)=\phi(t)-\phi(t-\tau_{\rm d}).
\label{eq:DSHI_phase_difference}
\end{equation}
In the Fourier domain,
\begin{equation}
\Delta\phi(f_{\rm F})=\left[1-e^{-i2\pi f_{\rm F}\tau_{\rm d}}\right]\phi(f_{\rm F}),
\end{equation}
where $f_{\rm F}$ is the Fourier frequency offset from the 80-MHz beat carrier.
The corresponding single-sided phase-noise power spectral densities (PSDs) therefore satisfy
\begin{equation}
S_{\Delta\phi}(f_{\rm F})=4\sin^2(\pi f_{\rm F}\tau_{\rm d})S_{\phi}(f_{\rm F}),
\label{eq:DSHI_phase_transfer}
\end{equation}
where $S_{\phi}(f_{\rm F})$ is the phase-fluctuation PSD of the test laser and $S_{\Delta\phi}(f_{\rm F})$ is the phase-difference PSD measured from the DSHI beat signal.
The instantaneous laser-frequency fluctuation is related to the phase fluctuation by
\begin{equation}
\delta\nu(t)=\frac{1}{2\pi}\frac{\mathrm d\phi(t)}{\mathrm dt},
\end{equation}
which gives
\begin{equation}
S_{\nu}(f_{\rm F})=f_{\rm F}^{2}S_{\phi}(f_{\rm F}).
\label{eq:DSHI_phase_to_frequency}
\end{equation}
Combining Eqs.~\eqref{eq:DSHI_phase_transfer} and \eqref{eq:DSHI_phase_to_frequency}, the laser frequency-noise PSD is obtained as~\cite{Camatel:08}
\begin{equation}
S_{\nu}(f_{\rm F})=\frac{f_{\rm F}^{2}S_{\Delta\phi}(f_{\rm F})}{4\sin^{2}(\pi f_{\rm F}\tau_{\rm d})}.
\label{eq:frequency_noise_conversion}
\end{equation}
The differential delay is
\begin{equation}
\tau_{\rm d}=\frac{n_{\rm g}\Delta L}{c}\simeq\frac{n_{\rm g}L}{c},
\label{eq:DSHI_delay}
\end{equation}
where $n_{\rm g}$ is the fiber group index, $\Delta L$ is the differential physical fiber length, $L\simeq20~\mathrm{m}$ is the fiber length in the delayed arm when the short-arm contribution is neglected, and $c$ is the speed of light in vacuum.

The phase-noise analyzer (PNA, Rohde \& Schwarz FSWP50) is used to characterize the phase fluctuations of the 80-MHz beat note.
In Eqs.~\eqref{eq:DSHI_phase_transfer} and \eqref{eq:frequency_noise_conversion}, $S_{\Delta\phi}$ denotes the single-sided phase-fluctuation PSD in units of $\mathrm{rad^2/Hz}$.
The intrinsic Lorentzian linewidth is $\Delta\nu_{\mathrm i}=\pi h_0$ for a single-sided white frequency-noise floor $S_\nu=h_0$~\cite{DiDomenico:10}.
The effective linewidth, including low-frequency technical noise, is evaluated using the $\beta$-separation-line method~\cite{DiDomenico:10}.
The $\beta$-separation line is
\begin{equation}
S_{\beta}(f_{\rm F})=\frac{8\ln2}{\pi^2}f_{\rm F},
\label{eq:beta_separation}
\end{equation}
and the frequency-noise area above this line is
\begin{equation}
A=\int_{1/T_0}^{\infty}\mathcal H\left[S_{\nu}(f_{\rm F})-S_{\beta}(f_{\rm F})\right]S_{\nu}(f_{\rm F})\,\mathrm df_{\rm F},
\label{eq:effective_linewidth_area}
\end{equation}
where $\mathcal H$ is the Heaviside function and $T_0$ is the effective observation time setting the low frequency cutoff.
In practical evaluation, the upper integration limit is set by the usable measurement bandwidth.
The corresponding effective linewidth is approximated by
\begin{equation}
\Delta\nu_\beta\simeq\sqrt{8\ln2\,A}.
\label{eq:effective_linewidth}
\end{equation}
Here, $S_{\nu}$ is the single-sided frequency-noise PSD in units of $\mathrm{Hz^2/Hz}$, so that $A$ has units of $\mathrm{Hz^2}$ and $\Delta\nu_\beta$ has units of Hz.
The $1/\pi$ integral-linewidth convention is defined by the reverse phase-noise integral~\cite{Isichenko:24},
\begin{equation}
\int_{\Delta\nu_{\mathrm{int}}}^{\infty}\frac{S_\nu(f_{\rm F})}{f_{\rm F}^2}\,\mathrm df_{\rm F}=\frac{1}{\pi}.
\label{eq:integral_linewidth}
\end{equation}
Both this quantity and the $\beta$-separation linewidth depend on the measured noise spectrum and usable bandwidth.

For the Allan-deviation measurement, the laser under test is heterodyned with a fiber-comb-stabilized Toptica laser (reference laser).
The Toptica laser is amplified by an EDFA and frequency doubled in a CPLN waveguide before being combined with the laser under test on a photodetector.
The resulting beat-frequency time series is recorded using a frequency counter (Tektronix FCA 3100).
We denote the instantaneous frequencies of the laser under test and the reference laser by $\nu_{\rm T}(t)$ and $\nu_{\rm ref}(t)$, respectively.
The measured beat frequency is
\begin{equation}
\nu_{\rm b}(t)=\left|\nu_{\rm T}(t)-\nu_{\rm ref}(t)\right|.
\label{eq:beat_frequency}
\end{equation}
The beat note remains far from zero frequency during the measurement, so the absolute-value operation does not alter the measured small frequency fluctuations.

Let $\nu_{\rm b,i}$ denote a sequence of $N_{\rm s}$ uniformly spaced beat-frequency samples acquired at the sampling rate $F_{\rm s}$.
The basic sampling interval is
\begin{equation}
T_{\rm s}=\frac{1}{F_{\rm s}}.
\end{equation}
For an integer averaging factor $m$, the corresponding averaging time is $T_{\rm a}=mT_{\rm s}$, and the $m$-sample average beginning at the $k$th sample is
\begin{equation}
\overline{\nu}_{\rm b,k}^{(m)}=\frac{1}{m}\sum_{i=0}^{m-1}\nu_{\rm b,k+i}.
\label{eq:beat_frequency_average}
\end{equation}
For uniformly sampled, dead-time-free frequency data, the fully overlapping absolute beat-frequency Allan deviation is~\cite{Riley:08}
\begin{equation}
\sigma_{\nu}(T_{\rm a})=\left[\frac{1}{2(N_{\rm s}-2m+1)}\sum_{k=1}^{N_{\rm s}-2m+1}\left(\overline{\nu}_{\rm b,k+m}^{(m)}-\overline{\nu}_{\rm b,k}^{(m)}\right)^2\right]^{1/2}.
\label{eq:absolute_allan_deviation}
\end{equation}
The corresponding fractional Allan deviation is
\begin{equation}
\sigma_y(T_{\rm a})=\frac{\sigma_{\nu}(T_{\rm a})}{\nu_{\rm opt}},
\qquad
\nu_{\rm opt}=\frac{c}{\lambda_{\rm opt}},
\label{eq:fractional_allan_deviation}
\end{equation}
where $\nu_{\rm opt}$ is the nominal optical frequency of the test laser and $\lambda_{\rm opt}$ is its nominal wavelength.

If the frequency fluctuations of the tested and reference lasers are statistically uncorrelated, their Allan variances contribute approximately as
\begin{equation}
\sigma_{\nu,\rm b}^{2}(T_{\rm a})\simeq\sigma_{\nu,\rm T}^{2}(T_{\rm a})+\sigma_{\nu,\rm ref}^{2}(T_{\rm a}).
\label{eq:allan_variance_sum}
\end{equation}
When $\sigma_{\nu,\rm ref}\ll\sigma_{\nu,\rm T}$ over the averaging-time range considered, the measured beat-frequency Allan deviation therefore represents the frequency instability of the laser under test to a good approximation.
Otherwise, it represents the combined relative-frequency instability of the two lasers.

\vspace{0.5cm}
\section{TOF imaging, optical depth, and atom-number calculation}

At the end of each experimental sequence, the laser beams and the bias magnetic field are abruptly switched off.
The atoms expand freely for a 25-ms time of flight (TOF) and are then repumped to the state $\left|5^2S_{1/2},F=2\right\rangle$ for absorption imaging.
A weak probe pulse resonant with the cycling transition $\left|5^2S_{1/2},F=2\right\rangle\rightarrow\left|5^2P_{3/2},F'=3\right\rangle$ is applied for $50~\upmu\mathrm{s}$.

Three successive camera exposures separated by 300 ms are recorded.
The first image, $A(x,y)$, is acquired with both the atoms and probe light present. 
The second image, $P(x,y)$, records the probe field after the atoms have been removed. 
The third image, $D(x,y)$, records the dark background without probe light.
Representative raw images are shown in Supplementary Fig.~\ref{Fig:S12}a--c.

\begin{figure*}[b!]
\renewcommand{\figurename}{Supplementary Figure}
\centering
\includegraphics{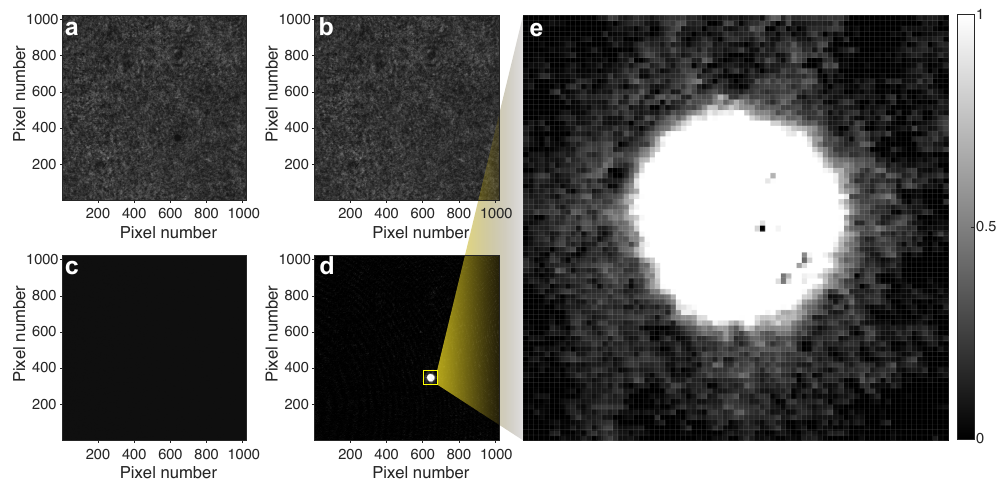}
\caption{
\textbf{Absorption-imaging procedure and optical-depth reconstruction}.
\textbf{a,}
Raw camera image $A(x,y)$ recorded with both the atoms and probe light present.
\textbf{b,}
Reference probe image $P(x,y)$ recorded after the atoms have been removed.
\textbf{c,}
Dark-background image $D(x,y)$ recorded without probe light.
\textbf{d,}
Optical-depth distribution reconstructed from panels \textbf{a--c} using Eq.~\eqref{eq:imaging_OD}.
\textbf{e,}
Enlarged view of the boxed region in \textbf{d} with the mesh guiding the pixels.
}
\label{Fig:S12}
\end{figure*}

After subtracting the dark background, the optical depth is obtained from the Beer--Lambert law~\cite{Beer:1852} as
\begin{equation}
\mathrm{OD}(x,y)= - \ln\left[\frac{A(x,y)-D(x,y)}{P(x,y)-D(x,y)}\right].
\label{eq:imaging_OD}
\end{equation}
The subtraction of $D(x,y)$ removes the background signal, while normalization by $P(x,y)-D(x,y)$ accounts for the spatially nonuniform probe intensity.
The resulting optical-depth distribution is shown in Supplementary Fig.~\ref{Fig:S12}d and e.

The optical depth is related to the two-dimensional atomic column density $n_{\mathrm{2D}}(x,y)$ by
\begin{equation}
\mathrm{OD}(x,y)=\sigma(\Delta_\mathrm{img})n_{\mathrm{2D}}(x,y),
\label{eq:imaging_column_density}
\end{equation}
where $\sigma(\Delta_\mathrm{img})$ is the detuning-dependent scattering cross-section.
For a probe intensity well below the saturation intensity,
\begin{equation}
\sigma(\Delta_\mathrm{img})=\frac{\sigma_0}{1+4(\Delta_\mathrm{img}/\Gamma_\mathrm{img})^2},
\label{eq:imaging_cross_section}
\end{equation}
where $\sigma_0$ is the resonant scattering cross-section, $\Delta_\mathrm{img}$ is the probe detuning, and $\Gamma_\mathrm{img}$ is the natural linewidth.
For equally populated Zeeman sublevels of the $F=2$ ground-state manifold, the Clebsch--Gordan-weighted resonant cross-section is
\begin{equation}
\sigma_0=\frac{7}{15}\frac{3\lambda_\mathrm{img}^2}{2\pi}=1.356\times10^{-9}~\mathrm{cm^2}.
\label{eq:imaging_resonant_cross_section}
\end{equation}
Here, $\lambda_\mathrm{img}\simeq 780.24$ nm is the imaging wavelength.
The factor $7/15$ is the line strength averaged over the Zeeman sublevels.

The total atom number is obtained by integrating the column density over the image plane,
\begin{equation}
N_\mathrm{atom}=\frac{1}{\sigma(\Delta_\mathrm{img})}\iint \mathrm{OD}(x,y)\,dx\,dy.
\label{eq:imaging_atom_number}
\end{equation}
For the discrete pixels, this becomes
\begin{equation}
N_\mathrm{atom}=\frac{A_{\mathrm{pix}}}{\sigma(\Delta_\mathrm{img})}\sum_{i,j}\mathrm{OD}_{ij},
\label{eq:imaging_atom_number_discrete}
\end{equation}
where
\begin{equation}
A_{\mathrm{pix}}=\left(\frac{d_{\mathrm{pix}}}{M}\right)^2
\end{equation}
is the object-plane area corresponding to one camera pixel.
Here, $d_{\mathrm{pix}}=13~\upmu\mathrm{m}$ is the physical camera-pixel size and $M=4.98$ is the calibrated imaging magnification.

\vspace{0.5cm}
\section{Atom loss and coherence dynamics}

To quantify atom loss, ultracold atoms are adiabatically loaded into the Raman lattice, and the total atom number $N_\mathrm{atom}$ is measured as a function of hold time $\tau$.
As shown in Supplementary Fig.~\ref{Fig:S11}a, the atom-number decay is described by a biexponential function
\begin{equation}
N_\mathrm{atom}(\tau)=N_{\mathrm{f}}\exp\left(-\frac{\tau}{\tau_{\mathrm{f}}}\right)+N_{\mathrm{s}}\exp\left(-\frac{\tau}{\tau_{\mathrm{s}}}\right),
\end{equation}
where $N_{\mathrm f}$ and $N_{\mathrm s}$ are the amplitudes of the fast- and slow-decaying components, respectively.
The fit yields decay time constants of $\tau_{\mathrm f}=1.4$ ms and $\tau_{\mathrm s}=53.2$ ms.
The fast-decaying component accounts for approximately 24 \% of the initial atom population, which may be mainly due to shallow traps and lattice defects. 

To characterize the coherence of the 782.25 nm laser in atomic physics, long-term KD diffraction and Rabi oscillation are measured in Supplementary Fig.~\ref{Fig:S11}b, c.
At a lattice depth of $V_{L} = 3.9~E_\mathrm{r}$, the population fraction $P_{\pm1}$ in the first-order diffracted momentum states undergoes more than ten resolved oscillations as the KD pulse duration $\tau$ is varied up to $800.0~\upmu\rm{s}$.
A damped sinusoidal fit yields a time to the first KD population maximum of $27.4~\upmu\rm{s}$ and a coherence lifetime of 1.0 ms.
In the Rabi oscillation, the spin-up population $P_{\uparrow}$ oscillates as the Raman pulse duration $\tau$ is varied.
At a Raman coupling energy of $\Omega = 0.4~E_\mathrm{r}$, a damped-sinusoidal fit yields a $\pi$-pulse duration of 0.3 ms and a coherence time of 1.6 ms. 
The maximum spin-transfer efficiency is approximately 0.8.

\begin{figure*}[t!]
\renewcommand{\figurename}{Supplementary Figure}
\centering
\includegraphics{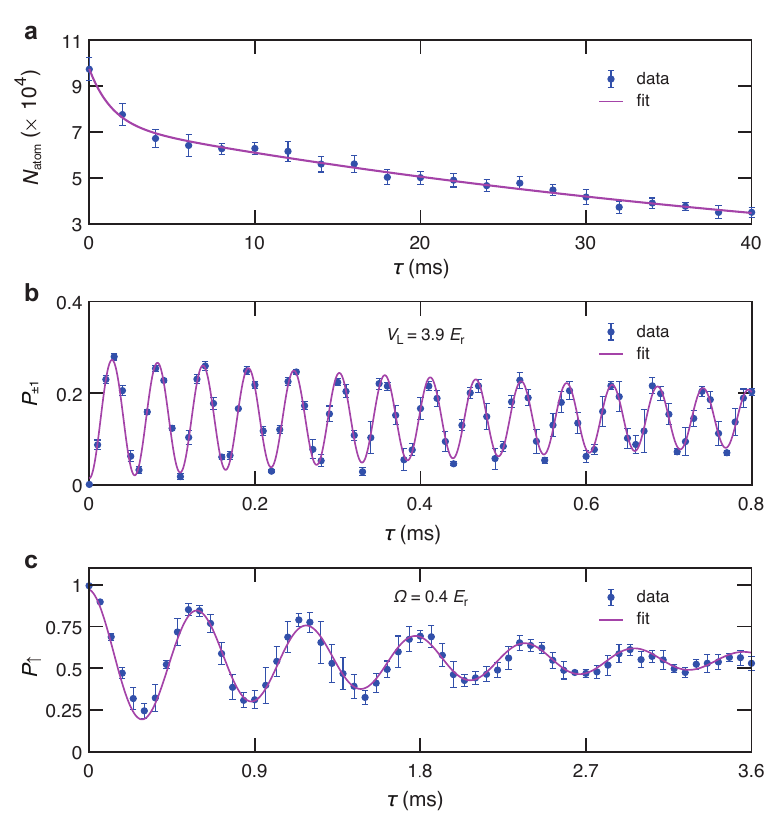}
\caption{
\textbf{Atom loss and coherence dynamics}.
\textbf{a,}
The total atom number $N_\mathrm{atom}$ as a function of hold time $\tau$ in the Raman lattice. The solid curve shows a biexponential fit, yielding fast and slow decay time constants of 1.4 ms and 53.2 ms, respectively.
\textbf{b,}
Population fraction $P_{\pm1}$ in the first-order diffracted momentum states as a function of the KD pulse duration $\tau$ at $V_{\mathrm L}=3.9~E_{\mathrm r}$.
\textbf{c,}
Spin-up population fraction $P_{\uparrow}$ as a function of the Raman pulse duration $\tau$ at a Raman coupling energy of $\Omega=0.4~E_{\mathrm r}$.
}
\label{Fig:S11}
\end{figure*}

\vspace{0.5cm}
\section{Estimation of laser-induced atomic heating}

We estimate the photon-recoil heating induced by the lattice (CW1) and Raman (CW2) fields using the experimentally calibrated lattice depth and Raman coupling energy together with the measured optical powers.
Because the beam waists and local intensities at the atomic cloud cannot be measured directly inside the vacuum cell, we introduce effective correction factors that account for propagation loss, the actual beam waist at the atoms, beam displacement, and aberrations introduced by the glass cell.

At $\lambda_0=782.25$ nm, the scalar dipole potential and spontaneous-scattering rate are written as~\cite{Grimm:00}
\begin{equation}
U=C_{\rm d}I,\qquad \Gamma_{\rm sc}=C_{\rm sc}I,
\end{equation}
with $C_{\rm d}=-1.134\times10^{-34}~\mathrm{J\,m^2\,W^{-1}}$ and $C_{\rm sc}=7.278\times10^{-6}~\mathrm{m^2\,J^{-1}}$ for the $^{87}$Rb D-line parameters. 
The nominal single-pass intensities inferred from the measured powers and nominal $1/e^2$ beam radii are
\begin{equation}
I_{\rm f}^{(\rm nom)}=\frac{2P_{\rm CW1}}{\pi w_{\rm CW1}^2},\qquad I_{\rm CW2}^{(\rm nom)}=\frac{2P_{\rm CW2}}{\pi w_{\rm CW2}^2}.
\end{equation}
For $P_{\rm CW1}=20$ mW, $P_{\rm CW2}=10$ mW, $w_{\rm CW1}=154~\upmu\mathrm{m}$, and $w_{\rm CW2}=300~\upmu\mathrm{m}$, these values are
\begin{equation}
I_{\rm f}^{(\rm nom)}=5.37\times10^5~\mathrm{W\,m^{-2}},\qquad I_{\rm CW2}^{(\rm nom)}=7.07\times10^4~\mathrm{W\,m^{-2}}.
\end{equation}
We relate the actual local intensities at the atoms to these nominal estimates through
\begin{equation}
I_{\rm f}^{(\rm atom)}=\mathcal C_1 I_{\rm f}^{(\rm nom)},\qquad I_{\rm CW2}^{(\rm atom)}=\mathcal C_2 I_{\rm CW2}^{(\rm nom)},
\label{eq:local_intensity_corrections}
\end{equation}
where $\mathcal C_1$ and $\mathcal C_2$ are effective local-intensity correction factors that account for propagation loss, the actual beam waist at the atoms, beam displacement, and glass-cell-induced aberrations.
Because changes in the actual waist can either increase or decrease the local intensity relative to the nominal estimate, $\mathcal C_1$ and $\mathcal C_2$ are not restricted to be smaller than unity.
Let $I_{\rm b}^{(\rm atom)}$ denote the local intensity of the retroreflected CW1 field at the atoms and define the local return ratio as $R=I_{\rm b}^{(\rm atom)}/I_{\rm f}^{(\rm atom)}$.
We further introduce a coherent-overlap factor $0<\eta_{\rm L}\leq1$ that accounts for polarization, spatial-mode, and wavefront mismatch between the forward and retroreflected CW1 fields.
The lattice intensity is then
\begin{equation}
I_{\rm L}(x)=I_{\rm f}^{(\rm atom)}+I_{\rm b}^{(\rm atom)}+2\eta_{\rm L}\sqrt{I_{\rm f}^{(\rm atom)}I_{\rm b}^{(\rm atom)}}\cos(2k_0x-2\phi_{\rm L}),
\end{equation}
and the experimentally calibrated lattice depth is
\begin{equation}
V_{\rm L}=4|C_{\rm d}|\eta_{\rm L}\mathcal C_1 I_{\rm f}^{(\rm nom)}\sqrt{R}.
\label{eq:lattice_depth_general}
\end{equation}
For the Raman transition, the single-pass $\sigma$--$\pi$ coupling is
\begin{equation}
\frac{\Omega}{2}=\eta_{\rm R}|\beta_{\rm R}C_{\rm d}|\sqrt{I_{\rm f}^{(\rm atom)}I_{\rm CW2}^{(\rm atom)}}=\eta_{\rm R}|\beta_{\rm R}C_{\rm d}|\sqrt{\mathcal C_1\mathcal C_2 I_{\rm f}^{(\rm nom)}I_{\rm CW2}^{(\rm nom)}},
\label{eq:raman_general}
\end{equation}
where $0< \eta_{\rm R}\le 1$ accounts for deviations from the ideal Raman polarization and spatial-mode geometry, and $\beta_{\rm R}= 0.2233$ for the $^{87}$Rb D-line parameters.
For the representative experimental values $V_{\rm L}=4.2~E_{\rm r}$ and $\Omega=0.7~E_{\rm r}$, it is convenient to define two experimentally constrained effective factors,
\begin{equation}
q_{\rm L}\equiv\frac{V_{\rm L}}{4|C_{\rm d}|I_{\rm f}^{(\rm nom)}}=\mathcal C_1\eta_{\rm L}\sqrt{R}=0.0429,\qquad
q_{\rm R}\equiv\frac{\Omega}{2|\beta_{\rm R}C_{\rm d}|\sqrt{I_{\rm f}^{(\rm nom)}I_{\rm CW2}^{(\rm nom)}}}=\eta_{\rm R}\sqrt{\mathcal C_1\mathcal C_2}=0.1763.
\label{eq:qL_qR_constraint}
\end{equation}
The quantities $q_{\rm L}$ and $q_{\rm R}$ therefore constrain combinations of the local-intensity corrections and optical-mode overlaps rather than determining $\mathcal C_{1,2}$, $R$, $\eta_{\rm L}$, and $\eta_{\rm R}$ independently.
For atoms localized near an intensity maximum, the harmonic-ground-state approximation gives
\begin{equation}
\mathcal A(s)\equiv\left\langle\cos(2k_0x-2\phi_{\rm L})\right\rangle\simeq\exp\left(-\frac{1}{\sqrt{s}}\right),\qquad s=\frac{V_{\rm L}}{E_{\rm r}}.
\end{equation}
For $s=4.2$, $\mathcal A(s)=0.6139$.
The spatially averaged lattice intensity is therefore
\begin{equation}
\overline I_{\rm L}=I_{\rm f}^{(\rm nom)}\left[\mathcal C_1(1+R)+2q_{\rm L}\mathcal A(s)\right].
\label{eq:lattice_average_general}
\end{equation}
The corresponding lattice-induced spontaneous-scattering rate is
\begin{equation}
\overline{\Gamma}_{\rm sc,L}=C_{\rm sc}I_{\rm f}^{(\rm nom)}\left[\mathcal C_1(1+R)+2q_{\rm L}\mathcal A(s)\right]\simeq\left[0.206+3.907\,\mathcal C_1(1+R)\right]~\mathrm{s^{-1}}.
\label{eq:lattice_scattering_constrained}
\end{equation}
Since $\eta_{\rm L}\leq1$, Eq.~\eqref{eq:qL_qR_constraint} requires $R\geq\left(q_{\rm L}/\mathcal C_1\right)^2$.
For a specified $\mathcal C_1$, the minimum lattice-scattering rate is obtained for $\eta_{\rm L}=1$ and $R=(q_{\rm L}/\mathcal C_1)^2$, giving
\begin{equation}
\overline{\Gamma}_{\rm sc,L}^{(\rm min)}(\mathcal C_1)=C_{\rm sc}\frac{V_{\rm L}}{4|C_{\rm d}|}\left[\frac{\mathcal C_1}{q_{\rm L}}+\frac{q_{\rm L}}{\mathcal C_1}+2\mathcal A(s)\right].
\label{eq:lattice_scattering_minimum}
\end{equation}
The CW2 scattering rate is determined by its actual local intensity,
\begin{equation}
\overline{\Gamma}_{\rm sc,R}=C_{\rm sc}I_{\rm CW2}^{(\rm atom)}=C_{\rm sc}\mathcal C_2I_{\rm CW2}^{(\rm nom)}\simeq0.515\,\mathcal C_2~\mathrm{s^{-1}}.
\label{eq:raman_scattering_corrected}
\end{equation}
Equation~\eqref{eq:qL_qR_constraint} together with $\eta_{\rm R}\leq1$ further requires
\begin{equation}
\mathcal C_1\mathcal C_2\geq q_{\rm R}^2.
\label{eq:raman_constraint}
\end{equation}
The total spontaneous-scattering rate is therefore
\begin{equation}
\overline{\Gamma}_{\rm sc,tot}\simeq0.206+3.907\,\mathcal C_1(1+R)+0.515\,\mathcal C_2~\mathrm{s^{-1}},
\label{eq:total_scattering_corrected}
\end{equation}
with $R$, $\mathcal C_1$, and $\mathcal C_2$ constrained by Eqs.~\eqref{eq:qL_qR_constraint} and \eqref{eq:raman_constraint}.
For fixed $\mathcal C_1$ and $\mathcal C_2$, minimizing over the lattice return ratio gives
\begin{equation}
\overline{\Gamma}_{\rm sc,tot}^{(\rm min)}(\mathcal C_1,\mathcal C_2)=0.206+3.907\left(\mathcal C_1+\frac{q_{\rm L}^2}{\mathcal C_1}\right)+0.515\,\mathcal C_2~\mathrm{s^{-1}}.
\label{eq:total_scattering_minimum_xi}
\end{equation}
Within the recoil-heating model, each spontaneous-scattering event deposits approximately $2E_{\rm r}$ of motional energy~\cite{Grimm:00}.
The corresponding equivalent temperature-conversion rate is
\begin{equation}
\dot T_{\rm eq}=\frac{2E_{\rm r}}{3k_{\rm B}}\overline{\Gamma}_{\rm sc,tot}=0.120\,\overline{\Gamma}_{\rm sc,tot}~\upmu\mathrm{K\,s^{-1}},
\label{eq:recoil_heating_general}
\end{equation}
where $\overline{\Gamma}_{\rm sc,tot}$ is expressed in $\mathrm{s^{-1}}$.
For $\mathcal C_1=\mathcal C_2=1$, the previous nominal-intensity estimate is recovered, with $\overline{\Gamma}_{\rm sc,tot}^{(\rm min)}=4.635~\mathrm{s^{-1}}$ and
\begin{equation}
\dot T_{\rm eq}^{(\rm min)}=0.556~\upmu\mathrm{K\,s^{-1}}.
\end{equation}
This value is therefore conditional on the nominal power- and waist-derived intensities being representative of the local intensities at the atoms.

The quantity $\dot T_{\rm eq}$ is an equivalent energy-conversion rate and should not be directly interpreted as the instantaneous temperature increase of a partially condensed or nonequilibrium Bose gas.
The measured $\sim 53.2$-ms slow decay refers to the total atom number and therefore includes loss mechanisms in addition to photon-recoil heating.
Additional technical heating may arise from mechanical motion of the retroreflection optics, beam-pointing and optical-path-length fluctuations, and intensity or overlap fluctuations of the lattice beams.
Lattice-position noise near a motional resonance and lattice-depth noise near twice the trap frequency can efficiently excite higher vibrational or Bloch-band states~\cite{Savard:97}.
Differential CW1--CW2 frequency and phase noise can also modulate the Raman detuning and coupling phase, while background-gas collisions, and density-dependent inelastic processes may contribute to the observed atom-number decay.


\vspace{0.5cm}
\section{Adiabatic loading and unloading}
\begin{figure*}[b!]
\renewcommand{\figurename}{Supplementary Figure}
\centering
\includegraphics{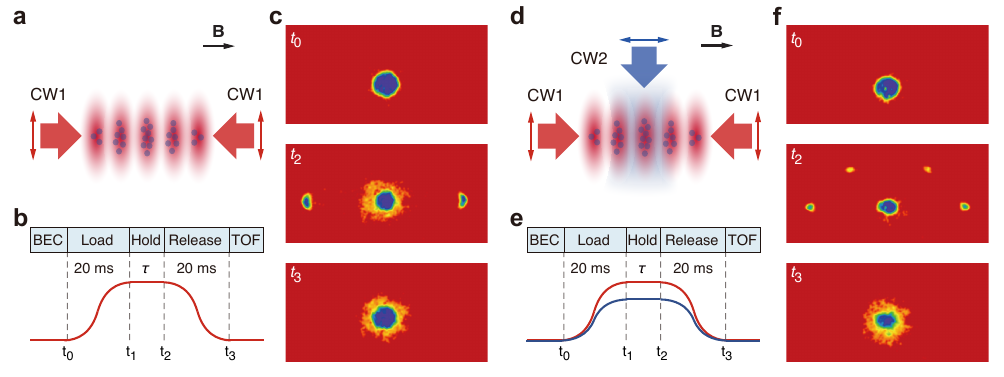}
\caption{
\textbf{Adiabatic loading and unloading}.
\textbf{a, d,}
Optical configurations for the pancake optical lattice (panel \textbf{a}) and Raman lattice (panel \textbf{d}).
\textbf{b, e,}
Time sequence of the lattice depth (red) and Raman coupling energy (blue). 
Both quantities are ramped up and down in 20 ms and held for $\tau=1$ ms.
\textbf{c, f,}
TOF absorption images for the pancake lattice (panel \textbf{c}) and Raman lattice (panel \textbf{f}) configurations. 
Images recorded at $t_0$, $t_2$ and $t_3$ are shown from top to bottom.
}
\label{Fig:S10}
\end{figure*} 
In the configurations of the pancake lattice and the Raman lattice (Supplementary Fig.~\ref{Fig:S10}a and d), the laser intensities are ramped up over 20 ms, held for $\tau = 1$ ms, and ramped down over 20 ms (Supplementary Fig.~\ref{Fig:S10}b and e).
After evolving through various spin-momentum states, the condensate returns to its initial state, albeit with some atomic heating.
TOF absorption images recorded at $t_0$, $t_2$ and $t_3$ are shown from top to bottom in Supplementary Fig.~\ref{Fig:S10}c and f.
Supplementary Fig.~\ref{Fig:S10}c and f correspond to the pancake lattice and the Raman-lattice configurations, respectively.
The images recorded at $t_0$ show the initial condensate.
The lattice depths are ramped to $4.5~E_{\mathrm r}$ and $3.8~E_{\mathrm r}$ for the pancake lattice and Raman-lattice configurations, respectively. 
The Raman coupling energy in the Raman-lattice configuration is $\Omega=0.4\,E_{\mathrm r}$.
At $t_2$, the end of the hold stage, the atoms occupy the momentum components associated with the corresponding optical configuration.
After the optical fields have been ramped down, the images acquired at $t_3$ show the recovery of a dominant zero-momentum condensate component. 
At $t_3$, the hot-atom fractions are 31.3\% and 39.3\% for the pancake lattice and Raman-lattice configurations.
A sufficient number of atoms remain in a condensed state after restoration.
Given that most experiments are conducted on this timescale, the 782.25 nm laser is capable of supporting research in ultracold atomic physics.

\vspace{0.5cm}
\section{Cold-atom control by microcomb without MTS locking}
To examine the role of atomic-referenced frequency stabilization, we repeat the Raman-lattice experiment with the MTS feedback disabled while keeping the optical configuration and experimental sequence unchanged.
Without MTS locking, the frequency noise of the pump laser is several orders of magnitude higher than that in the stabilized case, as shown in Fig.~3f in the main text. 
Besides, disabling the MTS lock also increases fluctuations of the optical frequency, Raman detuning, and optical phase, degrading cold-atom control.
Under this condition, the TOF images exhibit reduced reproducibility, including asymmetric diffraction patterns and, in some experimental shots, substantially distort atomic distributions accompanied by severe atom heating and loss, as shown in Supplementary Fig.~\ref{Fig:woMTS}.

\begin{figure*}[h!]
\renewcommand{\figurename}{Supplementary Figure}
\centering
\includegraphics{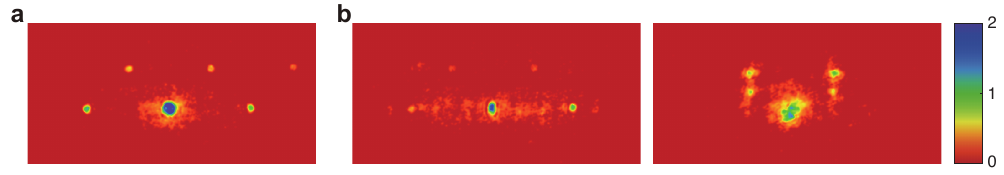}
\caption{
\textbf{Raman lattice driven by microcomb without MTS locking}.
\textbf{a,} 
Representative example of asymmetric atomic diffraction.
\textbf{b,} 
Representative examples of severe atomic heating and loss.
}
\label{Fig:woMTS}
\end{figure*}

\pagebreak
\vspace{0.5cm}
\section{Comparison with commercial lasers for cold atom control}

Cold-atom experiments require optical sources with different combinations of frequency stability, linewidth, output power, wavelength band, and channel count.
Here we compare the microcomb engine with representative ECDL, Ti:sapphire, and frequency-converted fiber lasers commonly used in cold-atom experiments. 
Supplementary Table~\ref{tab:commercial_lasers} summarizes the principal performance metrics including the output power, the linewidth, the stability, the wavelength access, and the source head size.

\begin{table*}[h!]
\renewcommand{\tablename}{Supplementary Table}
\caption{
\textbf{Comparison of the microcomb engine with representative
commercial laser systems.}
The microcomb-engine and TOPTICA DL pro performance metrics
were measured in our laboratory.
The PreciLasers and M Squared specifications are provided by
the manufacturers, except for the frequency-stability data
taken from the literature.
$\Delta\nu_i$ and $\Delta\nu_\beta$ denote the intrinsic and
$\beta$-separation linewidths defined in Supplementary Note~8.
The microcomb-engine $\beta$-separation linewidth uses a 100-Hz lower
integration limit.
The PreciLasers Allan deviations are estimated from the
free-running FL-SF-780-0.5-CW results in Ref.~\cite{YuB:25}.
The M Squared Allan deviations are estimated from the
free-running, frequency-doubled 461-nm output of a
SolsTiS-PSX-1600-F in Ref.~\cite{Couturier:18}. 
Allan deviations quoted at different averaging times are not
direct point-by-point comparisons.
Volumes refer to the indicated modules or laser heads.
IL, injection locked.
EDFA, erbium-doped fiber amplifier.
SHG, second-harmonic generation.
}
\label{tab:commercial_lasers}
\centering
\footnotesize
\setlength{\tabcolsep}{3.0pt}
\renewcommand{\arraystretch}{1.16}

\begin{tabular*}{\textwidth}{
    @{\extracolsep{\fill}}lcccc@{}
}
\hline\hline
\makecell{\textbf{Source /}\\\textbf{architecture}}
& \makecell{\textbf{Output}\\\textbf{power}}
& \makecell{\textbf{Linewidth /}\\\textbf{stability}}
& \makecell{\textbf{Wavelength}\\\textbf{band}}
& \makecell{\textbf{Source head /}\\\textbf{modules}}
\\
\hline

\makecell{Microcomb engine /\\hybrid-integrated}
& \makecell{102 mW\\from the IL channel}
& \makecell{
    $\Delta\nu_i=2.6$ kHz;\\
    $\Delta\nu_\beta=23.6$ kHz;\\
    $\sigma_y(33\,\mathrm{s})=1.6\times10^{-11}$
  }
& 775--785 nm
& \makecell{
    Microcomb $\simeq 0.18$ L;\\
    IL CW$\simeq 0.05$ L
  }
\\[6pt]
\hline
\makecell{TOPTICA DL pro /\\ECDL}
& 65 mW
& \makecell{
    $\Delta\nu_i=411.1$ Hz;\\
    $\Delta\nu_\beta=78.5$ kHz;\\
    $\sigma_y(8\,\mathrm{s})=4.9\times10^{-10}$;\\
    $\sigma_y(66\,\mathrm{s})=2.8\times10^{-9}$
  }
& 770--810 nm
& \makecell{Laser head $\simeq 1.5$ L}
\\[6pt]
\hline
\makecell{
    PreciLasers\\
    FL-SF-780-X-CW /\\
    fiber laser
  }
& \makecell{0.2--20 W,\\depending on model}
& \makecell{
    $<2$ kHz, 100-$\upmu$s integration;\\
    $\sigma_y(10\,\mathrm{s})\approx4\times10^{-10}$~\cite{YuB:25};\\
    $\sigma_y(100\,\mathrm{s})\approx5\times10^{-10}$~\cite{YuB:25}
  }
& \makecell{0.8 nm\\at 780 nm}
& \makecell{
    Pump $\simeq 1.2$ L;\\
    EDFA $\simeq 19.7$ L;\\
    SHG $\simeq 2.2$ L
  }
\\[6pt]
\hline
\makecell{
    M Squared\\
    SolsTiS SRX-1600 /\\
    Ti:sapphire
  }
& \makecell{$>1.6$ W\\around 780 nm}
& \makecell{
    $<50$ kHz, rms linewidth;\\
    $\sigma_y(10\,\mathrm{s})\approx5\times10^{-9}$~\cite{Couturier:18};\\
    $\sigma_y(100\,\mathrm{s})\approx3\times10^{-8}$~\cite{Couturier:18}
  }
& 725--875 nm
& \makecell{
    Pump $+$ laser head\\
    $\simeq 16.6$ L
  }
\\

\hline\hline
\end{tabular*}
\end{table*}

\begin{figure*}[h!]
\renewcommand{\figurename}{Supplementary Figure}
\centering
\includegraphics{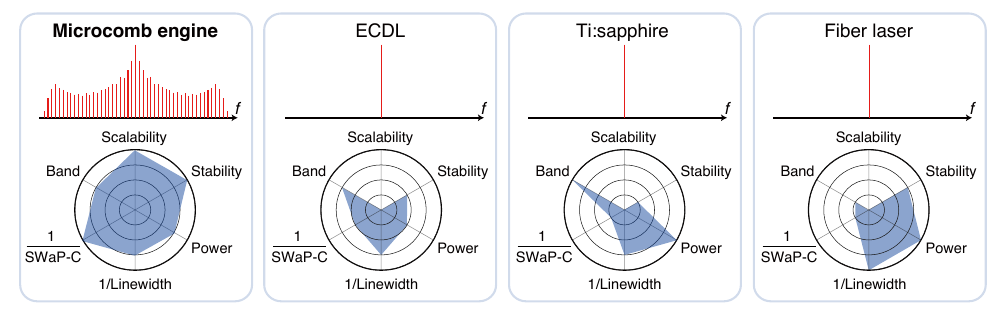}
\caption{\textbf{Qualitative comparison of laser-source architectures for cold-atom experiments.}
The upper sketches distinguish simultaneous multi-frequency emission of our microcomb engine from single-frequency emission of the other commercial CW lasers.
The lower diagrams compare multi-frequency scalability (Scalability), frequency stability (Stability), output power (Power), linewidth (1/Linewidth), SWaP-C (1/SWaP-C), and spectral band coverage (Band) using ordinal levels.}
\label{fig:commercial_laser_compare}
\end{figure*}

Supplementary Fig.~\ref{fig:commercial_laser_compare} summarizes the four source architectures using six qualitative attributes: multi-frequency scalability, frequency stability, output power, linewidth, SWaP-C (size, weight, power and cost), and spectral band coverage.
The radial levels are ordinal and are intended only to visualize architectural trade-offs. 
The polygon area is not a measured figure of merit, but a qualitative comparison of the source architectures.
Our microcomb engine has a unique advantage of multi-frequency scalability and a combined performance advantage over the other laser sources.

\pagebreak
\vspace{1cm}
\section*{Supplementary References}
\bigskip
\bibliographystyle{apsrev4-1}
\bibliography{bibliography}